\documentclass[12pt]{article}
\usepackage[title]{appendix}
\usepackage{rotating, lscape}
\usepackage{float}
\usepackage{graphicx, color}
\usepackage{threeparttable, booktabs, longtable, threeparttablex, tabularx, multirow, pdflscape}

\usepackage[utf8]{inputenc}
\usepackage[T1]{fontenc}
\usepackage{lmodern}

\usepackage{csquotes}
\usepackage[style=apa, backend=biber, natbib=true]{biblatex}
\DeclareNameWrapperFormat{labelname}{\ifcitation{\bibhyperref{#1}}{#1}}
\makeatletter
\renewcommand*{\cite}{\@ifnextchar[{\parencite}{\textcite}}
\makeatother 

\usepackage{amssymb,amsmath,mathrsfs,amsthm, bbm}
\usepackage{accents}

\usepackage{microtype}
\usepackage{courier}
\usepackage{afterpage}
\usepackage{caption}
\usepackage{amsmath}
\usepackage{setspace}
\usepackage[bottom]{footmisc}

\RequirePackage[font=small,format=plain,labelfont=bf,textfont=it]{caption}
\usepackage{chngcntr}
\usepackage{threeparttable}
\usepackage{amsthm}
\usepackage{ragged2e}
\usepackage{subcaption}

\newtheorem{assumption}{Assumption}

\usepackage{textcomp}
\usepackage{tkz-graph}
\usetikzlibrary{shapes.geometric}

\usetikzlibrary{arrows.meta,shapes.geometric}
\tikzstyle{VertexStyle}=[
  shape=ellipse,
  minimum width=7ex,
  minimum height=5ex,
  draw
]
\tikzstyle{EdgeStyle}=[{-Stealth[{length=8pt, width=5pt}]}, thick]

\usepackage{hyperref}
\usepackage{xcolor}
\hypersetup{
  colorlinks   = true, 
  urlcolor     = blue, 
  linkcolor    = blue, 
  citecolor   = blue 
}

\usepackage{amsmath}
\usepackage{graphicx,psfrag,epsf}
\usepackage{enumerate}
\usepackage{amsthm}
\usepackage{pdflscape}

\theoremstyle{plain}
\newtheorem{proposition}{Proposition}

\usepackage{url} 
\AtBeginDocument{}

\newcommand{\blind}{1}

\usepackage[hmargin=1in,vmargin=1in]{geometry}

\allowdisplaybreaks 

\begin{document}

\def\spacingset#1{\renewcommand{\baselinestretch}%
{#1}\small\normalsize} \spacingset{1}


\if1\blind
{
  \title{\bf Difference-in-Differences for Mediation Analysis Using Double Machine Learning}
  \author{Martin Huber\hspace{.2cm}\\
    \small{University of Fribourg, Dept.\ of Economics}\vspace{.2cm}\\
      Sarina Oberh\"{a}nsli \hspace{.2cm}\\
    \small{University of Fribourg, Dept.\ of Economics}}
  \maketitle
} \fi

\if0\blind
{
  \bigskip
  \bigskip
  \bigskip
  
  \medskip
} \fi

\bigskip
\begin{abstract}
We propose a difference-in-differences (DiD) framework with mediation for possibly multivalued discrete or continuous treatments and mediators, aimed at identifying the direct effect of the treatment on the outcome (net of effects operating through the mediator), the indirect effect via the mediator, and the joint effects of treatment and mediator, consistent with the framework of dynamic treatment effects. Identification relies on a conditional parallel trends assumption imposed on the mean potential outcome across treatment and mediator states, or (depending on the causal parameter) additionally on the mean potential outcomes and potential mediator distributions across treatment states. We propose ATET estimators for repeated cross sections and panel data within the double/debiased machine learning framework, which allows for data-driven control of covariates. We investigate the finite-sample performance of the proposed methods in a simulation study and illustrate our approach in an empirical application to the U.S. National Longitudinal Survey of Youth, estimating the direct effect of health care coverage on general health as well as the indirect effect operating through routine checkups.\end{abstract}

\noindent%
{\it Keywords:}  Difference-in-differences, mediation, causal inference, machine learning, parallel trends, controlled direct effect, dynamic treatment effects, natural direct effect, natural indirect effect.\vspace{.6cm}
\begin{spacing}{1} 
{\footnotesize We have benefited from the comments by Timo Schenk and Zachary Shahn as well as conference participants at EuroCIM 2026, IAAE 2026, IPDC 2026, and the Bristol Econometric Study Group Conference 2026.
\noindent
{\it Addresses for correspondence:} Martin Huber, Sarina Oberh\"{a}nsli, University of Fribourg, Bd.\ de P\'{e}rolles 90, 1700 Fribourg, Switzerland; martin.huber@unifr.ch, sarinajoy.oberhaensli@unifr.ch. }
\end{spacing}
\vfill
	\thispagestyle{empty}
	\newpage
	\setcounter{page}{1}

\spacingset{1.45} 
\section{Introduction}
\label{sec:intro}

Difference-in-differences (DiD) \citep{Snow1855, Ashenfelter78} is among the most popular methods for treatment evaluation, that is, for assessing the impact of a treatment on an outcome of interest in observational studies, provided that outcomes can be observed both before and after the introduction of the treatment. In such studies, the outcomes of treated individuals observed before and after treatment typically do not allow researchers to directly infer the treatment effect, due to confounding time trends in the treated individuals’ counterfactual outcomes — those that would have occurred in the absence of treatment. The DiD approach addresses this issue by using a control group that does not receive the treatment and by invoking a parallel trends assumption, which requires that the treated group’s unobserved counterfactual outcomes under non-treatment follow the same mean outcome trend as observed in the control group. This permits identifying the average treatment effect on the treated (ATET).  While the standard DiD setup focuses on the evaluation of the total effect of a treatment, researchers are often also interested in the causal mechanisms through which the effect operates, such as indirect effects transmitted via intermediate variables (so-called mediators) or the treatment's direct effect, net of effects transmitted through mediators. Furthermore, one may wish to study the joint effects of specific treatment and mediator values, which corresponds to dynamic treatment effects obtained by setting the treatment and mediator to specific levels.

Building on this motivation, our study proposes a new DiD framework for causal mediation and dynamic treatment effect analysis that accommodates (possibly multivalued) discrete or continuous treatments and mediators. Depending on the form of the imposed parallel trends assumption, the framework allows the assessment of average direct and indirect treatment effects, as well as joint effects of the treatment and the mediator. Identification of dynamic treatment effects --- such as the joint effect of treatment and mediator, or the (controlled) direct effect of the treatment when fixing the mediator at a specific value --- relies on a conditional parallel trends assumption imposed on mean potential outcomes under no treatment and no mediation, or under specific nonzero treatment or mediator values, conditional on observed covariates. The assumption requires that, conditional on covariates, the mean potential outcomes of the treatment group (characterized by specific values of the treatment and the mediator) would have followed the same average trend as those of the control group not receiving the treatment and mediator (or exposed to a different treatment--mediator combination than the treatment group), had the treatment group been assigned the same treatment and mediator values as the control group.

A further class of causal parameters studied in causal mediation analysis are natural direct and indirect effects \citep{RoGr92, Pearl01}, which are defined in terms of potential mediator values as functions of the treatment, rather than fixed mediator values. For example, the natural indirect effect captures the effect of the treatment on the outcome operating through treatment-induced changes in the mediator. We show that additional assumptions permit identification of such natural effects, by imposing conditional parallel trends in addition to those across treatment–mediator combinations. These are imposed either on mean potential outcomes across treatment groups, or on mean potential outcomes within subgroups defined by the potential mediator under the counterfactual treatment together with the distribution of the mediator across treatment groups.

We propose flexible DiD estimators for both repeated cross sections, where subjects differ across time periods, and panel data, where the same subjects are repeatedly observed over time. The estimators build on the double/debiased machine learning (DML) framework of \cite{Chetal2018} to control for covariates in a data-driven manner, which is particularly advantageous when the number of potential control variables is large relative to the sample size. To this end, we estimate the conditional mean outcome and treatment models using machine learning methods and employ these nuisance estimates as plug-ins in doubly robust (DR) score functions \citep[e.g.,][]{Robins+94, RoRo95}, which we adapt to the DiD setting for estimating direct and indirect effects.

We show that our estimators satisfy the \cite{Neyman1959}-orthogonality condition, implying that they are relatively robust (i.e., first-order insensitive) to estimation errors in the nuisance parameters under suitable regularity conditions, such as approximate sparsity when lasso regression is used for nuisance estimation \citep{Bellonietal2014}. Following \citet{Chetal2018}, we further employ cross-fitting to ensure that nuisance parameters and score functions are not estimated on the same subsamples, thereby mitigating overfitting bias. Furthermore, we present simulation evidence demonstrating favorable finite-sample performance of our estimators for sample sizes in the order of several thousand observations. Finally, we  provide an empirical application that revisits the setting of \cite{farbmacher2022causal}, in which the direct effect of health care coverage on general health is disentangled from the indirect effect operating through routine checkups, using data from the National Longitudinal Survey of Youth 1997 (NLSY97). Even though the point estimates of the total, direct, and indirect effects suggest a health-improving effect, we find no statistically significant evidence of a short-term effect of health care coverage on general health among individuals who obtain coverage, either through routine checkups or through other causal mechanisms.

Our study contributes to a growing literature on the semi- and nonparametric DiD-based identification of causal effects, which avoids potential misspecification errors of classical linear two-way fixed effects models (see, for instance, the discussion in \cite{GoodmanBacon2018} and \cite{cha19}, among others). In contrast to much of the existing work, we consider not only binary but also multivalued (and possibly continuous) treatments, as in \cite{callaway2024difference} and \cite{dechaisemartin2023differenceindifferences}, who, however, do not focus on mediation analysis or on controlling for observed covariates. In this paper, we invoke a conditional parallel trends assumption within a semiparametric DiD framework, as considered in \cite{abadie2005} for ATET estimation with binary treatments.

We study identification in both panel data and repeated cross-section settings. For the latter, our approach can accommodate covariate distributions that vary over time within treatment groups, whereas many existing methods require them to be time-invariant \citep{Hong2013}, including the DiD estimators of \cite{abadie2005} (inverse probability weighting), \cite{SantAnnaZhao2018} (DR estimation), and \cite{Chang2020} (DML). A notable exception is the DML approach of \cite{Zimmert2018}, which, however, focuses on the ATET of binary treatments rather than on mediation analysis. Further related work includes \cite{zhang2025}, who considers DML-based DiD estimation of the ATET for continuous versus no treatment, and \cite{haddad2024}, who extend the framework to the evaluation of time-varying treatments and comparisons of strictly positive treatment doses (as also considered by \cite{dechaisemartin2023differenceindifferences}, though without covariate adjustment). In contrast, we consider (potentially non-binary) treatments and mediators to assess direct and indirect effects on the treated, rather than the ATET alone.

Our paper is also related to recent work on DiD methods for mediation analysis. \cite{Deuchertetal2019} consider a binary, random treatment and a binary, nonrandom mediator to identify direct and indirect effects in specific subpopulations defined by potential mediator states, based on parallel trends assumptions between these subpopulations. Their results show that natural direct and indirect effects for the total or treated population require specific parallel trends assumptions to hold across multiple subpopulations. In this paper, we impose parallel trends assumptions strong enough to identify natural direct and indirect effects among the treated, but conditional on covariates, for which we adjust in a data-adaptive manner using machine learning, and allow for non-binary treatments and mediators. \cite{Schenk2024} assumes both the treatment and mediator to be nonrandom and discusses the identification of natural direct and indirect effects among the treated based on parallel trends assumptions that hold only across some subpopulations defined by potential mediators. This, however, requires additional restrictions, such as monotonicity of the mediator in treatment. By imposing stronger parallel trends assumptions across groups defined by treatment and potential mediator values, we do not require monotonicity, which also makes identification feasible for non-binary treatments and mediators.

More recently, \cite{deng2026did} identify natural direct and indirect effects among the treated for a binary nonrandom treatment in two-period panel data. They impose parallel trends across treatment groups conditional on the potential mediator under non-treatment, together with a sequential ignorability assumption that rules out unobserved confounders jointly affecting the treatment and the mediator. In contrast, we impose parallel trends across groups defined by observed treatment and mediator values. We combine them either with parallel trends across treatments, or with parallel trends across treatment groups conditional on the potential mediator under the counterfactual treatment, similar to \cite{deng2026did}. In the latter case, we replace sequential ignorability by parallel trends in the mediator distribution. \cite{ban2026did} focus on a binary nonrandom treatment and a time-varying mediator in panel data. Ruling out time-varying unobserved confounding of the mediator and the outcome, they identify the natural direct effect among the treated by DiD comparisons conditional on pre- and post-treatment mediator values. Identification of the indirect effect requires additional assumptions, such as stability of the average mediator effect over time. Since our parallel trends are imposed across groups defined by post-treatment mediator values, the mediator need not be observed prior to treatment. \cite{andersen2026} also consider two-period panel data and target natural effects among the treated, but rely on a linear structural model, under which the natural indirect effect has a product representation. Unlike these three studies, we allow for non-binary treatments. \cite{hsia2025mediation} target controlled direct as well as natural direct and indirect effects in the total population, which requires parallel trends assumptions to hold across treatment groups at every treatment level. For continuous treatments and mediators, they rely on linear two-way fixed effects models without treatment--mediator interactions, and they do not condition the parallel trends on covariates. In contrast, we allow for such interactions and impose conditional parallel trends. Unlike all of these studies, we further consider repeated cross sections and use DML for data-adaptive covariate adjustment.

Finally, our paper is related to the literature on dynamic treatment effects \citep[see, e.g.,][]{Ro86, RoHeBr00}, of which controlled direct effects are a special case. Within this literature, DR identification \citep{Tranetal2019} and DML estimation \citep{LewisSyrgkanis2020, bodory2022evaluating, bradic2024high} have been studied under selection-on-observables assumptions, while our paper focuses on DiD methods that invoke parallel trends assumptions. Among DiD approaches, \cite{blackwell2022difference} consider controlled direct effects under a random treatment and a nonrandom mediator, whereas our framework allows both to be nonrandom. \citet{shahn2022structural} propose a DR and DML approach for dynamic treatment effects, including controlled direct effects, based on a different DR score than the one considered here. They focus on parametric structural outcome models to assess potentially heterogeneous treatment effects over time. \citet{renson2023identifying} propose a DR approach based on targeted maximum likelihood estimation \citep{VanderLaanRubin2006}. Both \citet{shahn2022structural} and \citet{renson2023identifying} consider panel data only and focus on dynamic treatment effects, whereas we also cover repeated cross sections as well as natural direct and indirect effects.

The remainder of this paper is organized as follows. Section~\ref{sec:id}  discusses the DiD-based identification of dynamic treatment effects and controlled direct effects in  repeated cross-sectional data. Section~\ref{sec:id2} focuses on the identification of natural direct and indirect effects in causal mediation analysis in repeated cross-sectional data. Section~\ref{paneldata} considers identification in panel data. Section~\ref{sec:meth} proposes estimators based on DML with cross-fitting. Section~\ref{sim} provides a simulation study for the repeated cross-section and panel data cases.  Section~\ref{appl} presents an empirical application that decomposes the effect of health care coverage on general health into a direct effect and an indirect effect operating through routine checkups. Section~\ref{sec:conc} concludes.

\section{Identification of dynamic treatment and controlled direct effects in repeated cross sections}\label{sec:id}

This section discusses DiD-based identification of dynamic treatment effects and controlled direct effects in repeated cross-sectional data. We consider a discretely distributed treatment and mediator, denoted by $D$ and $M$, respectively. Furthermore, let $T$ denote the time period, where $T=0$ is the baseline period prior to treatment and mediator assignment, while $T=1$ is a post-treatment and post-mediator period in which the effect on the outcome is measured. We use capital letters to refer to random variables and lowercase letters for specific realizations of these variables. $Y_t$ denotes the observed outcome in period $T=t$, while $Y_t(d,m)$ represents the potential outcome in period $t$ when hypothetically setting the treatment to $D=d$ and the mediator to $M=m$; see, e.g., \cite{Neyman23} and \cite{Rubin74} for more discussion on the potential outcome framework. Finally, we denote by $X$ observed covariates that may serve as control variables. 

Throughout the paper, we impose the stable unit treatment value assumption (SUTVA), implying that a subject's potential outcomes are not affected by the treatment or mediators of others; see, for instance, the discussion in \cite{Rubin80} and \cite{Cox58}. In repeated cross sections, we assume that, conditional on $D$ and $X$, the joint distribution of potential outcomes and potential mediators does not depend on $T$. This allows us to state the parallel trends and no-anticipation assumptions below without conditioning on $T$ and implies that, conditional on $D$, $M$, and $X$, the distribution of potential outcomes does not depend on $T$ either. The distributions of the treatment and covariates may nevertheless change across periods, as in \cite{Zimmert2018}.



In our discussion, we refer to the treatment group as those subjects receiving a specific treatment value $D=d$ and mediator value $M=m$. We consider the identification of the average treatment effect on the treated (ATET) for this treatment–mediator combination, which is defined by comparing the observed average outcomes in the treatment group to the average potential outcomes for the same group under counterfactual treatment and mediator values $d'$ and $m'$. Formally, the ATET of treatment and mediator states $D=d, M=m$ versus $D=d', M=m'$ in the treatment group with $D=d, M=m$ in the post-treatment period $T=1$ is given by
\begin{align}\label{ATETdef}
\Delta_{d,m}(d,m,d',m') = E[Y_1(d,m) - Y_1(d',m') \mid D=d, M=m, T=1].
\end{align}
For instance, for a binary treatment and mediator and for $d=m=1$, $d'=m'=0$, the ATET $\Delta_{1,1}(1,1,0,0) = E[Y_1(1,1) - Y_1(0,0) \mid D=1, M=1, T=1]$ corresponds to the average effect of receiving both the treatment and the mediator versus receiving neither, among the group that actually receives the treatment and the mediator and is observed in the post-treatment period. More generally, for $d \neq d'$ and $m \neq m'$, the effect in \eqref{ATETdef} corresponds to the joint effect of the treatment and mediator on the outcome among those with $D=d$ and $M=m$. This framework permits the consideration of dynamic treatment effects, i.e., the effects of different sequences of treatments and mediators, as, for example, in \cite*{Ro86} and \cite*{RoHeBr00} (which rely on sequential selection-on-observables assumptions, whereas our study relies on parallel trends assumptions). In contrast, if we fix the mediator value at $m=m'$, the resulting ATET corresponds to the controlled direct effect; see, e.g., \cite{Pearl01}. For instance, $\Delta_{1,0}(1,0,0,0) = E[Y_1(1,0) - Y_1(0,0) \mid D=1, M=0, T=1]$ represents the controlled direct effect when fixing the mediator at $m=m'=0$.

For identification of the ATET, we impose the following assumptions related to \cite{Lechner2010} for binary treatments, but now imposed with respect to a combination of discrete treatment and mediator variables:

\begin{assumption} {\bf (Conditional parallel trends across treatment--mediator combinations):}\label{ass1}\\
$E[Y_1(d',m')-Y_0(d',m')|D=d,M=m,X]=E[Y_1(d',m')-Y_0(d',m')|D=d',M=m',X]$.
\end{assumption}

Assumption \ref{ass1} imposes parallel trends in the mean potential outcome under the treatment–mediator combination $D=d', M=m'$ across the treatment group with $D=d, M=m$ on the one hand and the control group with $D=d', M=m'$ on the other hand, conditional on covariates $X$. For binary treatment and mediator variables with $d=m=1$ and $d'=m'=0$, this implies $E[Y_1(0,0)-Y_0(0,0)|D=1,M(1)=1,X]=E[Y_1(0,0)-Y_0(0,0)|D=0,M(0)=0,X]$, where we have used the fact that, by the observational rule, $M = M(1)$ conditional on $D=1$, while $M = M(0)$ conditional on $D=0$. Therefore, the parallel trends assumption is imposed not only across groups with treatment values $D=1$ and $D=0$, but also across groups defined by potential mediators under treatment ($M(1)=1$) and under non-treatment ($M(0)=0$). It is worth emphasizing that these are generally distinct subgroups in terms of how the mediator responds to treatment, an issue that fits into the causal framework of principal stratification, see \cite{frangakis2002principal}, and is closely related to compliance in instrumental variable contexts, see \cite{Angrist+96}: The subpopulation with $M(1)=1$ consists of subjects complying with the treatment in the sense that they take the mediator as a result of being treated, characterized by $M(0)=0, M(1)=1$, as well as subjects who always take the mediator, characterized by $M(0)=M(1)=1$. In contrast, the subpopulation with $M(0)=0$ includes complying subjects ($M(0)=0, M(1)=1$) and subjects who never take the mediator ($M(0)=M(1)=0$). 

We note that this specific type of parallel trends assumption, $E[Y_1(0,0)-Y_0(0,0)|D=1,M(1)=1,X]=E[Y_1(0,0)-Y_0(0,0)|D=0,M(0)=0,X]$, is commonly imposed in the DiD literature on multiple treatment periods, including staggered treatment adoption across groups. See, for instance, the discussions in \cite*{borusyak2024revisiting}, \cite*{CallawaySantAnna2018}, \cite*{GoodmanBacon2018}, \cite*{cha19}, and \cite*{sun2021estimating}. 
Indeed, the multiple-treatment-period setup can be interpreted through the lens of our mediation framework by defining $D$ as treatment adoption in an earlier period and $M$ as treatment adoption in a later period, while typically allowing for additional treatment periods beyond these two. See also \cite{shahn2022structural} for a discussion of staggered and non-staggered DiD designs through the lens of dynamic treatment effects.

For this reason, Assumption \ref{ass1} implicitly imposes parallel trends across (mediator) always- and never-takers. Depending on the context and the values of $d, d', m, m'$ considered, our parallel trends assumptions are more stringent than those considered for causal mediation in \cite{Deuchertetal2019} under random treatment and nonrandom mediator, or in \cite{Schenk2024} under nonrandom treatment and mediator, who tailor parallel trends assumptions to specific subpopulations defined by potential mediator states as a function of treatment. Consequently, identification of particular direct and indirect effects in \cite{Deuchertetal2019} and \cite{Schenk2024} may require additional assumptions, such as monotonicity of the mediator in treatment, e.g., $\Pr(M(1)-M(0) \geq 0 \mid D = 1) = 1$.
Such monotonicity is not imposed here, due to the stronger parallel trends assumption across subgroups defined by potential mediators (and treatments). This parallel trends assumption imposes restrictions on the selection bias arising from unobserved confounders that jointly affect the treatment/mediator and the outcome, which can cause treatment and control groups defined by different treatment–mediator combinations, $D=d, M=m$ and $D=d', M=m'$, to have different average potential outcomes under control, $Y_t(d', m')$. Specifically, this selection bias must be constant across time periods $t=0$ and $t=1$, conditional on $X$.

\begin{assumption} \clubpenalty=10000 \widowpenalty=10000 {\bf (No anticipation of effect on outcome):}\label{ass2}\\
$ E[Y_0(d,m)|D=d,M=m,X]-E[Y_0(d',m')|D=d,M=m,X]=0$.
\end{assumption}

Assumption \ref{ass2} requires that, conditional on $X$, the ATET is zero in the baseline period. This rules out, on average, any anticipation effects among treated units in period $T=0$ with respect to the treatment and mediator to come. In other words, since the treatment and mediator have not yet been realized, they cannot, on average, induce behavioral adjustments in the treated group that affect pre-treatment outcomes in anticipation of future treatment or mediator values.

\begin{assumption}{\bf (Common support):}\label{ass3}\\
$\Pr(D=d, M=m, T=1|X, (D,M,T)\in\{ (d^*,m^*,t^*),(d,m,1)\})<1$  for $(d^*,m^*,t^*)\in \{(d,m,0),(d',m',1),(d',m',0)\}$. 
\end{assumption}

Assumption \ref{ass3} requires that, for any subject with $D=d,M=m$ in the post-treatment period $T=1$, there exist, in terms of $X$, comparable subjects with $D=d,M=m$ in the baseline period, with $D=d',M=m'$ in the post-treatment period, and with $D=d',M=m'$ in the baseline period. In other words, for every combination of covariates observed among treated and mediated units in the post-treatment period, there must exist comparable observations in all other three treatment/mediator–time cells. This common support condition ensures that the counterfactual outcome distributions for the target population with $D=d,M=m$ in the post-treatment period can be recovered from observable data in the other cells.

\begin{assumption} {\bf (Exogenous covariates):}\label{ass4}\\
$X(d^*,m^*)=X$ for all $d^*,m^*$ in the support of $D,M$.
\end{assumption}

Assumption \ref{ass4} imposes that the covariates are not causally affected by the treatment or the mediator. This can be a delicate restriction if the covariates include post-treatment variables (as is often the case in repeated cross sections); see, for instance, the discussion in \cite{caetano2022difference}. Controlling for post-treatment $X$ generally introduces bias if $X$ is affected by $D$ and/or $M$, and $X$ either affects $Y$, is correlated with unobservables affecting $Y$, or both. Assumption \ref{ass4} rules out such bias by requiring that potential covariates are not a function of the treatment or mediator, but this assumption should be carefully scrutinized in practice whenever one conditions on post-treatment covariates.

Under these assumptions, the mean potential outcome under non-treatment $D=d'$ and mediator state $M=m'$ is identified in the post-treatment period among the group receiving treatment doses $D=d$ and $M=m$, conditional on $X$:
\begin{align}\label{DiDobs}
&E[Y_1(d',m')|D=d,M=m,X]\notag\\ 
=&E[Y_0(d',m')|D=d,M=m,X]+E[Y_1(d',m')|D=d,M=m,X]-E[Y_0(d',m')|D=d,M=m,X]\notag\\
=&E[Y_0(d,m)|D=d,M=m,X]+E[Y_1(d',m')|D=d,M=m,X]-E[Y_0(d',m')|D=d,M=m,X]\notag\\
=&E[Y_0(d,m)|D=d,M=m,X]+E[Y_1(d',m')|D=d',M=m',X]\notag\\
-&E[Y_0(d',m')|D=d',M=m',X]\notag\\
=&E[Y_0|D=d,M=m,X]+E[Y_1|D=d',M=m',X]-E[Y_0|D=d',M=m',X].
\end{align}
The first equality in \eqref{DiDobs} follows from subtracting and adding $E[Y_0(d',m')|D=d,M=m,X]$ and the second from Assumption \ref{ass2}. The third equality follows from Assumption \ref{ass1}. The fourth equality follows from the fact that $Y_t=Y_t(d,m)$ conditional on $D=d$ and $M=m$, which is known as the observational rule. Finally, note that by Assumption~\ref{ass4}, $X$ is not causally affected by either $D$ or $M$, implying that conditioning on $X$ does not block any causal effect of $D$ or $M$ on $Y$.

It follows that $E[Y_1(d',m')|D=d,M=m,T=1]$ can be obtained by averaging the expression in the last line of \eqref{DiDobs} over $X$ conditional on $D=d, M=m$ in the post-treatment period $T=1$. This step requires satisfaction of Assumption \ref{ass3} (common support):
\begin{align}\label{DiDobs2}
&E[Y_1(d',m')|D=d,M=m,T=1]=E[E[Y_0|D=d,M=m,X]|D=d,M=m, T=1]\notag\\
&+\{E[E[Y_1|D=d',M=m',X]-E[Y_0|D=d',M=m',X]|D=d,M=m, T=1]\}
\end{align}
Furthermore, we have by the observational rule that 
\begin{align}\label{obsrule}
E[Y_1(d,m)|D=d,M=m,T=1]=E[Y_1|D=d,M=m].
\end{align}

For notational convenience, we henceforth denote the conditional mean outcome by $\mu_{d,m}(t,X)=E[Y_t \mid D=d,M=m,X]$. Furthermore, let $\Pi_{d,m,t}=\Pr(D=d,M=m,T=t)$ denote the joint probability of treatment, mediator, and period, and $\rho_{d,m,t}(X)=\Pr(D=d,M=m,T=t\mid X)$ the corresponding conditional probability given $X$, also known as propensity score. Let $I{\cdot}$ denote the indicator function, which equals one if its argument is satisfied and zero otherwise. For the moment, assume that the treatment $D$ and mediator $M$ are discrete. Taking the difference between \eqref{obsrule} and the average of \eqref{DiDobs2} then identifies the ATET of treatment and mediator states $D=d, M=m$ versus $D=d', M=m'$ in the group with $D=d, M=m$ in the post-treatment period, as given in equation \eqref{ATETdef}:
\begin{align}\label{effect}
&\Delta_{d,m}(d,m,d',m')=E[Y_1|D=d,M=m]\notag\\
&-E[\mu_{d,m}(0,X)+\mu_{d',m'}(1,X)-\mu_{d',m'}(0,X)|D=d,M=m, T=1]\\
&=E\left[  \frac{I\{D=d\}\cdot I\{M=m\}\cdot T \cdot [Y_T-\{\mu_{d,m}(0,X)+\mu_{d',m'}(1,X)-\mu_{d',m'}(0,X)\}]}{\Pi_{d,m,1}}\right].\notag
\end{align}

As a methodological contribution, we propose an alternative, doubly robust (DR) expression for the ATET which is numerically equivalent to  \eqref{effect} under our identifying assumptions and, in contrast to \eqref{effect}, satisfies \cite{Neyman1959}-orthogonality. This property makes it particularly attractive for combination with machine learning methods for the estimation of the conditional mean outcomes and propensity scores. The DR identification result is provided in Proposition \ref{prop:DRidentification}. 
\begin{proposition}\label{prop:DRidentification}
Under Assumptions \ref{ass1}--\ref{ass4}, the ATET is identified by the following expression, which satisfies Neyman orthogonality: 
\begin{align}\label{DiDidentDR}
&\Delta_{d,m}(d,m,d',m')\notag\\
&=E\left[  \frac{I\{D=d\}\cdot I\{M=m\}\cdot T \cdot [Y_T-\{\mu_{d,m}(0,X)+\mu_{d',m'}(1,X)-\mu_{d',m'}(0,X)\}]}{\Pi_{d,m,1}}\right.\notag\\
&- \frac{(Y_T-\mu_{D,M}(T,X))\cdot \rho_{d,m,1}(X)}{\Pi_{d,m,1}} \cdot \left\{\frac{I\{D=d\}\cdot I\{M=m\}\cdot (1-T)}{ \rho_{d,m,0}(X)}\right.\notag\\
&+ \left.\left.\frac{I\{D=d'\}\cdot I\{M=m'\}\cdot  T}{ \rho_{d',m',1}(X)} - \frac{I\{D=d'\}\cdot I\{M=m'\}\cdot (1-T)}{ \rho_{d',m',0}(X)} \right\}   \right].
\end{align}
\end{proposition}
\noindent Appendix \ref{Neymanrepeated} provides formal proofs that equation \eqref{DiDidentDR} identifies the ATET and satisfies Neyman orthogonality.

Strengthening Assumptions \ref{ass1}--\ref{ass4} to not only hold for the population with $D=d, M=m$, but also in analogous manner for the population with $D=d',M=m'$ permits identifying the average treatment effect (ATE) in the total population in the post-treatment period, $\Delta(d,m,d',m')=E[Y_1(d,m)-Y_1(d',m')|T=1]$. Note, however, that the symmetric assumptions required for identification imply that average effects are homogeneous across both groups conditional on $X$. Appendix \ref{ATEdef} provides the corresponding DR identification results for repeated cross sections and panel data.

\section{Identification of natural direct and indirect effects in repeated cross sections}\label{sec:id2}

While the last section focused on the identification of controlled direct effects as well as  dynamic treatment effects, this section discusses the disentangling of the total ATET into natural direct and indirect effects defined in terms of potential mediators as a function of the treatment \citep{RoGr92, Pearl01}, as aimed for in causal mediation analysis; see, for instance, \cite{Huber2019} for a survey. The motivation is that while canonical DiD has been developed for assessing the average effect of some treatment $D$ on the treated population, we are frequently interested in decomposing this ATET into various causal mechanisms, such as the average indirect effect operating through a mediator $M$. For this reason, we now consider the (total) ATET of treatment $D$ in the post-treatment period, defined as
\begin{align}\label{effecta}
 &\Delta_{d}(d,d',M(d),M(d'))\notag\\
 &=E[Y_1(d,M(d))-Y_1(d',M(d'))|D=d, T=1] =E[Y_1(d)-Y_1(d')|D=d, T=1],
\end{align}
where the second equality follows from the conventional definition of a potential outcome as a function of $D$ only, such that $Y_1(d,M(d))=Y_1(d)$, as typically considered in the canonical DiD literature.

By adding and subtracting $Y_1(d',M(d))$ within the expectation in \eqref{effecta}, it can be easily seen that the ATET can be decomposed into a natural direct effect, defined as
\begin{align}\label{effect2}
&\Delta_{d}(d,d',M(d),M(d))=E[Y_1(d,M(d))-Y_1(d',M(d))|D=d, T=1],
\end{align}
and a natural indirect effect, defined as
\begin{align}\label{effect2a}
\Delta_{d}(d',d',M(d),M(d'))&=E[Y_1(d',M(d))-Y_1(d',M(d'))|D=d, T=1].
\end{align}
While $E[Y_1(d,M(d))|D=d,T=1]=E[Y_1|D=d]$ is directly observed in the data, identification of the ATET and the natural effects hinges on the identification of the counterfactuals
\begin{align}
&E[Y_1(d',M(d'))|D=d,T=1],\label{counterfactual1}\\
&E[Y_1(d',M(d))|D=d,T=1]\notag\\
&=\sum_{m \in \mathcal{M}} E[Y_1(d',m)| D=d, M(d)=m,T=1]\cdot \Pr(M(d)=m|D=d, T=1),\label{counterfactual2}
\end{align}
where $\mathcal{M}$ denotes the support of $M$, which is here assumed to be discrete. 

The identification of the counterfactual in equation \eqref{counterfactual2} requires that Assumption \ref{ass1} holds for potential outcomes $Y_1(d',m)$ across treatment groups $D=d$ and $D=d'$ when fixing the mediator at any value $m$ in the support of $M(d)$ among the group with $D=d, T=1$: $E[Y_1(d',m)-Y_0(d',m)|D=d,M=m,X]=E[Y_1(d',m)-Y_0(d',m)|D=d',M=m,X]$. This condition permits identification of $E[Y_1(d',m)|D=d,M(d)=m,T=1]$, while $\Pr(M(d)=m \mid D=d, T=1)$ is directly identified from the observed data. Under this assumption, $E[Y_1(d',M(d)) \mid D=d,T=1]$ is given by
\begin{align}\label{mediationcounterfactual}
&E[Y_1(d',M(d))|D=d, T=1]\notag\\&=\sum_{m \in \mathcal{M}} E[Y_1(d',m)| D=d, M(d)=m,T=1]\cdot \Pr(M(d)=m|D=d, T=1)\notag\\
&=\sum_{m \in \mathcal{M}} E[Y_1(d',m)|D=d, M=m,T=1]\cdot \Pr(M=m|D=d, T=1)\\
&=\sum_{m \in \mathcal{M}} E[E[Y_1(d',m)|D=d, M=m,X]|D=d, M=m, T=1]\cdot \Pr(M=m|D=d,T=1)\notag\\
&=\sum_{m \in \mathcal{M}} E[ \mu_{d,m}(0,X)+\mu_{d',m}(1,X)-\mu_{d',m}(0,X)|D=d, M=m, T=1]\cdot \Pr(M=m|D=d,T=1)\notag\\
&= E[ \mu_{d,M}(0,X)+\mu_{d',M}(1,X)-\mu_{d',M}(0,X)|D=d, T=1],\notag
\end{align}
where the second equality follows from the observational rule, the third from the law of iterated expectations, the fourth from Assumption \ref{ass1} if it holds for all values $m$ in the support of $M(d)$, and the fifth from integrating over the distribution of the observed mediator $M$ given $D=d$ and $T=1$.

A DR expression for the counterfactual in equation~\eqref{mediationcounterfactual}, which satisfies Neyman orthogonality (as formally shown in Appendix~\ref{Neymanrepeated}), is given in Proposition~\ref{prop:DRidentificationcount}. It involves the joint probabilities of treatment and period, $P_{d,t}=\Pr(D=d,T=t)$, and the corresponding propensity scores given the mediator and covariates, $\pi_{d,t}(M,X)=\Pr(D=d,T=t\mid M,X)$.
\begin{proposition}\label{prop:DRidentificationcount}
Under Assumptions \ref{ass1}--\ref{ass4}, holding for any $m$ in the support of $M(d)$ conditional on $D=d$ and $T=1$, the counterfactual $E[Y_1(d',M(d))|D=d, T=1]$ is identified by the following expression, which satisfies Neyman orthogonality:
\begin{align}
E[Y_1(d',M(d))|D=d,T=1]&=E\left[ \frac{I\{D=d\}\cdot T \cdot [\mu_{d,M}(0,X)+\mu_{d',M}(1,X)-\mu_{d',M}(0,X)]}{P_{d,1}}\right.\notag\\
&+ \frac{(Y_T-\mu_{D,M}(T,X))\cdot \pi_{d,1}(M,X)}{P_{d,1}} \cdot \left\{\frac{I\{D=d\} \cdot (1-T)}{ \pi_{d,0}(M,X)}\right.\notag\\
&+ \left.\left.\frac{I\{D=d'\} \cdot  T}{ \pi_{d',1}(M,X)} - \frac{I\{D=d'\} \cdot (1-T)}{ \pi_{d',0}(M,X)} \right\} \right].\label{noprobidentification2}
\end{align}
\end{proposition}

The identification of the mean potential outcome in \eqref{counterfactual1}, that is, the counterfactual under treatment $d'$, requires a different parallel trends assumption than Assumption \ref{ass1} (for identifying the counterfactual under treatment $d'$ and mediator $m'$). The assumption is now imposed with respect to treatment only (rather than both treatment and mediator), and is formalized in the following assumption.
\begin{assumption} {\bf (Conditional parallel trends across treatments):}\label{ass5}\\
$E[Y_1(d',M(d'))-Y_0(d',M(d'))|D=d,X]=E[Y_1(d',M(d'))-Y_0(d',M(d'))|D=d',X].$
\end{assumption}
It is worth noting that, in contrast to Assumption \ref{ass1}, the validity of Assumption \ref{ass5} implies that those confounders affecting treatment and outcome, which are differenced out by a DiD approach, do not directly affect the mediator $M$ other than through the treatment $D$. If there were unobserved variables affecting both $D$ and $M$, the assumption would fail because differencing mean outcomes over time cannot account for these confounders.

To see this, consider the following example. Suppose the outcome in some time period $T$ is a (possibly unknown) function, denoted by $\mathcal{F}_1$, of the observed variables $D$, $M$, and $X$, which may interact with time $T$, and an additively separable, time-constant function, $\mathcal{F}_2$, of time-invariant unobservables:
\[
Y_T = \mathcal{F}_1(D, M, X, T) + \mathcal{F}_2(U).
\]
Furthermore, assume that the treatment is a function of $X$ and $U$, 
\[
D = \mathcal{F}_D(X, U),
\]
while the mediator is a function of $D$ and $X$, 
\[
M = \mathcal{F}_M(D, X).
\]
Random, time-varying, and additively separable error terms could be added to the models for $Y_T$, $D$, and $M$, but they are omitted for simplicity.

The outcome model implies that differencing mean potential outcomes under treatment value $d'$ across periods conditional on $X=x$ eliminates the time-constant component $\mathcal{F}_2(U)$:
\[
E[Y_1(d',M(d')) - Y_0(d',M(d')) \mid X=x] 
= E[\mathcal{F}_1(d', M(d'), x, 1) - \mathcal{F}_1(d', M(d'), x, 0)].
\]
For this reason, Assumption \ref{ass5} holds because $U$, while affecting $D$, does not affect $E[\mathcal{F}_1(d', M(d'), x, 1) - \mathcal{F}_1(d', M(d'), x, 0)]$, as the potential mediator $M(d') = \mathcal{F}_M(d', X)$ is not a function of $U$. However, changing the mediator model to $M = \mathcal{F}_M(D, X, U)$ would violate Assumption \ref{ass5}, as the potential mediator becomes $M(d') = \mathcal{F}_M(d', X, U)$. In this case, $U$ influences both the mean potential outcome difference
\[
E[Y_1(d',M(d')) - Y_0(d',M(d')) \mid X=x]  = E[\mathcal{F}_1(d', \mathcal{F}_M(d', x, U), x, 1) - \mathcal{F}_1(d', \mathcal{F}_M(d', x, U), x, 0)],
\]
through the mediator, and also the treatment, such that it acts as a confounder of the treatment and the difference in mean potential outcomes.

Figure \ref{fig:dags} illustrates the two cases by means of directed acyclic graphs (DAGs; see, e.g., \citealp{Pearl00}). In both panels, the observed covariates $X$ affect the treatment, the mediator, and the outcome, while the treatment $D$ affects the outcome both directly and indirectly through the mediator $M$. The outcome node represents the difference $Y_1 - Y_0$ rather than the outcome level. Since the unobservable $U$ (dotted node) enters $Y_T$ through the time-constant, additively separable component $\mathcal{F}_2(U)$, its direct effect on the outcome is eliminated by differencing, such that there is no arrow from $U$ to $Y_1 - Y_0$. In panel (a), $U$ affects $D$ but not $M$, which corresponds to the setting with mediator model $M = \mathcal{F}_M(D,X)$, under which Assumption~\ref{ass5} holds. In panel (b), $U$ additionally affects $M$. This is the case when the mediator model is $M = \mathcal{F}_M(D,X,U)$, under which Assumption~\ref{ass5} is violated. Assumption~\ref{ass1}, however, continues to hold in panel (b). This is because Assumption~\ref{ass1} fixes the mediator at a specific value $m'$, such that the mean potential outcome difference $E[Y_1(d',m') - Y_0(d',m') \mid X=x] = \mathcal{F}_1(d',m',x,1) - \mathcal{F}_1(d',m',x,0)$ does not depend on $U$, irrespective of how $U$ affects selection into treatment and mediator states. Consequently, the dynamic treatment and controlled direct effects, as well as the natural direct effect, remain identified in panel (b), whereas the total ATET and the natural indirect effect require either Assumption~\ref{ass5} or Assumptions~\ref{ass7}, \ref{ass8}, and \ref{assPTmed} introduced below.

\begin{figure}[t]
    \centering
    \caption{Causal models with a time-invariant unobservable $U$}
    \begin{subfigure}[b]{0.48\textwidth}
        \centering
        \scalebox{0.85}{
        \begin{tikzpicture}[scale=0.4]
            \SetGraphUnit{1}
            \Vertex[x=5,y=0]{D}
            \Vertex[x=11,y=0]{M}
            \Vertex[x=0,y=3,style={dotted}]{U}
            \node[draw,ellipse,minimum width=8ex,minimum height=5ex]
                (Y) at (18,0) {$Y_1 - Y_0$};
            \Vertex[x=3,y=-5]{X}

            \Edge(D)(M)
            \Edge(M)(Y)
            \Edge(X)(D)
            \Edge(X)(M)
            \Edge(X)(Y)

            \draw[{-Stealth[{length=8pt, width=5pt}]}, thick, dotted]
                (U) -- (D);
            \draw[{-Stealth[{length=8pt, width=5pt}]}, thick]
                (D) to[bend left=50] (Y);
        \end{tikzpicture}
        }
        \caption{Assumptions \ref{ass1} and \ref{ass5} hold.}
        \label{fig:dag_a}
    \end{subfigure}
    \hfill
    \begin{subfigure}[b]{0.48\textwidth}
        \centering
        \scalebox{0.85}{
        \begin{tikzpicture}[scale=0.4]
            \SetGraphUnit{1}
            \Vertex[x=5,y=0]{D}
            \Vertex[x=11,y=0]{M}
            \Vertex[x=0,y=3,style={dotted}]{U}
            \node[draw,ellipse,minimum width=8ex,minimum height=5ex]
                (Y) at (18,0) {$Y_1 - Y_0$};
            \Vertex[x=3,y=-5]{X}

            \Edge(D)(M)
            \Edge(M)(Y)
            \Edge(X)(D)
            \Edge(X)(M)
            \Edge(X)(Y)

            \draw[{-Stealth[{length=8pt, width=5pt}]}, thick, dotted]
                (U) -- (D);
            \draw[{-Stealth[{length=8pt, width=5pt}]}, thick, dotted]
                (U) to[bend left=50] (M);
            \draw[{-Stealth[{length=8pt, width=5pt}]}, thick]
                (D) to[bend left=50] (Y);
        \end{tikzpicture}
        }
        \caption{Assumption \ref{ass1} holds, Assumption \ref{ass5} fails.}
        \label{fig:dag_b}
    \end{subfigure}
    \label{fig:dags}
\end{figure}

When representing the potential outcome only as a function of $d'$, i.e.,\ $Y_t(d')=Y_t(d',M(d'))$, we see that Assumption \ref{ass5} corresponds to the parallel trends assumption in the canonical DiD literature on ATET identification, $E[Y_1(d')-Y_0(d')|D=d,X]=E[Y_1(d')-Y_0(d')|D=d',X]$.The identification of $E[Y_1(d',M(d'))|D=d,X]$ can be shown by following analogous steps as in equation \eqref{DiDobs}. Denoting by $\mu_{d}(t,X)=E[Y_t|D=d,X]$, we have that under Assumptions \ref{ass2}--\ref{ass5} satisfied for the distribution of $M(d')$ among $D=d, T=1$,
\begin{align}\label{standarddid}
E[Y_1(d',M(d'))|D=d,X]&=\mu_{d}(0,X)+\mu_{d'}(1,X)-\mu_{d'}(0,X),\textrm{ and therefore,}\notag\\
E[Y_1(d',M(d'))|D=d, T=1]&=E[\mu_{d}(0,X)+\mu_{d'}(1,X)-\mu_{d'}(0,X)|D=d, T=1].
\end{align}
Let $\pi_{d,t}(X)=\Pr(D=d,T=t|X)$ denote the joint propensity score of treatment and period given $X$. A DR expression of \eqref{standarddid} is given in Proposition \ref{prop:regularoutcome}. For $d=1$ and $d'=0$, the expression for the counterfactual $E[Y_1(d',M(d'))|D=d, T=1]$ in the proposition is equivalent to that derived in \cite{Zimmert2018}, who shows that it satisfies Neyman orthogonality (and for completeness, identification based on expression \eqref{standarddid} and Neyman orthogonality is also demonstrated in Appendix \ref{Neymanrepeated}).
\begin{proposition}\label{prop:regularoutcome}
Under Assumptions \ref{ass2}--\ref{ass5} holding for any $m$ in the support of $M(d')$ conditional on $D=d$ and $T=1$, the counterfactual $E[Y_1(d',M(d'))|D=d, T=1]$ is identified by the following expression, which satisfies Neyman orthogonality:
\begin{align}\label{DiDidentDRsimple}
E[Y_1(d',M(d'))|D=d, T=1]&=E\left[  \frac{I\{D=d\}\cdot T \cdot [\mu_{d}(0,X)+\mu_{d'}(1,X)-\mu_{d'}(0,X)]}{P_{d,1}}\right.\notag\\
&+  \frac{(Y_T-\mu_{D}(T,X))\cdot \pi_{d,1}(X)}{P_{d,1}} \cdot \left\{\frac{I\{D=d\}\cdot (1-T)}{ \pi_{d,0}(X)}\right.\notag\\
&+ \left.\left.\frac{I\{D=d'\}\cdot  T}{ \pi_{d',1}(X)} - \frac{I\{D=d'\}\cdot (1-T)}{ \pi_{d',0}(X)} \right\}   \right].
\end{align}
\end{proposition}

As an alternative to Assumption \ref{ass5}, we may impose a parallel trends assumption on mean potential outcomes within subpopulations defined by the potential mediator under treatment $d'$, which permits identifying the counterfactual outcome trend of the treated. This is combined with an additional parallel trends assumption on the mediator that permits identifying $\Pr(M(d')=m|D=d,T=1)$. The latter assumption pertains not only to the conditional mean of the mediator but to its entire distribution for all $m$ in the support of $M(d')$ given $D=d,T=1$, see \cite*{CALLAWAY2018395} and \cite*{CallawayLi} for related distributional parallel trend conditions. The approach requires that the past value of the mediator can be observed in the pre-treatment period. For this reason, we introduce additional notation and denote by $M_0$ the mediator in the pre-treatment period ($T=0$), while $M$ continues to denote the post-treatment mediator through which $D$ may affect the post-treatment outcome $Y_t$. The parallel trends assumption on the conditional distribution of the mediator is formalized as follows:
\begin{assumption} {\bf (Conditional distributional parallel trends in the mediator):}\label{ass7}\\
$\Pr(M(d') = m|D=d,X)-\Pr(M_0(d') = m|D=d,X)$\\$=\Pr(M(d') = m|D=d',X)-\Pr(M_0(d') = m|D=d',X)$.
\end{assumption}
In addition to parallel trends with respect to the mediator, we also need to impose a no-anticipation assumption on the mediator values in the pre-treatment period, analogous to Assumption \ref{ass2} for the outcomes:
\begin{assumption} {\bf (No anticipation of effect on mediator):}\label{ass8}\\
$ M_0(d) = M_0$ for any $d$ in the support of $D$ among the treated.
\end{assumption}
Finally, we impose parallel trends in the mean potential outcome under treatment $d'$ and mediator value $m$ among subjects whose potential mediator under $d'$ equals $m$:
\begin{assumption} {\bf (Conditional parallel trends across treatments within potential mediator strata):}\label{assPTmed}\\
$E[Y_1(d',m)-Y_0(d',m)|D=d,M(d')=m,X]$\\$=E[Y_1(d',m)-Y_0(d',m)|D=d',M(d')=m,X]$.
\end{assumption}
Assumption \ref{assPTmed} requires that, conditional on $X$, the mean potential outcome under treatment $d'$ and mediator value $m$ follows the same trend in the treated group and in the group with $D=d'$, among subjects whose mediator would equal $m$ under treatment $d'$. In the group with $D=d'$, the observational rule implies $M=M(d')$. The right-hand side therefore corresponds to the observed outcome trend among subjects with $D=d'$ and $M=m$. Among the treated, $M(d')$ is a potential mediator that generally differs from the observed mediator $M=M(d)$. For this reason, Assumption \ref{assPTmed} is distinct from Assumption \ref{ass1}, which is imposed within the treated subpopulation with $M(d)=m$. Assumption \ref{assPTmed} holds in both panels of Figure \ref{fig:dags}, because the time-constant term $\mathcal{F}_2(U)$ is eliminated when differencing potential outcomes over time, irrespective of how $U$ affects $D$ and $M$.

If the pre-treatment value of the mediator is zero (or more generally, the same) for everyone, Assumption \ref{ass7} collapses to a standard selection-on-observables assumption with respect to the mediator, as discussed in \cite{Im04}. Specifically, in that case, we have $\Pr(M_0(D) = 0| D, X) = 1$ and $\Pr(M_0(D) \neq 0 | D, X) = 0$, so that Assumption \ref{ass7} reduces to $\Pr(M(d') = m|D=d,X)=\Pr(M(d') = m|D=d',X)$. Together with Assumption \ref{assPTmed}, this implies Assumption \ref{ass5}. To see this, note that $E[Y_1(d',M(d'))-Y_0(d',M(d'))|D=d,X]=\sum_{m \in \mathcal{M}}E[Y_1(d',m)-Y_0(d',m)|D=d,M(d')=m,X]\cdot\Pr(M(d')=m|D=d,X)$. Under Assumptions \ref{assPTmed} and \ref{ass7}, both factors equal their counterparts in the group with $D=d'$. Since conditional on $X$ the distribution of $M(d')$ cannot depend on unobserved variables that jointly affect the treatment and mediator, scenarios such as the one above with $U$ entering both $D=\mathcal{F}_D(X,U)$ and $M=\mathcal{F}_M(D,X,U)$ are ruled out. In this case, jointly imposing Assumptions \ref{ass7} and \ref{assPTmed} adds no identifying content to Assumption \ref{ass5}.

The newly introduced assumptions permit the identification of the mean counterfactual outcome based on the following approach:
\begin{align}
&E[Y_1(d',M(d'))|D=d,T=1]=E\big[E[Y_1(d',M(d'))|D=d,X]\,\big|\,D=d,T=1\big]\notag\\
&=E\big[E[Y_0(d',M(d'))|D=d,X]+E[Y_1(d',M(d'))-Y_0(d',M(d'))|D=d,X]\,\big|\,D=d,T=1\big]\notag\\
&=E\Big[E[Y_0(d,M(d))|D=d,X]+\sum_{m \in \mathcal{M}} E[Y_1(d',m)-Y_0(d',m)|M(d')=m, D=d,X]\notag\\
&\times \Pr(M(d')=m|D=d,X)\,\Big|\, D=d, T=1\Big]\notag\\
&=E\Big[E[Y_0(d,M(d))|D=d,X]+\sum_{m \in \mathcal{M}} E[Y_1(d',m)-Y_0(d',m)|M(d')=m, D=d',X]\notag\\
&\times \{\Pr(M_0(d')=m|D=d,X)+\Pr(M(d')=m|D=d',X)\notag\\
&-\Pr(M_0(d')=m|D=d',X)\}\,\Big|\,D=d, T=1\Big]\notag\\
&=E\Big[\mu_{d}(0,X)+\sum_{m \in \mathcal{M}}\{\mu_{d',m}(1,X)-\mu_{d',m}(0,X)\}\cdot\{\Pr(M_0=m|D=d,X)\label{doubleparalleltrends}\\
&+\Pr(M=m|D=d',X)-\Pr(M_0=m|D=d',X)\}\,\Big|\,D=d, T=1\Big],\notag
\end{align}
where the first equality follows from the law of iterated expectations and the second from adding and subtracting $E[Y_0(d',M(d'))|D=d,X]$. The third equality follows from Assumption \ref{ass2} imposed with respect to the treatment, $E[Y_0(d',M(d'))|D=d,X]=E[Y_0(d,M(d))|D=d,X]$, as for Proposition \ref{prop:regularoutcome}. It also uses the law of iterated expectations with respect to the potential mediator $M(d')$, since $Y_t(d',M(d'))=Y_t(d',m)$ if $M(d')=m$. The fourth equality follows from Assumptions \ref{assPTmed} and \ref{ass7} (parallel trends for mean potential outcomes within potential mediator strata and for potential mediators, respectively). The last equality follows from Assumption \ref{ass8} (no anticipation of the effect on the mediator) and from the observational rule. The latter implies $Y_t=Y_t(d,M(d))$ conditional on $D=d$, as well as $M=M(d')$ and $Y_t=Y_t(d',m)$ conditional on $D=d'$ and $M=m$.

We henceforth denote by $\nu_{d,m}(0,X)=\Pr(M_0=m\mid D=d,X)$ and $\nu_{d,m}(1,X)=\Pr(M=m\mid D=d,X)$ the conditional mediator probabilities in the pre- and post-treatment periods, respectively, and note that $M_T=M_0$ if $T=0$ and $M_T=M$ if $T=1$. Proposition \ref{prop:DRidentificationcountdoubl} provides the DR expression for \eqref{doubleparalleltrends}, and Appendix \ref{Neymanrepeated} shows its Neyman orthogonality.
\begin{proposition}\label{prop:DRidentificationcountdoubl}
Under Assumptions \ref{ass2}--\ref{ass4} and \ref{ass7}--\ref{assPTmed} holding for any $m$ in the support of $M(d')$ conditional on $D=d$ and $T=1$, the counterfactual $E[Y_1(d',M(d'))|D=d,T=1]$ is identified by the following expression, which satisfies Neyman orthogonality:
\begin{align}\label{DiDidentDRsimple2doubl}
&E[Y_1(d',M(d'))|D=d,T=1]=E\Bigg[\frac{I\{D=d\}\cdot T}{P_{d,1}}\cdot\Big\{\mu_{d}(0,X)+\sum_{m \in \mathcal{M}}[\mu_{d',m}(1,X)-\mu_{d',m}(0,X)]\notag\\
&\times[\nu_{d,m}(0,X)+\nu_{d',m}(1,X)-\nu_{d',m}(0,X)]\Big\}\!+\!\frac{(Y_T-\mu_{D}(T,X))\cdot \pi_{d,1}(X)\cdot I\{D=d\}\cdot (1-T)}{P_{d,1}\cdot \pi_{d,0}(X)}\notag\\
&+\sum_{m \in \mathcal{M}}[\nu_{d,m}(0,X)+\nu_{d',m}(1,X)-\nu_{d',m}(0,X)]\cdot\frac{(Y_T-\mu_{D,M}(T,X))\cdot \pi_{d,1}(X)}{P_{d,1}}\notag\\
&\times \left(\frac{I\{D=d'\}\cdot I\{M=m\}\cdot  T}{ \rho_{d',m,1}(X)} - \frac{I\{D=d'\}\cdot I\{M=m\}\cdot (1-T)}{ \rho_{d',m,0}(X)} \right)\notag\\
&+\sum_{m \in \mathcal{M}}[\mu_{d',m}(1,X)-\mu_{d',m}(0,X)]\cdot\frac{(I\{M_T=m\}-\nu_{D,m}(T,X))\cdot \pi_{d,1}(X)}{P_{d,1}}\notag\\
&\times \left(\frac{I\{D=d\}\cdot (1-T)}{ \pi_{d,0}(X)}+\frac{I\{D=d'\}\cdot  T}{ \pi_{d',1}(X)} - \frac{I\{D=d'\}\cdot (1-T)}{ \pi_{d',0}(X)} \right)\Bigg].
\end{align}
\end{proposition}
The first two terms in \eqref{DiDidentDRsimple2doubl} correspond to the DR expression for the pre-treatment mean outcome of the treated, $E[\mu_{d}(0,X)|D=d,T=1]$, and the first term additionally contains the plug-in expression for the counterfactual outcome trend. The third term corrects for estimation errors in the conditional mean outcomes $\mu_{d',m}(t,X)$ and is weighted by the identified potential mediator probabilities. The fourth term corrects for estimation errors in the conditional mediator probabilities and is weighted by the outcome trends $\mu_{d',m}(1,X)-\mu_{d',m}(0,X)$.

As a final remark in this section, we note that all previous results derived for discrete treatments and mediators can be extended to the case of continuous treatments and mediators. To this end, each indicator function for treatment values $d$ and $d'$ as well as mediator values $m$ and $m'$ is replaced by a kernel weighting function with a bandwidth approaching zero to achieve identification. Consider, for instance, a continuous treatment $D$ taking value $d$, while maintaining a discretely distributed mediator. We denote by $\omega(D;d,h)$ a weighting function that depends on the distance between $D$ and the reference value $d$, as well as on a non-negative bandwidth parameter $h$: $\omega(D;d,h)=\frac{1}{h}\mathcal{K}\left(\frac{D-d}{h}\right)$, where $\mathcal{K}$ denotes a well-behaved kernel function. The closer $h$ is to zero, the less weight is assigned to observations with larger discrepancies between $D$ and $d$.

Furthermore, note that the previously defined (conditional) probabilities, such as $\Pi_{d,m,t}$ and $\rho_{d,m,t}(X)$, correspond to (conditional) density functions when the treatment is continuous.  As discussed in \citet{fan1996estimation}, the conditional treatment density --- also referred to as the generalized propensity score in \citet{ImaivanDyk2004} and \citet{HiranoImbens2005} --- can be expressed as  $\rho_{d,m,t}(X)=\lim_{h \rightarrow 0} E[\omega(D;d,h) \cdot I\{M=m\} \cdot I\{T=t\}|X]$. Furthermore, $\Pi_{d,m,t}=\lim_{h \rightarrow 0} E[\omega(D;d,h) \cdot I\{M=m\} \cdot I\{T=t\}]$. 

Applying these modifications, for instance, to the DR expression in equation \eqref{DiDidentDR} yields
\begin{align}\label{DiDidentDRcont}
&\Delta_{d,m}(d,m,d',m')\notag\\
&=E\left[  \frac{\omega(D;d, h)\cdot I\{M=m\}\cdot T \cdot [Y_T-\{\mu_{d,m}(0,X)+\mu_{d',m'}(1,X)-\mu_{d',m'}(0,X)\}]}{\Pi_{d,m,1}}\right.\notag\\
&- \frac{(Y_T-\mu_{D,M}(T,X))\cdot \rho_{d,m,1}(X)}{\Pi_{d,m,1}} \cdot \left\{\frac{\omega(D;d, h)\cdot I\{M=m\}\cdot (1-T)}{ \rho_{d,m,0}(X)}\right.\notag\\
&+ \left.\left.\frac{\omega(D;d', h)\cdot I\{M=m'\}\cdot  T}{ \rho_{d',m',1}(X)} - \frac{\omega(D;d', h)\cdot I\{M=m'\}\cdot (1-T)}{ \rho_{d',m',0}(X)} \right\}   \right].
\end{align}
If the mediator is continuous as well, analogous modifications apply to $I\{M=m\}$ and $I\{M=m'\}$, which are then replaced by kernel weighting functions. 

\section{Identification in panel data}\label{paneldata}

This section discusses identification in panel data under the identifying assumptions proposed in the previous sections. Our results differ from the DR approaches for panel data proposed by \citet{shahn2022structural}, who focus on parametric structural models, and \citet{renson2023identifying}, who develop a method based on targeted maximum likelihood estimation. Panel data permit taking outcome differences within subjects over time, such that we now consider conditional means of within-subject differences in outcomes over time, defined as $\mu_{d',m'}(X)=E[Y_1-Y_0|D=d',M=m',X]$, instead of differences in time-specific conditional means, $\mu_{d',m'}(1,X)-\mu_{d',m'}(0,X)=E[Y_1|D=d',M=m',X]-E[Y_0|D=d',M=m',X]$, as in the repeated cross-section setting. Furthermore, as the same individuals are observed in both time periods $T=1$ and $T=0$, it follows that the treatment and mediator distributions are independent of $T$, both unconditionally and conditional on covariates $X$, the distribution of which does not change over time either. For this reason, $\Pi_{d,m,1}=\Pr(D=d,M=m,T=1)$ is replaced by $\Pi_{d,m}=\Pr(D=d,M=m)$ and $\rho_{d,m,1}(X)=\Pr(D=d,M=m,T=1|X)$ by $\rho_{d,m}(X)=\Pr(D=d,M=m|X)$.

We first reconsider the identification of the treatment effect in equation \eqref{ATETdef}, which (as conditioning on $T=1$ is unnecessary) simplifies to $\Delta_{d,m}(d,m,d',m')=E[Y_1(d,m)-Y_1(d',m')|D=d,M=m]$. Not conditioning on $T$ and taking within-subject differences allows simplifying the previous identification result provided in equation \eqref{DiDidentDR} to the DR expression in Proposition \ref{prop:DRidentificationpanel}, with the proof of Neyman orthogonality given in Appendix \ref{Neymanpanel}. We note that this proposition is closely related to the conventional DR expression for ATET evaluation of discrete treatments under the selection-on-observables assumption (see, for example, Section 4.6 in \cite{huber2023causal}), with the difference that here, outcome differences (rather than outcome levels) and indicators for both the treatment and the mediator (rather than the treatment alone) enter the expression.

\begin{proposition}\label{prop:DRidentificationpanel}
Under Assumptions \ref{ass1}--\ref{ass4}, the ATET is identified by the following expression, which satisfies Neyman orthogonality: 
\begin{align}\label{DiDidentDRpanel}
\Delta_{d,m}(d,m,d',m')&=E\left[  \frac{I\{D=d\}\cdot I\{M=m\}\cdot [Y_1-Y_0-\mu_{d',m'}(X)]}{\Pi_{d,m}}\right.\notag\\
&- \left.\frac{I\{D=d'\}\cdot I\{M=m'\} \cdot [Y_1-Y_0-\mu_{D,M}(X)]\cdot \rho_{d,m}(X)}{\Pi_{d,m}\cdot \rho_{d',m'}(X)}  \right].
\end{align}
 \end{proposition}

Likewise, the DR identification result of Proposition~\ref{prop:DRidentificationcount}, which identifies the counterfactual $E[Y_1(d', M(d)) \mid D=d]$ (where conditioning on $T=1$ is now omitted), simplifies in panel data to the expression in Proposition \ref{prop:DRidentificationcountpanel}.
\begin{proposition}\label{prop:DRidentificationcountpanel}
Under Assumptions \ref{ass1}--\ref{ass4}, holding for any $m$ in the support of $M(d)$ conditional on $D=d$, the counterfactual $E[Y_1(d',M(d))|D=d]$ is identified by the following expression, which satisfies Neyman orthogonality:
\begin{align}\label{DiDidentDRcounterfacpanel2}
&E[Y_1(d',M(d))|D=d]\notag\\
&= E\left[ \frac{I\{D=d\}\cdot [Y_0+\mu_{d',M}(X)]}{\Pr(D=d)}+ \frac{I\{D=d'\}\cdot[Y_1-Y_0-\mu_{D,M}(X)]\cdot \pi_{d}(M,X)}{\Pr(D=d)\cdot \pi_{d'}(M,X)}  \right],
\end{align}
where $\pi_{d}(M,X)=\Pr(D=d|M,X)$.
\end{proposition}

Furthermore, the DR expression in Proposition \ref{prop:regularoutcome} for identifying the counterfactual $E[Y_1(d', M(d')) \mid D=d]$ (where conditioning on $T=1$ is now omitted) simplifies, as $P_{d,t}=\Pr(D=d,T=t)$ is replaced by $P_d=\Pr(D=d)$ and $\pi_{d,t}(X)=\Pr(D=d,T=t\mid X)$ by $\pi_d(X)=\Pr(D=d\mid X)$. Denoting $\mu_{d'}(X)=E[Y_1 - Y_0 \mid D=d', X]$, we obtain the expression in Proposition \ref{prop:regularoutcomepanel}, with the proof of Neyman orthogonality provided in Appendix~\ref{Neymanpanel}. This result is analogous to the DR expression, for instance, provided in \citet{SantAnnaZhao2018}, although in that case it pertains to the ATET rather than to the counterfactual outcome, as considered here.
\begin{proposition}\label{prop:regularoutcomepanel}
Under Assumptions \ref{ass2}--\ref{ass5} holding for any $m$ in the support of $M(d')$ conditional on $D=d$, the counterfactual $E[Y_1(d',M(d'))|D=d]$ is identified by the following expression, which satisfies Neyman orthogonality:
\begin{align}\label{DiDidentDRsimplepanel}
E[Y_1(d',M(d'))|D=d]&=E\left[  \frac{I\{D=d\}\cdot [Y_0+\mu_{d'}(X)]}{\Pr(D=d)}\right.\notag\\
&+\left.  \frac{I\{D=d'\}\cdot [Y_1-Y_0-\mu_{D}(X)]\cdot \pi_{d}(X)}{\Pr(D=d) \cdot \pi_{d'}(X)}\right].
\end{align}
\end{proposition}

Finally, we consider the panel data version of the result in Proposition \ref{prop:DRidentificationcountdoubl} for the identification of $E[Y_1(d',M(d'))\mid D=d]$ based on an alternative set of assumptions (including a parallel trends assumption on the mediator). To this end, we denote by $\nu_{d',m}(X)=E[ I\{M=m\}-I\{M_0=m\}\mid D=d',X]$ the difference in the conditional probability that the mediator takes value $m$ between post- and pre-treatment periods, given treatment and covariates. Furthermore, $\nu_{d,m}(0,X)=\Pr(M_0=m\mid D=d,X)$ is defined as in Section \ref{sec:id2}. Following the same steps as in \eqref{doubleparalleltrends}, the counterfactual is identified by
\begin{align}\label{doubleparalleltrendspanel}
E[Y_1(d',M(d'))|D=d]=E\Big[Y_0+\sum_{m \in \mathcal{M}}\mu_{d',m}(X)\cdot\{\nu_{d,m}(0,X)+\nu_{d',m}(X)\}\,\Big|\,D=d\Big].
\end{align}
Proposition \ref{prop:DRidentificationcountdoublpanel} provides the DR expression, and Appendix \ref{Neymanpanel} shows its Neyman orthogonality.
\begin{proposition}\label{prop:DRidentificationcountdoublpanel}
Under Assumptions \ref{ass2}--\ref{ass4} and \ref{ass7}--\ref{assPTmed} holding for any $m$ in the support of $M(d')$ conditional on $D=d$, the counterfactual $E[Y_1(d',M(d'))|D=d]$ is identified by the following expression, which satisfies Neyman orthogonality:
\begin{align}\label{DiDidentDRsimple2doublpanel}
	&E[Y_1(d',M(d'))|D=d]=E\Bigg[ \frac{I\{D=d\}\cdot  \Big[Y_0+\sum_{m \in \mathcal{M}}\mu_{d',m}(X)\cdot\{I\{M_0=m\}+\nu_{d',m}(X)\}\Big]}{\Pr(D=d)}\notag\\
	&+\sum_{m \in \mathcal{M}}\{\nu_{d,m}(0,X)+\nu_{d',m}(X)\}\cdot \frac{I\{D=d'\}\cdot I\{M=m\}\cdot[Y_1-Y_0-\mu_{D,M}(X)]\cdot \pi_{d}(X)}{\Pr(D=d)\cdot \rho_{d',m}(X)}  \notag\\
	&+\sum_{m \in \mathcal{M}}\mu_{d',m}(X)\cdot \frac{I\{D=d'\}\cdot[I\{M=m\}-I\{M_0=m\}-\nu_{D,m}(X)]\cdot \pi_{d}(X)}{\Pr(D=d)\cdot \pi_{d'}(X)}  \Bigg].
\end{align}
\end{proposition}
The first term in \eqref{DiDidentDRsimple2doublpanel} contains the pre-treatment outcome of the treated and the plug-in expression for the counterfactual outcome trend, in which $I\{M_0=m\}$ takes the place of $\nu_{d,m}(0,X)$. The second term corrects for estimation errors in the conditional mean outcome differences $\mu_{d',m}(X)$ and is weighted by the identified potential mediator probabilities. The third term corrects for estimation errors in $\nu_{d',m}(X)$ and is weighted by $\mu_{d',m}(X)$.

\section{Estimation}
\label{sec:meth}

In this section, we outline DiD estimation within the DML framework using cross-fitting, following \cite{Chetal2018}. The approach is based on the sample analogs of the DR identification results in Sections \ref{sec:id} to \ref{paneldata}. We focus on estimating the ATET $\Delta_{d,m}(d,m,d',m')$ defined in equation \eqref{ATETdef}, using the sample analog of equation \eqref{DiDidentDR} from Proposition \ref{prop:DRidentification}. Estimation for other DR results proceeds analogously using the corresponding normalized sample analogs; hence, we omit a detailed discussion.

Let $\mathcal{W} = \{W_i \mid 1 \leq i \leq n\}$, with $W_i = (Y_{i,T}, D_i, M_i, X_i, T_i)$ for $i=1,\ldots,n$, denote an i.i.d.\ sample of size $n$. The nuisance parameters, comprising conditional mean outcomes and propensity scores, are
\begin{align}
\eta = (\mu_{d,m}(0,X), \mu_{d',m'}(0,X), \mu_{d',m'}(1,X), \rho_{d,m,1}(X), \rho_{d,m,0}(X), \rho_{d',m',1}(X), \rho_{d',m',0}(X)), \notag
\end{align}
with corresponding estimates
\begin{align}
\hat{\eta} = (\hat\mu_{d,m}(0,X), \hat\mu_{d',m'}(0,X), \hat\mu_{d',m'}(1,X), \hat\rho_{d,m,1}(X), \hat\rho_{d,m,0}(X), \hat\rho_{d',m',1}(X), \hat\rho_{d',m',0}(X)). \notag
\end{align}

We estimate $\Delta_{d,m}(d,m,d',m')$ using the following DML algorithm with cross-fitting to avoid overfitting, where within-group weights --- defined by treatment state and time period --- are normalized to sum to one:\vspace{15pt}\newline
  \textbf{DML algorithm:}
\begin{enumerate}
\item Split $\mathcal{W}$ into $K$ subsamples. For each subsample $k$, let $n_k$ denote its size, $\mathcal{W}_k$ the set of observations in the subsample, and $\mathcal{W}_k^{C}$ the complement set of all observations not in $\mathcal{W}_k$.
\item For each $k$, use $\mathcal{W}_k^{C}$ to estimate the nuisance parameters $\eta$ via machine learning, and predict these nuisance parameters in $\mathcal{W}_k$. Denote the predictions by $\hat{\eta}^k$.

\item Stack fold-specific estimates $\hat{\eta}^1, \ldots, \hat{\eta}^K$ to form a matrix of nuisance estimates for the full sample:
\[
\hat{\eta} = \begin{pmatrix} \hat{\eta}^1 \\ \vdots \\ \hat{\eta}^K \end{pmatrix}.
\]

\item Plug the nuisance estimates into the normalized sample analog of equation \eqref{DiDidentDR} to obtain an estimate of the ATET $\Delta_{d,m}(d,m,d',m')$, denoted by $\hat{\Delta}_{d,m}(d,m,d',m')$:
\begin{align} 
&\hat{\Delta}_{d,m}(d,m,d',m') = \notag\\
&\sum_{i=1}^{n} \Bigg[ 
\frac{I\{D_i=d\} I\{M_i=m\} T_i \big[Y_{i,T}-\hat\mu_{d,m}(0,X_i)-\hat\mu_{d',m'}(1,X_i)+\hat\mu_{d',m'}(0,X_i)\big]}{\sum_{i=1}^{n} I\{D_i=d\} I\{M_i=m\} T_i} \notag\\
&\quad - \frac{(Y_{i,T}-\hat\mu_{d,m}(0,X_i)) \hat\rho_{d,m,1}(X_i) I\{D_i=d\} I\{M_i=m\} (1-T_i)/\hat\rho_{d,m,0}(X_i)}{\sum_{i=1}^{n} \hat\rho_{d,m,1}(X_i) I\{D_i=d\} I\{M_i=m\} (1-T_i)/\hat\rho_{d,m,0}(X_i)} \notag\\
&\quad - \frac{(Y_{i,T}-\hat\mu_{d',m'}(1,X_i)) \hat\rho_{d,m,1}(X_i) I\{D_i=d'\} I\{M_i=m'\} T_i/\hat\rho_{d',m',1}(X_i)}{\sum_{i=1}^{n} \hat\rho_{d,m,1}(X_i) I\{D_i=d'\} I\{M_i=m'\} T_i/\hat\rho_{d',m',1}(X_i)} \notag\\
&\quad + \frac{(Y_{i,T}-\hat\mu_{d',m'}(0,X_i)) \hat\rho_{d,m,1}(X_i) I\{D_i=d'\} I\{M_i=m'\} (1-T_i)/\hat\rho_{d',m',0}(X_i)}{\sum_{i=1}^{n} \hat\rho_{d,m,1}(X_i) I\{D_i=d'\} I\{M_i=m'\} (1-T_i)/\hat\rho_{d',m',0}(X_i)}
\Bigg]. \notag
\end{align}
\end{enumerate}

Under specific regularity conditions --- in particular, that the nuisance parameter estimators converge at rate $o(n^{-1/4})$ to the true models --- DML-based effect or counterfactual estimators achieve $\sqrt{n}$-consistency and asymptotic normality for discrete treatments and mediators, analogous to \citet{Chetal2018} for ATET and ATE with binary treatments. For example, lasso regression can be used to estimate the nuisance functions, and the required convergence rate is attainable under approximate sparsity, as shown in \citet{Bellonietal2014}. When treatments and/or mediators are continuous and kernel smoothing is applied as in equation \eqref{DiDidentDRcont}, convergence rates are slower than $\sqrt{n}$ due to the local nature of the estimation. Nonetheless, asymptotic normality can still be achieved under appropriate regularity conditions; see, for instance, \citet{zhang2025} and \citet{haddad2024} for asymptotic results on DiD-based DML with continuous treatments.



\section{Simulation study}\label{sim}

This section presents a simulation study to assess the finite-sample behavior of the proposed methods for repeated cross sections and panel data. We first focus on natural direct and indirect effects in repeated cross sections and consider the following data-generating process (DGP):
\begin{eqnarray*}
X_j &=& 0.5\cdot T+Q_j\textrm{ for }j \in \{1,...,p\},\quad X=(X_1,...,X_p),\\
D &=& I \{X\beta+0.5\cdot U + V_d>0\},\\
M &=& X\beta+0.5\cdot D + V_m,\\
Y_T &=& X\beta+ (1+D+M+D\cdot M)\cdot T + U+W,\\
Q_j,U,V_d,V_m,W &\sim& \mathcal{N}(0,1)\textrm{ independent of each other},\\
T&\sim& \textrm{Bernoulli}(0.5).
\end{eqnarray*}

$X$ is a vector of $p=100$ covariates. Each covariate $X_j$, $j\in\{1,...,p\}$, depends on a random error term $Q_j$ and a binary time index $T$, where $T=0$ denotes the pre-treatment period and $T=1$ denotes the post-treatment period, each occurring with equal probability. The binary treatment $D$ is determined by the observed covariates $X$ and two unobserved terms, $U$ and $V_d$. The mediator $M$ is continuous and is a function of $X$, $D$, and the unobserved term $V_m$. The outcome $Y_T$ depends on $X$, $T$, and unobserved terms $U$ and $W$; in the post-treatment period, it additionally depends on $D$, $M$, and their interaction. The impact of $X$ on $D$, $M$, and $Y_T$ is captured by the coefficient vector $\beta$. The $j$th element of $\beta$ is set to $0.4/j^2$, implying a quadratic decay in covariate importance such that covariates with lower indices $j$ exert stronger confounding. The unobserved variable $U$ enters both the treatment and outcome equations and therefore acts as a confounder of the treatment--outcome relationship; this confounding is addressed by our proposed approach. Importantly, $M$ does not depend on $U$, ensuring satisfaction of Assumption~\ref{ass5}. All error terms $Q_j$, $U$, $V_d$, $V_m$, and $W$ are independently drawn from a standard normal distribution.

We consider two sample sizes, $n=2000$ and $n=8000$, and generate 1000 independent simulation replications. The parameters of interest are the total effect, the natural direct effect, and the natural indirect effect, as defined in equations~\eqref{effecta}, \eqref{effect2}, and \eqref{effect2a}, respectively. These effects depend on counterfactual outcomes identified in Propositions~\ref{prop:DRidentificationcount} and \ref{prop:regularoutcome}. Specifically, the counterfactuals are estimated using normalized sample analogs of equations~\eqref{noprobidentification2} and \eqref{DiDidentDRsimple}.

To estimate these effects, we use the \textit{didmedDML} function, which is available in the \textit{causalweight} package by \citet{BodoryHuber2018} for \textsf{R} \citep{Rcore2025}. We employ four-fold cross-fitting and estimate all nuisance parameters by cross-validated lasso regression, with logit specifications for the propensity scores $\pi_{d,t}(X)$ and $\pi_{d,t}(M,X)$ and linear specifications for the conditional mean outcomes $\mu_d(t,X)$ and $\mu_{d,M}(t,X)$. To safeguard against too influential observations in inverse probability weighting, we impose a trimming rule based on the estimated propensity scores. Observations in the subgroups $(d,0)$, $(d',1)$, and $(d',0)$ are discarded whenever either $\pi_{d,t}(X)$ or $\pi_{d,t}(M,X)$ falls below a pre-specified threshold, which we set to 0.05 in all simulations. Inference is conducted using an asymptotic approximation. Standard errors are computed based on the estimated variance of the score function associated with the estimator.

The upper panel of Table \ref{tab:sim} provides the simulation results for estimating the (total) ATET and decomposing it into the natural direct and indirect effects. The true ATET equals $\Delta_1 (1,0,M(1),M(0))=E[Y_1(1)-Y_1(0)\mid D=1, T=1]=1.5+E[M(1)\mid D=1,T=1]=2.419$, with a natural direct effect of $\Delta_1 (1,0,M(1),M(1))=E[Y_1(1,M(1))-Y_1(0,M(1))\mid D=1, T=1]=1+E[M(1)\mid D=1,T=1]=1.919$ and a natural indirect effect of $\Delta_1 (0,0,M(1),M(0))=E[Y_1(0,M(1))-Y_1(0,M(0))\mid D=1, T=1]=E[M(1)-M(0)\mid D=1,T=1]=0.5$. For each estimator and sample size, the table reports the bias (`bias'), standard deviation (`std'), root mean squared error (`rmse'), average standard error (`avse'), and the coverage rate of the 95\% confidence intervals (`cover').

\begin{table}[t]
		\begin{center}
        \begin{footnotesize}
        \renewcommand{\arraystretch}{1.5}
			\caption{Simulation results for repeated cross sections}
			\label{tab:sim}
			\begin{tabular}{c|ccccc|ccccc}
				\hline\hline
    &\multicolumn{10}{c}{natural direct and indirect effects}\\
        &\multicolumn{5}{c|}{$n=2000$}&\multicolumn{5}{c}{$n=8000$}\\
				     estimator & bias & std & rmse & avse & cover &   bias & std & rmse &avse & cover\\ 
				\hline 
$\hat\Delta_1(1,0,M(1),M(1))$ & 0.049 & 0.264 & 0.268 & 0.247 & 0.926 & 0.024 & 0.139 & 0.141 & 0.128 & 0.925 \\ 
  $\hat\Delta_1(0,0,M(1),M(0))$ & 0.029 & 0.137 & 0.140 & 0.125 & 0.925 & 0.025 & 0.073 & 0.077 & 0.069 & 0.925 \\ 
  $\hat\Delta_1(1,0,M(1),M(0))$ & 0.079 & 0.266 & 0.277 & 0.256 & 0.927 & 0.049 & 0.138 & 0.147 & 0.130 & 0.924 \\
\hline
     &\multicolumn{10}{c}{dynamic treatment and controlled direct effects}\\
        &\multicolumn{5}{c|}{$n=2000$}&\multicolumn{5}{c}{$n=8000$}\\
				estimator & bias & std & rmse & avse & cover &  
    bias & std & rmse &avse & cover\\ 
				\hline
  $\hat\Delta_{1,1}(1,1,0,0)$ & 0.049 & 0.318 & 0.322 & 0.306 & 0.944 & 0.033 & 0.167 & 0.171 & 0.154 & 0.923 \\ 
  $\hat\Delta_{1,0}(1,0,0,0)$ & 0.090 & 0.321 & 0.333 & 0.311 & 0.937 & 0.075 & 0.162 & 0.178 & 0.148 & 0.895 \\ 
  $\hat\Delta_{1,1}(1,1,0,1)$ & 0.008 & 0.288 & 0.288 & 0.265 & 0.927 & 0.010 & 0.145 & 0.146 & 0.133 & 0.923 \\ 
     \hline
			\end{tabular}
		\end{footnotesize}
        \end{center}
		\par
		{\scriptsize Notes: columns `bias',  `std', `rmse', `avse', and `cover' provide the bias of the estimator, its standard deviation, its root mean squared error, the average of the standard error, and the coverage rate of the 95\% confidence intervals, respectively. $\hat\Delta_1(1,0,M(1),M(1))$ is the natural direct effect, $\hat\Delta_1(0,0,M(1),M(0))$ is the natural indirect effect and $\hat\Delta_1(1,0,M(1),M(0))$ is the total effect. $\hat\Delta_{d,m}(d,m,d',m')$ is the dynamic treatment effect of $(d,m)$ versus $(d',m')$, which is a controlled direct effect if $m=m'$.}
	\end{table}

For $n=2000$, all estimators exhibit only small finite-sample bias. A direct comparison in the table suggests that the natural indirect effect is least biased; however, because its true effect size is smaller than those of the natural direct and total effects, its bias is more pronounced relative to the magnitude of the true effect. Root mean squared errors are moderate across all estimators. Regarding statistical inference, the average standard errors tend to be slightly smaller than the corresponding empirical standard deviations, resulting in mild undercoverage of the 95\% confidence intervals. This indicates that statistical inference is slightly optimistic in finite samples. When the sample size is increased to $n=8000$, the standard deviations are approximately halved, which suggests that the estimators converge to the true effects at an $n^{-1/2}$ rate.

Next, we focus on dynamic treatment effects and controlled direct effects in repeated cross sections. To this end, we slightly modify the DGP. In contrast to the previous setting with a continuous mediator, we now assume that the mediator is binary. Specifically, the mediator equation in the DGP is replaced by
\[
    M = I\{X\beta+0.5\cdot D+V_m>0\}.
\]
This modification ensures that the mediator levels $M=m$ and $M=m'$ used to define the dynamic treatment and controlled direct effects occur with positive probability. All other components of the DGP remain unchanged. 

As before, we consider sample sizes of $n=2000$ and $n=8000$ and generate 1000 independent simulation replications. The effect of interest is the ATET defined in equation~\eqref{ATETdef}. Estimation is based on the normalized sample analog of equation~\eqref{DiDidentDR} in Proposition \ref{prop:DRidentification}. We examine three scenarios that differ in the values at which $d$, $d'$, $m$, and $m'$ are fixed. Specifically, we estimate (i) the ATET of receiving both the treatment and the mediator versus receiving neither, (ii) the controlled direct effect under $m=m'=0$, and (iii) the controlled direct effect under $m=m'=1$.

We estimate the effects with the \textit{dyndidDML} function of the \textit{causalweight} package \citep{BodoryHuber2018}. The procedure mirrors that used for the natural direct and indirect effects. We again employ four-fold cross-fitting with nuisance parameters estimated via cross-validated lasso, using logit specifications for the propensity scores and linear specifications for the conditional mean outcomes. The trimming rule and statistical inference are implemented as before. 

The lower panel of Table \ref{tab:sim} provides the simulation results for estimating the ATET at different values of $d$, $d'$, $m$, and $m'$. The true effects equal $\Delta_{1,1}(1,1,0,0) = E[Y_1(1,1) - Y_1(0,0) \mid D=1, M=1, T=1]=4-1=3$, $\Delta_{1,0}(1,0,0,0) = E[Y_1(1,0) - Y_1(0,0) \mid D=1, M=0, T=1]=2-1=1$ and $\Delta_{1,1}(1,1,0,1) = E[Y_1(1,1) - Y_1(0,1) \mid D=1, M=1, T=1]=4-2=2$ in the three scenarios considered.

The results are broadly similar to those for the natural direct and indirect effects in the upper panel. For $n=2000$, all estimators again exhibit small finite-sample bias, although their standard deviations and RMSEs are somewhat larger in absolute terms than those of the natural effects. Coverage rates remain slightly below 95\%, reflecting that the average standard errors tend to be slightly smaller than the standard deviations. For $n=8000$, the standard deviations are approximately half as large as those at $n=2000$, as expected for estimators that converge to the true effects at an $n^{-1/2}$ rate. In terms of nearly all the performance metrics considered, the results for the controlled direct effect under $m=m'=0$ are somewhat worse than in the other two scenarios. This likely reflects that, under our DGP, the treatment–mediator combination $d=1$ and $m=0$ occurs least frequently, so support is weakest in this case.

We now turn to the case of panel data. As for repeated cross sections, we first focus on natural direct and indirect effects and consider a slightly modified version of the initial DGP in which $X$ only affects the outcome in the post-treatment period:
\begin{eqnarray*}
X &=& (X_1,...,X_p),\\
D &=& I\{X\beta+0.5\cdot U + V_d>0\},\\
M &=& X\beta+0.5\cdot D + V_m,\\
Y_t &=& (1+X\beta+D+M+D\cdot M)\cdot t + U+W_t \textrm{ for } t\in\{0,1\},\\
X_j,U,V_d,V_m,W_0,W_1 &\sim& \mathcal{N}(0,1)\textrm{ independent of each other and for }j \in \{1,...,p\}.
\end{eqnarray*}

The simulation setup is the same as for repeated cross sections with respect to sample sizes, the number of replications, the number of folds for cross-fitting, the specifications for nuisance parameter estimation, and inference. However, the counterfactual outcomes entering the total, natural direct, and natural indirect effects are now estimated using normalized sample analogs of equations~\eqref{DiDidentDRcounterfacpanel2} and \eqref{DiDidentDRsimplepanel} in Propositions~\ref{prop:DRidentificationcountpanel} and \ref{prop:regularoutcomepanel}. To estimate these effects, we use the \textit{didmedDMLpanel} function of the \textit{causalweight} package \citep{BodoryHuber2018}. Since each unit is observed in both periods, reweighting is no longer required across group--time cells but only within the control group $D=d'$. Trimming is therefore restricted to this group, again using a threshold of 0.05.

The upper panel of Table~\ref{tab:sim2} reports the results for the natural direct and indirect effects. The true ATET equals $\Delta_1(1,0,M(1),M(0))=E[Y_1(1)-Y_1(0)\mid D=1]=1.5+E[M(1)\mid D=1]=2.116$, with a natural direct effect of $\Delta_1(1,0,M(1),M(1))=E[Y_1(1,M(1))-Y_1(0,M(1))\mid D=1]=1+E[M(1)\mid D=1]=1.616$ and a natural indirect effect of $\Delta_1(0,0,M(1),M(0))=E[Y_1(0,M(1))-Y_1(0,M(0))\mid D=1]=E[M(1)-M(0)\mid D=1]=0.5$.

\begin{table}[t]
		\begin{center}
        \begin{footnotesize}
        \renewcommand{\arraystretch}{1.5}
			\caption{Simulation results for panel data}
			\label{tab:sim2}
			\begin{tabular}{c|ccccc|ccccc}
				\hline\hline
    &\multicolumn{10}{c}{natural direct and indirect effects}\\
        &\multicolumn{5}{c|}{$n=2000$}&\multicolumn{5}{c}{$n=8000$}\\
				estimator & bias & std & rmse & avse & cover &   bias & std & rmse &avse & cover\\ 
				\hline 
    $\hat\Delta_1(1,0,M(1),M(1))$ & 0.025 & 0.084 & 0.087 & 0.090 & 0.956 & 0.007 & 0.040 & 0.041 & 0.045 & 0.971 \\ 
    $\hat\Delta_1(0,0,M(1),M(0))$ & 0.005 & 0.055 & 0.056 & 0.051 & 0.932 & 0.002 & 0.027 & 0.027 & 0.027 & 0.947 \\ 
    $\hat\Delta_1(1,0,M(1),M(0))$ & 0.030 & 0.105 & 0.109 & 0.113 & 0.960 & 0.010 & 0.049 & 0.050 & 0.056 & 0.972 \\
    \hline
     &\multicolumn{10}{c}{dynamic treatment and controlled direct effects}\\
        &\multicolumn{5}{c|}{$n=2000$}&\multicolumn{5}{c}{$n=8000$}\\
				estimator & bias & std & rmse & avse & cover &  bias & std & rmse &avse & cover\\ 
	\hline
    $\hat\Delta_{1,1}(1,1,0,0)$ & 0.054 & 0.111 & 0.123 & 0.142 & 0.985 & 0.019 & 0.054 & 0.057 & 0.070 & 0.987 \\ 
    $\hat\Delta_{1,0}(1,0,0,0)$ & 0.014 & 0.108 & 0.109 & 0.125 & 0.974 & 0.004 & 0.052 & 0.052 & 0.061 & 0.982 \\ 
    $\hat\Delta_{1,1}(1,1,0,1)$ & 0.021 & 0.092 & 0.094 & 0.110 & 0.970 & 0.009 & 0.044 & 0.045 & 0.054 & 0.988 \\ 
    \hline
			\end{tabular}
		\end{footnotesize}
        \end{center}
		\par
		{\scriptsize Notes: columns `bias',  `std', `rmse', `avse', and `cover' provide the bias of the estimator, its standard deviation, its root mean squared error, the average of the standard error, and the coverage rate of the 95\% confidence intervals, respectively. $\hat\Delta_1(1,0,M(1),M(1))$ is the natural direct effect, $\hat\Delta_1(0,0,M(1),M(0))$ is the natural indirect effect and  $\hat\Delta_1(1,0,M(1),M(0))$ is the total effect. $\hat\Delta_{d,m}(d,m,d',m')$ is the dynamic treatment effect of $(d,m)$ versus $(d',m')$, which is a controlled direct effect if $m=m'$.}
	\end{table}

For $n=2000$, the bias of all estimators is small, amounting to less than 2\% of the respective true effect. Bias, standard deviation, and RMSE are all smaller than for repeated cross sections. Unlike in the repeated cross-section case, the average standard errors slightly exceed the standard deviations for the total and natural direct effects, resulting in mild overcoverage, while the natural indirect effect exhibits mild undercoverage. For $n=8000$, the biases decrease further and the standard deviations are again approximately halved, consistent with $n^{-1/2}$-convergence. The coverage rate of the natural indirect effect is close to the nominal level at this sample size, whereas the other two estimators remain slightly conservative.

Finally, we consider dynamic treatment and controlled direct effects in panel data. Analogously to the repeated cross-section case, we replace the mediator equation by $M = I\{X\beta+0.5\cdot D+V_m>0\}$, while all other components of the panel DGP remain unchanged. Estimation is now based on the normalized sample analog of equation~\eqref{DiDidentDRpanel} in Proposition~\ref{prop:DRidentificationpanel}. The corresponding estimator is provided by the \textit{dyndidDMLpanel} function of the \textit{causalweight} package \citep{BodoryHuber2018}. The three scenarios considered for $d$, $d'$, $m$, and $m'$ and all other implementation details are as in the corresponding analysis for repeated cross sections, except that trimming is again restricted to the control group, here defined by $D=d'$ and $M=m'$.

The lower panel of Table~\ref{tab:sim2} reports the results. The true effects remain 3, 1, and 2 in the three scenarios. For $n=2000$, the bias of all estimators is again small. The average standard errors exceed the standard deviations more markedly than for the natural direct and indirect effects, so that inference is conservative, with coverage rates between 97\% and 99\%. Increasing the sample size to $n=8000$ further reduces the biases and approximately halves the standard deviations. The latter indicates that the estimators converge to the true effects at an $n^{-1/2}$ rate. Coverage remains above the nominal 95\% level.

\section{Empirical application}\label{appl}

In this section, we illustrate our approach using an empirical application on health care coverage and health outcomes. We examine the causal effect of health care coverage ($D$) on general health ($Y$) and decompose this effect into an indirect effect operating through the incidence of routine checkups ($M$) and a direct effect capturing all remaining causal mechanisms. Our analysis uses data from the National Longitudinal Survey of Youth 1997 (NLSY97), a nationally representative U.S. sample of 8984 respondents who were ages 12--17 at the first interview in 1997.\footnote{For additional information on the NLSY97 sample, see U.S. Bureau of Labor Statistics \citeyearpar{nlsy97}.} We consider the same empirical setting as \cite{farbmacher2022causal}, which allows us to compare our findings with theirs, while our DiD framework implies a different sample, set of identifying assumptions, and target parameter.

The panel structure of the NLSY97 provides the temporal ordering of treatment, mediator, and outcome required by our approach. The treatment is a binary indicator equal to one if the respondent had any kind of health care coverage in 2006 and zero otherwise. The outcome variable is self-reported general health, recorded on a five-point ordinal scale with categories excellent, very good, good, fair, and poor, where higher values correspond to poorer health status. We observe the pre-treatment outcome in 2005 and the post-treatment outcome in 2008. The mediator captures whether the respondent visited a doctor for a routine checkup within the past 12 months and is measured in 2007, i.e., after treatment but before the post-treatment outcome. The control variables comprise a rich set of demographic, socioeconomic, household, and health-related characteristics, including interactions and higher-order terms. This set corresponds to that of \cite{farbmacher2022causal}, except that we exclude lagged outcomes and their interactions.

In our analysis, we focus on individuals without any form of health care coverage before 2006, so that all respondents are untreated in the pre-treatment period. The treatment thus corresponds to gaining coverage in 2006, such that our ATET refers to the effect on those who transition into coverage. We further require that respondents report their general health in both 2005 and 2008. This leaves a sample of 1020 respondents, of whom 245 gain coverage in 2006.

To illustrate the composition of the groups underlying our estimates, Table \ref{tab:desc} in Appendix~\ref{app:desc} presents descriptive statistics for selected variables from the set of nearly 1000 covariates. Treated and untreated respondents are broadly comparable across most of the reported characteristics. The same holds when comparing respondents by mediator status. Table~\ref{tab:desc} reports the same variables as \cite{farbmacher2022causal}, which allows for a direct comparison of the group composition. Because we focus on respondents without prior coverage, the groups in our sample are considerably more balanced in observed characteristics than in their sample, which also includes individuals covered before 2006.

Given that the NLSY97 is a panel dataset, we apply the panel-data version of our DR estimator, as implemented in the \textit{didmedDMLpanel} function of the \textit{causalweight} package by \cite{BodoryHuber2018}, to assess the total, natural direct, and natural indirect effects of health care coverage on self-reported general health. As in the simulations, our estimators are based on four-fold cross-fitting. The nuisance parameters are estimated by cross-validated lasso regression. Following \cite{farbmacher2022causal}, we set the trimming threshold to 0.02 (2\%). However, no observations are dropped due to this trimming rule.

Table \ref{tab:res} reports the estimated effects, standard errors, and $p$-values. The ATET is not statistically significant at any conventional level, nor are the natural direct and natural indirect effects. For individuals who obtain health care coverage, we thus find no evidence of a short-term effect of coverage on general health, either through routine checkups or via other causal pathways. The point estimates are negative throughout, corresponding to a health improvement, as lower values of the outcome indicate better health, but they are small relative to their standard errors. Almost all of the estimated total effect is attributed to the direct effect, while the indirect effect through routine checkups is close to zero. An additional analysis with routine checkups as the outcome suggests that health care coverage significantly increases their incidence by about 9 percentage points.\footnote{We estimate this effect with the ATETDML function of the \textit{causalweight} package by \cite{BodoryHuber2018}, using the change in the mediator between 2005 and 2007 as the outcome. As in our main analysis, we use the same set of covariates, estimate the nuisance parameters by cross-validated lasso regression, and set the number of folds used in cross-fitting to four. The estimated effect is 0.089 and statistically significant at the 5\% level ($p$=0.019).} Yet, the checkups induced by coverage do not appear to translate into measurably better self-reported health over the horizon considered.

\begin{table}[t]
\begin{center}
\caption{Total, natural direct and natural indirect effects on general health in 2008}
\label{tab:res}
\renewcommand{\arraystretch}{1.5}
\begin{tabular}{ccccc}
  \hline
 & effect & se & $t$-val & $p$-val \\ 
  \hline
  $\hat\Delta_1(1,0,M(1),M(0))$ & -0.089 & 0.083 & -1.067 & 0.286 \\ 
  $\hat\Delta_1(1,0,M(1),M(1))$ & -0.085 & 0.083 & -1.028 & 0.304 \\ 
  $\hat\Delta_1(0,0,M(1),M(0))$ & -0.003 & 0.010 & -0.341 & 0.733
  \\
   \hline
\end{tabular}
\end{center}
\par
{\scriptsize Notes: columns `effect',  `se', `$t$-val', and `$p$-val', provide the effect estimate, standard error, $t$-value and $p$-value, respectively. $\hat\Delta_1(1,0,M(1),M(0))$ is the total effect, $\hat\Delta_1(1,0,M(1),M(1))$ is the natural direct effect, and $\hat\Delta_1(0,0,M(1),M(0))$ is the natural indirect effect}.
\end{table}

The similar empirical setting allows us to compare our results to those of \cite{farbmacher2022causal}, who rely on selection on observables. Nevertheless, since they estimate effects for the overall population rather than for the treated, the comparison is not direct. Most of the natural direct and indirect effects reported in \cite{farbmacher2022causal} are not statistically significant. Their ATE, however, is significant at the 10\% or 5\% level, depending on the estimation approach, whereas our ATET is not. Our point estimates of the total and direct effects are larger in absolute value than theirs, but the corresponding standard errors are considerably larger. This lower precision is unsurprising given that we focus on a much smaller subsample of the NLSY97. In both applications, the indirect effect through routine checkups is much smaller than the direct effect. Overall, despite the different subsample, identification strategy, and target parameters, our findings point in the same direction as those of \cite{farbmacher2022causal}.

\section{Conclusion}\label{sec:conc}

This paper suggests a flexible difference-in-differences (DiD) framework for causal mediation and dynamic treatment effect analysis. By extending the DiD design beyond total treatment effects, our approach allows evaluating controlled direct effects, natural direct and indirect effects operating through mediators, and joint treatment–mediator (or dynamic treatment) effects. Identification of controlled direct effects and dynamic treatment effects is achieved under conditional parallel trends assumptions on mean potential outcomes across treatment--mediator states. For the identification of natural direct and indirect effects, we impose additional parallel trends assumptions on mean potential outcomes or mediator distributions across treatment states. 

On the estimation side, we propose doubly robust DiD estimators within the double machine learning framework for both repeated cross sections and panel data. By combining machine learning–based nuisance estimation with Neyman-orthogonal score functions and cross-fitting, our estimators allow for data-driven covariate adjustment while retaining valid large-sample inference under specific regularity conditions. We demonstrate the favorable finite-sample performance of our methods in a simulation study. We also provide an empirical application using data from the National Longitudinal Survey of Youth 1997 to disentangle the direct effect of health care coverage on general health from the indirect effect operating through routine checkups. Despite negative point estimates throughout, suggesting health improvements, we find no statistically significant short-term effect of health care coverage on general health among those who obtain coverage, either via routine checkups or other causal pathways.


\pagebreak


\appendix
\clearpage
\phantomsection
\section*{Appendix}
\addcontentsline{toc}{section}{Appendix}
\renewcommand{\thesubsection}{\Alph{subsection}}
\numberwithin{equation}{subsection}
\numberwithin{proposition}{subsection}
\numberwithin{table}{subsection}

{\small
\subsection{Identification of dynamic treatment effects in the total population}\label{ATEdef}

Our framework permits identifying the average treatment effect (ATE) in the total population in the post-treatment period when Assumptions \ref{ass1}--\ref{ass4} are strengthened to hold not only for the population with $D=d, M=m$, but in an analogous manner also for the population with $D=d',M=m'$. The ATE is defined as
\begin{align}\label{defeffectATE}
\Delta(d,m,d',m')=E[Y_1(d,m)-Y_1(d',m')|T=1].
\end{align}
This result follows because, under the stated assumptions applied symmetrically to both $(D=d,M=m)$ and $(D=d',M=m')$ cells, the conditional mean potential outcomes $E[Y_1(d,m)\mid X]$ and $E[Y_1(d',m')\mid X]$ are identified for all values of $X$ in the population. 

Assuming for instance that both $D$ and $M$ are binary, imposing Assumption \ref{ass1} symmetrically to $(D=1,M=1)$ and $(D=0,M=0)$ imposes parallel trends both without and with treatment and mediation, that is, $E[Y_1(0,0)-Y_0(0,0)|D=1,M=1,X]=E[Y_1(0,0)-Y_0(0,0)|D=0,M=0,X]$ and $E[Y_1(1,1)-Y_0(1,1)|D=1,M=1,X]=E[Y_1(1,1)-Y_0(1,1)|D=0,M=0,X]$, respectively. It is worth noting that these two assumptions jointly imply homogeneity in average effects. To see this, note that by subtracting the two parallel trend conditions, we obtain
\begin{align}\label{homeffects}
&E[Y_1(0,0)-Y_0(0,0)-Y_1(1,1)+Y_0(1,1)|D=1,M=1,X]\notag\\
&=E[Y_1(0,0)-Y_0(0,0)-Y_1(1,1)+Y_0(1,1)|D=0,M=0,X]\notag\\
&=E[Y_1(0,0)-Y_0(0,0)-Y_1(1,1)+Y_0(0,0)|D=1,M=1,X]\notag\\
&=E[Y_1(0,0)-Y_0(0,0)-Y_1(1,1)+Y_0(0,0)|D=0,M=0,X]\notag\\
&=E[Y_1(0,0)-Y_1(1,1)|D=1,M=1,X]=E[Y_1(0,0)-Y_1(1,1)|D=0,M=0,X],
\end{align}
where the second and third equalities follow from Assumption \ref{ass2}. Under this stronger set of assumptions, the ATE is identified by
\begin{align}\label{effectATE}
\Delta(d,m,d',m')&=E[\mu_{d,m}(1,X)-\mu_{d,m}(0,X)-\{\mu_{d',m'}(1,X)-\mu_{d',m'}(0,X)\}|T=1].
\end{align}

A corresponding DR expression for identifying the ATE that is Neyman orthogonal (as formally shown in Appendix~\ref{Neymanrepeated}) is given in Proposition \ref{prop:DRidentification2}, where $p_t(X) = \Pr(T = t \mid X)$ denotes the conditional period probability given covariates.

\begin{proposition}\label{prop:DRidentification2}
Under Assumptions \ref{ass1}--\ref{ass4} and assuming that these assumptions also hold when swapping  $D=d',M=m'$ and $D=d, M=m$, the ATE is identified by the following expression, which satisfies Neyman orthogonality: 
\begin{align}\label{DiDidentDRATE}
&\Delta(d,m,d',m')=E\left[  \frac{ T \cdot [\mu_{d,m}(1,X)-\mu_{d,m}(0,X)-\{\mu_{d',m'}(1,X)-\mu_{d',m'}(0,X)\}]}{\Pr(T=1)}\right.\\
&+ \frac{(Y_T-\mu_{D,M}(T,X))\cdot p_1(X)}{\Pr(T=1)} \cdot \left\{\frac{I\{D=d\}\cdot I\{M=m\}\cdot T }{ \rho_{d,m,1}(X)}-  \frac{I\{D=d\}\cdot I\{M=m\}\cdot (1-T)}{ \rho_{d,m,0}(X)}\right.\notag\\
&- \left.\left.\left(\frac{I\{D=d'\}\cdot I\{M=m'\}\cdot  T}{ \rho_{d',m',1}(X)} - \frac{I\{D=d'\}\cdot I\{M=m'\}\cdot (1-T)}{ \rho_{d',m',0}(X)}\right) \right\}   \right].\notag
\end{align}
\end{proposition}

We now turn to panel data, where the ATE is defined as $\Delta(d,m,d',m')=E[Y_1(d,m)-Y_1(d',m')]$ and the DR expression in Proposition \ref{prop:DRidentification2} simplifies. Since each subject is observed in both time periods, $\Pr(T=1)=\Pr(T=0)$ and $p_1(X)=\Pr(T=1\mid X)=\Pr(T=0\mid X)$, such that reweighting based on $T$, $\Pr(T=1)$, and $p_1(X)$ is no longer required. This yields the equation stated in Proposition \ref{prop:DRidentification2panel}, whose Neyman orthogonality is shown in Appendix \ref{Neymanpanel}. The equation is closely related to the standard DR expression for ATE identification under the selection-on-observables assumption, with the difference that here, outcome differences (rather than outcome levels) and indicators for both the treatment and the mediator (rather than the treatment alone) enter the expression.
\begin{proposition}\label{prop:DRidentification2panel}
Under Assumptions \ref{ass1}--\ref{ass4} and assuming that these assumptions also hold when swapping  $D=d',M=m'$ and $D=d, M=m$, the ATE is identified by the following expression, which satisfies Neyman orthogonality: 
\begin{align}\label{DiDidentDRATEpanel}
&\Delta(d,m,d',m')=E\bigg[ \left[\mu_{d,m}(X)-\mu_{d',m'}(X)\right] \Bigg.\\
&+ \left. \left[Y_1-Y_0-\mu_{D,M}(X)\right]\cdot \left(\frac{I\{D=d\}\cdot I\{M=m\}}{ \rho_{d,m}(X)}-  \frac{I\{D=d'\}\cdot I\{M=m'\}}{ \rho_{d',m'}(X)} \right)   \right].\notag
\end{align}
\end{proposition}

\subsection{Identification and Neyman orthogonality in repeated cross sections}\label{Neymanrepeated}

We first show that the score function underlying DR expression \eqref{DiDidentDR} for repeated cross sections satisfies Assumption 3.1(a)--(e) of \cite{Chetal2018}, implying that expression \eqref{DiDidentDR}  identifies the ATET $\Delta_{d,m}(d,m,d',m')=E[Y_1(d,m)-Y_1(d',m')|D=d,M=m,T=1]$ under Assumptions \ref{ass1} to \ref{ass4} and satisfies Neyman orthogonality. In the subsequent discussion, all bounds hold uniformly over all probability laws $P \in \mathcal{P}$, where $\mathcal{P}$ is the set of all possible probability laws, and we omit $P$ for brevity. We define the nuisance parameters 
$$\eta=(\mu_{d,m}(0,X),\mu_{d',m'}(0,X),\mu_{d',m'}(1,X),\rho_{d,m,1}(X),\rho_{d,m,0}(X), \rho_{d',m',1}(X),\rho_{d',m',0}(X))$$ 
and denote their true values by 
$$\eta^0=(\mu^0_{d,m}(0,X),\mu^0_{d',m'}(0,X),\mu^0_{d',m'}(1,X),\rho^0_{d,m,1}(X),\rho^0_{d,m,0}(X), \rho^0_{d',m',1}(X),\rho^0_{d',m',0}(X)).$$ 
We also define $\Psi_{d',m',t}=E[\mu_{d',m'}(t,X)|D=d,M=m, T=1]$ to be the average of the conditional mean $\mu_{d',m'}(t,X)$ given $D=d$, $M=m$, and $T=1$, in order to express the ATET provided in equation \eqref{effect} as a function of $\Psi_{d',m',t}$:
\begin{align}\label{nuisancepar_mod}
\Delta_{d,m}(d,m,d',m') &= E[Y_1|D=d,M=m] - \Psi_{d,m,0} - \Psi_{d',m',1} + \Psi_{d',m',0},
\end{align}
Therefore, considering Neyman-orthogonal functions for $\Psi_{d',m',t}$ yields a Neyman-orthogonal expression of the ATET.

Let us denote by $\phi_{d',m',t}$ the Neyman-orthogonal score function for $\Psi_{d',m',t}$, which corresponds to the following expression for $W=(Y,D,M,X,T)$:
\begin{align}\label{neymanscore_mod}
\phi_{d',m',t}(W,\eta,\Psi_{d',m',t}) &= \frac{I\{D=d\}\cdot I\{M=m\}\cdot T \cdot \mu_{d',m'}(t,X)}{\Pi_{d,m,1}} \\
&+ \frac{I\{D=d'\}\cdot I\{M=m'\}\cdot I\{T=t\}\cdot \rho_{d,m,1}(X)\cdot (Y_T - \mu_{d',m'}(t,X))}{\rho_{d',m',t}(X)\cdot \Pi_{d,m,1}} - \Psi_{d',m',t}.\notag
\end{align}
We note that the DR expression \eqref{DiDidentDR} for ATET identification is obtained by identifying $\Psi_{d',m',t}$ as the solution to the moment condition $E[\phi_{d',m',t}(W,\eta^0,\Psi_{d',m',t}) ]=0$ and plugging it into equation \eqref{nuisancepar_mod}. To show that $E[\phi_{d',m',t}(W,\eta^0,\Psi_{d',m',t})]=0$ holds, we consider the expectation of the second term in equation \eqref{neymanscore_mod}, evaluated at $\eta^0$, and note that it is zero. This follows from applying the law of iterated expectations and basic probability theory: 
\begin{align}
&E\left[\frac{I\{D=d'\}\cdot I\{M=m'\}\cdot I\{T=t\}\cdot \rho^0_{d,m,1}(X)\cdot (Y_T - \mu^0_{d',m'}(t,X))}{\rho^0_{d',m',t}(X)\cdot \Pi_{d,m,1}}\right]\notag\\
&=E\left[ E\left[\frac{\rho^0_{d,m,1}(X)}{\Pi_{d,m,1}} \cdot  \frac{I\{D=d'\}\cdot I\{M=m'\}\cdot I\{T=t\}\cdot (Y_T-\mu^0_{D,M}(T,X))}{\rho^0_{d',m',t}(X)}\Big| X\right]  \right]\notag\\
&=E\left[ \frac{\rho^0_{d,m,1}(X)}{\Pi_{d,m,1}} \cdot \frac{E\left[ I\{D=d'\}\cdot I\{M=m'\}\cdot I\{T=t\}\cdot (Y_T-\mu^0_{D,M}(T,X))| X\right] }{\rho^0_{d',m',t}(X)}  \right].\notag
\end{align}
We have by basic probability theory and iterated expectations that
\begin{align}
\frac{E [I\{D=d'\} \cdot I\{M=m'\} \cdot I\{T=t\} \cdot Y_T|X]}{\rho^0_{d',m',t}(X)} &= E[Y_t|D=d', M=m', X] = \mu^0_{d',m'}(t,X) \notag
\end{align}
and
\begin{align}
\frac{E [I\{D=d'\} \cdot I\{M=m'\} \cdot I\{T=t\} \cdot \mu^0_{d',m'}(t,X)|X]}{\rho^0_{d',m',t}(X)} &= \frac{\rho^0_{d',m',t}(X) \cdot \mu^0_{d',m'}(t,X)}{\rho^0_{d',m',t}(X)} = \mu^0_{d',m'}(t,X), \notag
\end{align}
and thus, the expression corresponds to 
\begin{align}
E \left[\frac{\rho^0_{d,m,1}(X)}{\Pi_{d,m,1}}\cdot[\mu^0_{d',m'}(t,X)-\mu^0_{d',m'}(t,X)]\right]=0.
\end{align}
For this reason, 
\begin{align}
E[\phi_{d',m',t}(W,\eta^0,\Psi_{d',m',t})] &= E\left[\frac{I\{D=d\}\cdot I\{M=m\}\cdot T \cdot \mu^0_{d',m'}(t,X)}{\Pi_{d,m,1}}\right] - \Psi_{d',m',t}\\
&= E\left[ \mu^0_{d',m'}(t,X) | D=d, M=m, T=1\right] - \Psi_{d',m',t} = \Psi_{d',m',t}-\Psi_{d',m',t}=0.\notag
\end{align}
The score function in equation \eqref{neymanscore_mod} thus satisfies Assumption 3.1(a) of \cite{Chetal2018}. It also satisfies Assumptions 3.1(b) and (c) of \cite{Chetal2018} because the score function is linear in $\Psi_{d',m',t}$ and the second Gateaux derivative $\eta \mapsto E[\phi_{d',m',t}(W,\eta,\Psi_{d',m',t})]$ is continuous. 

The Gateaux derivative of $E[\phi_{d',m',t}(W,\eta^0,\Psi_{d',m',t})]$ in the direction of $[\eta - \eta^0]$ is given by:
\begin{align}\label{derivscore_mod}
&\partial E[\phi_{d',m',t}(W,\eta^0,\Psi_{d',m',t})] [\eta-\eta^0]  \\
&=  E \left[\frac{I\{D=d\}\cdot I\{M=m\}\cdot T \cdot [\mu_{d',m'}(t,X) - \mu^0_{d',m'}(t,X)]}{\Pi_{d,m,1}}\right]\notag\\
&-  E \left[\frac{I\{D=d'\}\cdot I\{M=m'\}\cdot I\{T=t\}\cdot \rho^0_{d,m,1}(X) \cdot [\mu_{d',m'}(t,X) - \mu^0_{d',m'}(t,X)]}{\rho^0_{d',m',t}(X) \cdot \Pi_{d,m,1}}\right] \notag\\
&+  E \left[\frac{I\{D=d'\}\cdot I\{M=m'\}\cdot I\{T=t\}\cdot [Y_T - \mu^0_{d',m'}(t,X)]}{\rho^0_{d',m',t}(X)} \cdot \frac{[\rho_{d,m,1}(X) - \rho^0_{d,m,1}(X)]}{\Pi_{d,m,1}}\right] \notag\\
&-  E \left[\frac{I\{D=d'\}\cdot I\{M=m'\}\cdot I\{T=t\} \cdot \rho^0_{d,m,1}(X)\cdot [Y_T - \mu^0_{d',m'}(t,X)]}{\rho^0_{d',m',t}(X)\cdot \Pi_{d,m,1}} \cdot \frac{\rho_{d',m',t}(X) - \rho^0_{d',m',t}(X)}{\rho^0_{d',m',t}(X)}\right] = 0.\notag
\end{align}
To see that the Gateaux derivative is zero, we first consider the first term and apply the law of iterated expectations:
\begin{align}
&E \left[\frac{I\{D=d\}\cdot I\{M=m\}\cdot T \cdot [\mu_{d',m'}(t,X) - \mu^0_{d',m'}(t,X)]}{\Pi_{d,m,1}}\right] \\
&= E \left[E\left[\frac{I\{D=d\}\cdot I\{M=m\}\cdot T \cdot [\mu_{d',m'}(t,X) - \mu^0_{d',m'}(t,X)]}{\Pi_{d,m,1}}\Big|X\right]\right] \notag\\
& =E \left[\frac{E[I\{D=d\}\cdot I\{M=m\}\cdot T|X] \cdot [\mu_{d',m'}(t,X) - \mu^0_{d',m'}(t,X)]}{\Pi_{d,m,1}}\right] \notag\\
& = E\left[ \frac{\rho^0_{d,m,1}(X) \cdot [\mu_{d',m'}(t,X) - \mu^0_{d',m'}(t,X)]}{\Pi_{d,m,1}}\right].\notag
\end{align}
Concerning the second term, we have 
\begin{align}
 &- E \left[\frac{I\{D=d'\}\cdot I\{M=m'\}\cdot I\{T=t\} \cdot \rho^0_{d,m,1}(X) \cdot [\mu_{d',m'}(t,X) - \mu^0_{d',m'}(t,X)]}{\rho^0_{d',m',t}(X) \cdot \Pi_{d,m,1}}\right]\notag\\
& = - E \left[\frac{ \rho^0_{d',m',t}(X) \cdot \rho^0_{d,m,1}(X) \cdot [\mu_{d',m'}(t,X) - \mu^0_{d',m'}(t,X)]}{\rho^0_{d',m',t}(X) \cdot \Pi_{d,m,1}}\right]\notag\\
&= - E \left[\frac{\rho^0_{d,m,1}(X) \cdot [\mu_{d',m'}(t,X) - \mu^0_{d',m'}(t,X)]}{\Pi_{d,m,1}}\right].
\end{align}
Since the first and second terms involve identical expressions but with opposite signs, they cancel out.
Applying the law of iterated expectations to the third term, we obtain
\begin{align}
& E \left[\frac{I\{D=d'\}\cdot I\{M=m'\}\cdot I\{T=t\}\cdot [Y_T - \mu^0_{d',m'}(t,X)]}{\rho^0_{d',m',t}(X)} \cdot \frac{[\rho_{d,m,1}(X) - \rho^0_{d,m,1}(X)]}{\Pi_{d,m,1}}\right] \\
&= E \left[E \left[\frac{I\{D=d'\}\cdot I\{M=m'\}\cdot I\{T=t\}\cdot [Y_T - \mu^0_{d',m'}(t,X)]}{\rho^0_{d',m',t}(X)} \cdot \frac{[\rho_{d,m,1}(X) - \rho^0_{d,m,1}(X)]}{\Pi_{d,m,1}} \Bigg| X\right]\right] \notag\\
&= E \left[\frac{E\left[ I\{D=d'\}\cdot I\{M=m'\}\cdot I\{T=t\}\cdot [Y_T - \mu^0_{d',m'}(t,X)] \mid X \right]}{\rho^0_{d',m',t}(X)} \cdot \frac{[\rho_{d,m,1}(X) - \rho^0_{d,m,1}(X)]}{\Pi_{d,m,1}}\right].\notag
\end{align}
Concerning the first factor, we have by basic probability theory and iterated expectations that
\begin{align}
\frac{E [I\{D=d'\} \cdot I\{M=m'\} \cdot I\{T=t\} \cdot Y_T|X]}{\rho^0_{d',m',t}(X)} &= E[Y_t|D=d', M=m', X] = \mu^0_{d',m'}(t,X) \notag
\end{align}
and
\begin{align}
\frac{E [I\{D=d'\} \cdot I\{M=m'\} \cdot I\{T=t\} \cdot \mu^0_{d',m'}(t,X)|X]}{\rho^0_{d',m',t}(X)} &= \frac{\rho^0_{d',m',t}(X) \cdot \mu^0_{d',m'}(t,X)}{\rho^0_{d',m',t}(X)} = \mu^0_{d',m'}(t,X). \notag
\end{align}
Therefore, the third term corresponds to 
\begin{align}
E \left[  [\mu^0_{d',m'}(t,X) - \mu^0_{d',m'}(t,X)]  \cdot \frac{[\rho_{d,m,1}(X) - \rho^0_{d,m,1}(X)]}{\Pi_{d,m,1}}\right]=0.
\end{align}
Finally, we apply the same reasoning to the fourth term to obtain
\begin{align}
&- E \left[\frac{I\{D=d'\}\cdot I\{M=m'\}\cdot I\{T=t\} \cdot \rho^0_{d,m,1}(X)\cdot [Y_T - \mu^0_{d',m'}(t,X)]}{\rho^0_{d',m',t}(X)\cdot \Pi_{d,m,1}} \cdot \frac{\rho_{d',m',t}(X) - \rho^0_{d',m',t}(X)}{\rho^0_{d',m',t}(X)}\right]\notag\\
&=- E \left[\frac{E \left[I\{D=d'\}\cdot I\{M=m'\}\cdot I\{T=t\} \cdot \rho^0_{d,m,1}(X)\cdot [Y_T - \mu^0_{d',m'}(t,X)]|X\right]}{\rho^0_{d',m',t}(X)\cdot \Pi_{d,m,1}} \cdot \frac{\rho_{d',m',t}(X) - \rho^0_{d',m',t}(X)}{\rho^0_{d',m',t}(X)}\right]\notag\\
&= - E \left[\frac{ \rho^0_{d,m,1}(X)\cdot [\mu^0_{d',m'}(t,X) - \mu^0_{d',m'}(t,X)]}{ \Pi_{d,m,1}} \cdot \frac{\rho_{d',m',t}(X) - \rho^0_{d',m',t}(X)}{\rho^0_{d',m',t}(X)}\right]=0.
\end{align}
For this reason, $\partial E\left[\phi_{d',m',t}\left(W,\eta^0,\Psi_{d',m',t}\right)\right] \left[\eta-\eta^0\right]=0$, such that the Gateaux derivative of $E\left[\phi_{d',m',t}\left(W,\eta^0,\Psi_{d',m',t}\right)\right]$ is equal to zero, which satisfies Assumption 3.1(d) of \cite{Chetal2018}. The Gateaux derivative of $E[Y_1|D=d,M=m]$ is trivially equal to zero, too. For this reason, expression \eqref{nuisancepar_mod}, which is a linear combination of these terms, satisfies Neyman orthogonality. Assumption 3.1(e) of \cite{Chetal2018}, which states that the factor multiplying $\Psi_{d',m',t}$ in the definition of the score function $\phi_{d',m',t}$ must be bounded in expectation, holds trivially because the factor is a constant, specifically $-1$.

Next, we consider the counterfactual $E[Y_1(d',M(d))|D=d,T=1]$, which is identified by equation~\eqref{mediationcounterfactual} and can be expressed in terms of $\Psi_{d',M,t}$ as
\begin{align}\label{decomp_noprob}
    E\left[ Y_{1}\left( d',M(d) \right)\mid D = d,T = 1 \right] = \Psi_{d,M,0} + \Psi_{d',M,1} - \Psi_{d',M,0},
\end{align}
when defining $\Psi_{d',M,t} = E\left[ \mu_{d',M}(t,X) \mid D = d,T = 1 \right]$. Let us denote the nuisance parameters by
$$\eta = \left( \mu_{d,M}(0,X),\mu_{d',M}(1,X),\mu_{d',M}(0,X),\pi_{d,1}(M,X),\pi_{d,0}(M,X),\pi_{d',1}(M,X),\pi_{d',0}(M,X) \right)$$
and by $\phi_{d',M,t}$ the Neyman-orthogonal score function of $\Psi_{d',M,t}$, which is defined as:
\begin{align}\label{neymanscore_noprob}
\phi_{d',M,t}\left( W,\eta,\Psi_{d',M,t} \right)
  &= \frac{I\left\{ D = d \right\} \cdot T \cdot \mu_{d',M}(t,X)}{P_{d,1}}\\
  &+ \frac{I\left\{ D = d' \right\} \cdot I\left\{ T = t \right\} \cdot \pi_{d,1}(M,X) \cdot \left( Y_{T} - \mu_{d',M}(t,X) \right)}{\pi_{d',t}(M,X) \cdot P_{d,1}}- \Psi_{d',M,t}.\notag
\end{align}
We note that the DR expression~\eqref{noprobidentification2} is obtained by identifying $\Psi_{d',M,t}$ as the solution to the moment condition $E\left[ \phi_{d',M,t}\left( W,\eta^{0},\Psi_{d',M,t} \right) \right] = 0$ and plugging it into equation~\eqref{decomp_noprob}. To show that $E\left[ \phi_{d',M,t}\left( W,\eta^{0},\Psi_{d',M,t} \right) \right] = 0$ holds, we consider the expectation of the second term in equation~\eqref{neymanscore_noprob}, evaluated at $\eta^{0}$, and note that it is zero. This follows from applying the law of iterated expectations and basic probability theory:
\begin{align}
  &E\left[ \frac{I\left\{ D = d' \right\} \cdot I\left\{ T = t \right\} \cdot \pi_{d,1}^{0}(M,X) \cdot (Y_{T} - \mu_{d',M}^{0}(t,X))}{\pi_{d',t}^{0}(M,X) \cdot P_{d,1}} \right]\notag\\
  &= E\left[ \frac{\pi_{d,1}^{0}(M,X)}{P_{d,1}} \cdot \frac{E\left[ I\left\{ D = d' \right\} \cdot I\left\{ T = t \right\} \cdot (Y_{T} - \mu_{d',M}^{0}(t,X))| M,X \right]}{\pi_{d',t}^{0}(M,X)} \right].\notag
\end{align}
We have by basic probability theory and iterated expectations that
\begin{align}
\frac{E[ I\left\{ D = d' \right\} \cdot I\left\{ T = t \right\} \cdot Y_{T}\mid M,X]}{\pi_{d',t}^{0}(M,X)} &= E\left[ Y_t \mid D = d',M,X \right] = \mu_{d',M}^{0}(t,X) \notag
\end{align}
and
\begin{align}
\frac{E[I\left\{ D = d' \right\} \cdot I\left\{ T = t \right\} \cdot \mu_{d',M}^{0}(t,X)\mid M,X]}{\pi_{d',t}^{0}(M,X)} &= \frac{\pi_{d',t}^{0}(M,X) \cdot \mu_{d',M}^{0}(t,X)}{\pi_{d',t}^{0}(M,X)} = \mu_{d',M}^{0}(t,X), \notag
\end{align}
and thus, the expression corresponds to
\begin{align}
E\left[ \frac{\pi_{d,1}^{0}(M,X)}{P_{d,1}} \cdot [\mu_{d',M}^{0}(t,X) - \mu_{d',M}^{0}(t,X)] \right] = 0.
\end{align}
For this reason,
\begin{align}
E\left[ \phi_{d',M,t}\left( W,\eta^{0},\Psi_{d',M,t} \right) \right]
  &= E\left[ \frac{I\left\{ D = d \right\} \cdot T \cdot \mu_{d',M}^{0}(t,X)}{P_{d,1}} \right] - \Psi_{d',M,t}\\
  &= E\left[ \mu_{d',M}^{0}(t,X) \mid D = d,T = 1 \right] - \Psi_{d',M,t}= \Psi_{d',M,t} - \Psi_{d',M,t} = 0.\notag
\end{align}
The score function in equation \eqref{neymanscore_noprob} thus satisfies Assumption 3.1(a) of \cite{Chetal2018}. It also satisfies Assumptions 3.1(b) and (c) of \cite{Chetal2018} because the score function is linear in $\Psi_{d',M,t}$ and the second Gateaux derivative $\eta \mapsto E\left[ \phi_{d',M,t}\left( W,\eta,\Psi_{d',M,t} \right) \right]$ is continuous.

The Gateaux derivative of $E\left[ \phi_{d',M,t}\left( W,\eta^{0},\Psi_{d',M,t} \right) \right]$ in the direction of $\left[ \eta - \eta^{0} \right]$ is given by:
\begin{align}
    & \partial E\left[ \phi_{d',M,t}\left( W,\eta^{0},\Psi_{d',M,t} \right) \right]\left[ \eta - \eta^{0} \right]\\
    & = E\left[ \frac{I\left\{ D = d \right\} \cdot T \cdot [\mu_{d',M}(t,X) - \mu_{d',M}^{0}(t,X)]}{P_{d,1}} \right]\notag\\
    & - E\left[ \frac{I\left\{ D = d' \right\} \cdot I\left\{ T = t \right\} \cdot \pi_{d,1}^{0}(M,X) \cdot [\mu_{d',M}(t,X) - \mu_{d',M}^{0}(t,X)]}{\pi_{d',t}^{0}(M,X) \cdot P_{d,1}} \right]\notag\\
    & + E\left[ \frac{I\left\{ D = d' \right\} \cdot I\left\{ T = t \right\} \cdot [Y_{T} - \mu_{d',M}^{0}(t,X)]}{\pi_{d',t}^{0}(M,X)} \cdot \frac{\pi_{d,1}(M,X) - \pi_{d,1}^{0}(M,X)}{P_{d,1}} \right]\notag\\
    & - E\left[ \frac{I\left\{ D = d' \right\} \cdot I\left\{ T = t \right\} \cdot \pi_{d,1}^{0}(M,X) \cdot [Y_{T} - \mu_{d',M}^{0}(t,X)]}{\pi_{d',t}^{0}(M,X) \cdot P_{d,1}} \cdot \frac{\pi_{d',t}(M,X) - \pi_{d',t}^{0}(M,X)}{\pi_{d',t}^{0}(M,X)} \right] = 0.\notag
\end{align}
To see that the Gateaux derivative is zero, we first consider the first term and apply the law of iterated expectations:
\begin{align}
    & E\left[ \frac{I\left\{ D = d \right\} \cdot T \cdot [\mu_{d',M}(t,X) - \mu_{d',M}^{0}(t,X)]}{P_{d,1}} \right]\notag\\
    & = E\left[ E\left[ \frac{I\left\{ D = d \right\} \cdot T \cdot [\mu_{d',M}(t,X) - \mu_{d',M}^{0}(t,X)]}{P_{d,1}}\Big| M,X\right] \right]\notag\\
    & = E\left[ \frac{E\left[ I\left\{ D = d \right\} \cdot T\mid M,X\right] \cdot [\mu_{d',M}(t,X) - \mu_{d',M}^{0}(t,X)]}{P_{d,1}} \right]\notag\\
    & = E\left[ \frac{\pi_{d,1}^0(M,X) \cdot [\mu_{d',M}(t,X) - \mu_{d',M}^{0}(t,X)]}{P_{d,1}}\right].
\end{align}
Concerning the second term, we have
\begin{align}
  &- E\left[ \frac{I\left\{ D = d' \right\} \cdot I\left\{ T = t \right\} \cdot \pi_{d,1}^{0}(M,X) \cdot [\mu_{d',M}(t,X) - \mu_{d',M}^{0}(t,X) ]}{\pi_{d',t}^{0}(M,X) \cdot P_{d,1}} \right]\notag\\
  &= - E\left[ \frac{\pi_{d',t}^{0}(M,X) \cdot \pi_{d,1}^{0}(M,X) \cdot [\mu_{d',M}(t,X) - \mu_{d',M}^{0}(t,X)]}{\pi_{d',t}^{0}(M,X) \cdot P_{d,1}} \right]\notag\\
  &= - E\left[ \frac{\pi_{d,1}^{0}(M,X) \cdot [\mu_{d',M}(t,X) - \mu_{d',M}^{0}(t,X)]}{P_{d,1}}\right].
\end{align}
Since the first and second terms involve identical expressions but with opposite signs, they cancel out. Applying the law of iterated expectations to the third term, we obtain
\begin{align}
  &E\left[ \frac{I\left\{ D = d' \right\} \cdot I\left\{ T = t \right\} \cdot [Y_{T} - \mu_{d',M}^{0}(t,X)]}{\pi_{d',t}^{0}(M,X)} \cdot \frac{\pi_{d,1}(M,X) - \pi_{d,1}^{0}(M,X)}{P_{d,1}} \right]\\
  &= E\left[ E\left[ \frac{I\left\{ D = d' \right\} \cdot I\left\{ T = t \right\} \cdot [Y_{T} - \mu_{d',M}^{0}(t,X)]}{\pi_{d',t}^{0}(M,X)} \cdot \frac{\pi_{d,1}(M,X) - \pi_{d,1}^{0}(M,X)}{P_{d,1}} \Bigg| M,X \right] \right]\notag\\
  &= E\left[ \frac{E\left[ I\left\{ D = d' \right\} \cdot I\left\{ T = t \right\} \cdot [Y_{T} - \mu_{d',M}^{0}(t,X)]\mid M,X \right]}{\pi_{d',t}^{0}(M,X)} \cdot \frac{\pi_{d,1}(M,X) - \pi_{d,1}^{0}(M,X)}{P_{d,1}} \right].\notag
\end{align}
Concerning the first factor, we have by basic probability theory and iterated expectations that
\begin{align}
\frac{E[ I\left\{ D = d' \right\} \cdot I\left\{ T = t \right\} \cdot Y_{T}\mid M,X]}{\pi_{d',t}^{0}(M,X)} &= E\left[ Y_{t}\mid D = d',M,X \right] = \mu_{d',M}^{0}(t,X) \notag
\end{align}
and
\begin{align}
\frac{E[I\left\{ D = d' \right\} \cdot I\left\{ T = t \right\} \cdot \mu_{d',M}^{0}(t,X)\mid M,X]}{\pi_{d',t}^{0}(M,X)} &= \frac{\pi_{d',t}^{0}(M,X) \cdot \mu_{d',M}^{0}(t,X)}{\pi_{d',t}^{0}(M,X)} = \mu_{d',M}^{0}(t,X). \notag
\end{align}
Therefore, the third term corresponds to
\begin{align}
E\left[ [\mu_{d',M}^{0}(t,X) - \mu_{d',M}^{0}(t,X)] \cdot \frac{\pi_{d,1}(M,X) - \pi_{d,1}^{0}(M,X)}{P_{d,1}} \right] = 0.
\end{align}
Finally, we apply the same reasoning to the fourth term to obtain
\begin{align}
  &- E\left[ \frac{I\left\{ D = d' \right\} \cdot I\left\{ T = t \right\} \cdot \pi_{d,1}^{0}(M,X) \cdot [Y_{T} - \mu_{d',M}^{0}(t,X)] }{\pi_{d',t}^{0}(M,X) \cdot P_{d,1}} \cdot \frac{\pi_{d',t}(M,X) - \pi_{d',t}^{0}(M,X)}{\pi_{d',t}^{0}(M,X)} \right]\notag\\
&= -E\Biggl[\frac{E\left[ I\left\{ D = d' \right\} \cdot I\left\{ T = t \right\} \cdot \pi_{d,1}^{0}(M,X)
    \cdot [Y_{T} - \mu_{d',M}^{0}(t,X)]\mid M,X \right]}{\pi_{d',t}^{0}(M,X) \cdot P_{d,1}}\notag\\
&\qquad \cdot \frac{\pi_{d',t}(M,X) - \pi_{d',t}^{0}(M,X)}{\pi_{d',t}^{0}(M,X)}\Biggr]\notag\\
  &= - E\left[ \frac{\pi_{d,1}^{0}(M,X) \cdot [\mu_{d',M}^{0}(t,X) - \mu_{d',M}^{0}(t,X)]}{P_{d,1}} \cdot \frac{\pi_{d',t}(M,X) - \pi_{d',t}^{0}(M,X)}{\pi_{d',t}^{0}(M,X)} \right] = 0.
\end{align}
For this reason, $\partial E\left[ \phi_{d',M,t}\left( W,\eta^{0},\Psi_{d',M,t} \right) \right]\left[ \eta - \eta^{0} \right] = 0$,
such that the Gateaux derivative of $E\left[ \phi_{d',M,t}\left( W,\eta^{0},\Psi_{d',M,t} \right) \right]$ is equal to zero, which satisfies Assumption 3.1(d) of \cite{Chetal2018}. For this reason, expression \eqref{decomp_noprob}, which is a linear combination of such terms, satisfies Neyman orthogonality. Assumption 3.1(e) of \cite{Chetal2018}, which states that the factor multiplying $\Psi_{d',M,t}$ in the definition of the score function $\phi_{d',M,t}$ must be bounded in expectation, holds trivially because the factor is a constant, specifically $-1$.

In a next step, we consider the identification of the counterfactual $E[Y_1(d',M(d'))|D=d,T=1]$ based on equation \eqref{standarddid}, which corresponds to
\begin{align}\label{psi_simple}
E[Y_1(d',M(d'))|D=d,T=1]=\Psi_{d,0} + \Psi_{d',1} - \Psi_{d',0}
\end{align}
when defining $\Psi_{d',t}= E[\mu_{d'}(t,X)|D=d,T=1]$. Let us denote the nuisance parameters by
$$\eta=(\mu_{d}(0,X),\mu_{d'}(1,X),\mu_{d'}(0,X),\pi_{d,1}(X),\pi_{d,0}(X), \pi_{d',1}(X),\pi_{d',0}(X))$$ and by $\phi_{d',t}$ the Neyman-orthogonal score function of $\Psi_{d',t}$, which is defined as:
\begin{align}\label{neymanscore_modsimple}
\phi_{d',t}(W,\eta,\Psi_{d',t}) &= \frac{I\{D=d\}\cdot T \cdot \mu_{d'}(t,X)}{P_{d,1}} \\
&+ \frac{I\{D=d'\}\cdot I\{T=t\}\cdot \pi_{d,1}(X)\cdot (Y_T - \mu_{d'}(t,X))}{\pi_{d',t}(X)\cdot P_{d,1}} - \Psi_{d',t}.\notag
\end{align}
We note that the DR expression \eqref{DiDidentDRsimple} for the identification of the counterfactual $E[Y_1(d',M(d'))|D=d,T=1]$ is obtained by identifying $\Psi_{d',t}$ as the solution to the moment condition $E[\phi_{d',t}(W,\eta^0,\Psi_{d',t}) ]=0$ and plugging it into equation \eqref{psi_simple}. Showing that Assumptions 3.1(a)--(e) of \cite{Chetal2018} hold for \(\phi_{d',t}(W,\eta,\Psi_{d',t})\) proceeds analogously to the case of \(\phi_{d',M,t}(W,\eta,\Psi_{d',M,t})\) and is therefore omitted.

Next, we consider the alternative identification of the counterfactual based on equation \eqref{doubleparalleltrends}, which corresponds to
\begin{align}\label{psi_simple_doubl}
E[Y_1(d',M(d'))|D=d,T=1]=\Psi_{d,0}+\sum_{m \in \mathcal{M}}\Psi(m)
\end{align}
when defining $\Psi_{d,0}=E[\mu_{d}(0,X)|D=d,T=1]$ and $\Psi(m)=E[a_m(X)\cdot b_m(X)|D=d,T=1]$, with $a_m(X)=\mu_{d',m}(1,X)-\mu_{d',m}(0,X)$ and $b_m(X)=\nu_{d,m}(0,X)+\nu_{d',m}(1,X)-\nu_{d',m}(0,X)$. Identification and Neyman orthogonality of $\Psi_{d,0}$ follow from the previous results for $\Psi_{d',t}$ when replacing $d'$ by $d$ and setting $t=0$. For $\Psi(m)$, let us denote the nuisance parameters by
\begin{align}
\eta=(&\mu_{d',m}(1,X),\mu_{d',m}(0,X),\nu_{d,m}(0,X),\nu_{d',m}(1,X),\nu_{d',m}(0,X),\notag\\
&\rho_{d',m,1}(X),\rho_{d',m,0}(X),\pi_{d,1}(X),\pi_{d,0}(X),\pi_{d',1}(X),\pi_{d',0}(X))\notag
\end{align}
and their true values by $\eta^0$, whose elements carry the superscript $0$. Accordingly, $a^0_m(X)=\mu^0_{d',m}(1,X)-\mu^0_{d',m}(0,X)$ and $b^0_m(X)=\nu^0_{d,m}(0,X)+\nu^0_{d',m}(1,X)-\nu^0_{d',m}(0,X)$. We denote by $\phi(m)$ the Neyman-orthogonal score function of $\Psi(m)$, which corresponds to the expression in \eqref{DiDidentDRsimple2doubl} without the terms for $\Psi_{d,0}$ and with the indicators written out separately for each period $t\in\{0,1\}$:
\begin{align}\label{neymanscore_doubl}
&\phi(m)(W,\eta,\Psi(m))=\frac{I\{D=d\}\cdot T\cdot a_m(X)\cdot b_m(X)}{P_{d,1}}-\Psi(m)\notag\\
&+b_m(X)\cdot\sum_{t\in\{0,1\}}(2t-1)\cdot\frac{I\{D=d'\}\cdot I\{M=m\}\cdot I\{T=t\}\cdot \pi_{d,1}(X)\cdot (Y_T-\mu_{d',m}(t,X))}{\rho_{d',m,t}(X)\cdot P_{d,1}}\notag\\
&+a_m(X)\cdot\frac{I\{D=d\}\cdot (1-T)\cdot \pi_{d,1}(X)\cdot (I\{M_T=m\}-\nu_{d,m}(0,X))}{\pi_{d,0}(X)\cdot P_{d,1}}\notag\\
&+a_m(X)\cdot\sum_{t\in\{0,1\}}(2t-1)\cdot\frac{I\{D=d'\}\cdot I\{T=t\}\cdot \pi_{d,1}(X)\cdot (I\{M_T=m\}-\nu_{d',m}(t,X))}{\pi_{d',t}(X)\cdot P_{d,1}}.
\end{align}
The factor $(2t-1)$ equals $1$ for $t=1$ and $-1$ for $t=0$. DR expression \eqref{DiDidentDRsimple2doubl} is obtained by identifying $\Psi(m)$ as the solution to the moment condition $E[\phi(m)(W,\eta^0,\Psi(m))]=0$ for each $m$ and plugging it into equation \eqref{psi_simple_doubl}. To show that this moment condition holds, we first derive two sets of conditional moment restrictions. We have by basic probability theory and iterated expectations that
\begin{align}
E[I\{D=d'\}\cdot I\{M=m\}\cdot I\{T=t\}\cdot Y_T|X]&=\rho^0_{d',m,t}(X)\cdot E[Y_t|D=d',M=m,X]\notag\\
&=\rho^0_{d',m,t}(X)\cdot\mu^0_{d',m}(t,X),\notag\\
E[I\{D=d'\}\cdot I\{M=m\}\cdot I\{T=t\}\cdot\mu^0_{d',m}(t,X)|X]&=\rho^0_{d',m,t}(X)\cdot\mu^0_{d',m}(t,X),\notag
\end{align}
such that for $t\in\{0,1\}$,
\begin{align}\label{resid1_doubl}
E[I\{D=d'\}\cdot I\{M=m\}\cdot I\{T=t\}\cdot (Y_T-\mu^0_{d',m}(t,X))|X]=0.
\end{align}
In the same way, we have for $d^*\in\{d,d'\}$ that
\begin{align}
E[I\{D=d^*\}\cdot I\{T=t\}\cdot I\{M_T=m\}|X]&=\pi^0_{d^*,t}(X)\cdot \Pr(M_t=m|D=d^*,X)\notag\\
&=\pi^0_{d^*,t}(X)\cdot\nu^0_{d^*,m}(t,X),\notag\\
E[I\{D=d^*\}\cdot I\{T=t\}\cdot\nu^0_{d^*,m}(t,X)|X]&=\pi^0_{d^*,t}(X)\cdot\nu^0_{d^*,m}(t,X),\notag
\end{align}
such that for $(d^*,t)\in\{(d,0),(d',1),(d',0)\}$,
\begin{align}\label{resid2_doubl}
E[I\{D=d^*\}\cdot I\{T=t\}\cdot (I\{M_T=m\}-\nu^0_{d^*,m}(t,X))|X]=0.
\end{align}
Applying the law of iterated expectations to the second term in equation \eqref{neymanscore_doubl}, evaluated at $\eta^0$, and using \eqref{resid1_doubl}, we obtain
\begin{align}
&E\left[b^0_m(X)\cdot\sum_{t\in\{0,1\}}(2t-1)\cdot\frac{I\{D=d'\}\cdot I\{M=m\}\cdot I\{T=t\}\cdot \pi^0_{d,1}(X)\cdot (Y_T-\mu^0_{d',m}(t,X))}{\rho^0_{d',m,t}(X)\cdot P_{d,1}}\right]\notag\\
&=\sum_{t\in\{0,1\}}(2t-1)\cdot E\left[\frac{b^0_m(X)\cdot\pi^0_{d,1}(X)}{\rho^0_{d',m,t}(X)\cdot P_{d,1}}\right.\notag\\
&\qquad\quad\left.{}\cdot E[I\{D=d'\}\cdot I\{M=m\}\cdot I\{T=t\}\cdot (Y_T-\mu^0_{d',m}(t,X))|X]\vphantom{\frac{b^0_m(X)\cdot\pi^0_{d,1}(X)}{\rho^0_{d',m,t}(X)\cdot P_{d,1}}}\right]=0.\notag
\end{align}
Likewise, the expectations of the third and fourth terms evaluated at $\eta^0$ are zero by \eqref{resid2_doubl}, with $(d^*,t)=(d,0)$ for the third term and $(d^*,t)\in\{(d',1),(d',0)\}$ for the fourth term. For instance, for the third term,
\begin{align}
&E\left[a^0_m(X)\cdot\frac{I\{D=d\}\cdot (1-T)\cdot \pi^0_{d,1}(X)\cdot (I\{M_T=m\}-\nu^0_{d,m}(0,X))}{\pi^0_{d,0}(X)\cdot P_{d,1}}\right]\notag\\
&=E\left[\frac{a^0_m(X)\cdot\pi^0_{d,1}(X)}{P_{d,1}}\cdot \frac{E[I\{D=d\}\cdot (1-T)\cdot (I\{M_T=m\}-\nu^0_{d,m}(0,X))|X]}{\pi^0_{d,0}(X)}\right]=0.\notag
\end{align}
Concerning the first term, we obtain
\begin{align}
E[\phi(m)(W,\eta^0,\Psi(m))]&=E\left[\frac{I\{D=d\}\cdot T\cdot a^0_m(X)\cdot b^0_m(X)}{P_{d,1}}\right]-\Psi(m)\notag\\
&=E[a^0_m(X)\cdot b^0_m(X)|D=d,T=1]-\Psi(m)=\Psi(m)-\Psi(m)=0.\notag
\end{align}
The score function in equation \eqref{neymanscore_doubl} thus satisfies Assumption 3.1(a) of \cite{Chetal2018}. It also satisfies Assumptions 3.1(b) and (c) of \cite{Chetal2018} because the score function is linear in $\Psi(m)$ and the second Gateaux derivative $\eta \mapsto E[\phi(m)(W,\eta,\Psi(m))]$ is continuous.

The Gateaux derivative of $E[\phi(m)(W,\eta^0,\Psi(m))]$ in the direction of $[\eta - \eta^0]$ is given by:
\begin{align}\label{derivscore_doubl}
&\partial E[\phi(m)(W,\eta^0,\Psi(m))] [\eta-\eta^0]\notag\\
&=E\left[\frac{I\{D=d\}\cdot T\cdot [a_m(X)-a^0_m(X)]\cdot b^0_m(X)}{P_{d,1}}\right]+E\left[\frac{I\{D=d\}\cdot T\cdot a^0_m(X)\cdot [b_m(X)-b^0_m(X)]}{P_{d,1}}\right]\notag\\
&+E\left[\sum_{t\in\{0,1\}}(2t-1)\cdot\frac{[b_m(X)-b^0_m(X)]\cdot\pi^0_{d,1}(X)}{P_{d,1}}\right.\notag\\
&\qquad\quad\left.{}\cdot\frac{I\{D=d'\}\cdot I\{M=m\}\cdot I\{T=t\}\cdot (Y_T-\mu^0_{d',m}(t,X))}{\rho^0_{d',m,t}(X)}\right]\notag\\
&-E\left[\sum_{t\in\{0,1\}}(2t-1)\cdot\frac{b^0_m(X)\cdot\pi^0_{d,1}(X)\cdot [\mu_{d',m}(t,X)-\mu^0_{d',m}(t,X)]}{P_{d,1}}\right.\notag\\
&\qquad\quad\left.{}\cdot\frac{I\{D=d'\}\cdot I\{M=m\}\cdot I\{T=t\}}{\rho^0_{d',m,t}(X)}\right]\notag\\
&+E\left[\sum_{t\in\{0,1\}}(2t-1)\cdot\frac{b^0_m(X)\cdot[\pi_{d,1}(X)-\pi^0_{d,1}(X)]}{P_{d,1}}\right.\notag\\
&\qquad\quad\left.{}\cdot\frac{I\{D=d'\}\cdot I\{M=m\}\cdot I\{T=t\}\cdot (Y_T-\mu^0_{d',m}(t,X))}{\rho^0_{d',m,t}(X)}\right]\notag\\
&-E\left[\sum_{t\in\{0,1\}}(2t-1)\cdot\frac{b^0_m(X)\cdot\pi^0_{d,1}(X)\cdot[\rho_{d',m,t}(X)-\rho^0_{d',m,t}(X)]}{\rho^0_{d',m,t}(X)\cdot P_{d,1}}\right.\notag\\
&\qquad\quad\left.{}\cdot\frac{I\{D=d'\}\cdot I\{M=m\}\cdot I\{T=t\}\cdot (Y_T-\mu^0_{d',m}(t,X))}{\rho^0_{d',m,t}(X)}\right]\notag\\
&+E\left[[a_m(X)-a^0_m(X)]\cdot\left\{\frac{I\{D=d\}\cdot (1-T)\cdot \pi^0_{d,1}(X)\cdot (I\{M_T=m\}-\nu^0_{d,m}(0,X))}{\pi^0_{d,0}(X)\cdot P_{d,1}}\right.\right.\notag\\
&\qquad\left.\left.+\sum_{t\in\{0,1\}}(2t-1)\cdot\frac{I\{D=d'\}\cdot I\{T=t\}\cdot \pi^0_{d,1}(X)\cdot (I\{M_T=m\}-\nu^0_{d',m}(t,X))}{\pi^0_{d',t}(X)\cdot P_{d,1}}\right\}\right]\notag\\
&-E\left[a^0_m(X)\cdot\left\{\frac{I\{D=d\}\cdot (1-T)\cdot \pi^0_{d,1}(X)\cdot [\nu_{d,m}(0,X)-\nu^0_{d,m}(0,X)]}{\pi^0_{d,0}(X)\cdot P_{d,1}}\right.\right.\notag\\
&\qquad\left.\left.+\sum_{t\in\{0,1\}}(2t-1)\cdot\frac{I\{D=d'\}\cdot I\{T=t\}\cdot \pi^0_{d,1}(X)\cdot [\nu_{d',m}(t,X)-\nu^0_{d',m}(t,X)]}{\pi^0_{d',t}(X)\cdot P_{d,1}}\right\}\right]\notag\\
&+E\left[a^0_m(X)\cdot\left\{\frac{I\{D=d\}\cdot (1-T)\cdot (I\{M_T=m\}-\nu^0_{d,m}(0,X))}{\pi^0_{d,0}(X)}\right.\right.\notag\\
&\qquad\left.\left.+\sum_{t\in\{0,1\}}(2t-1)\cdot\frac{I\{D=d'\}\cdot I\{T=t\}\cdot (I\{M_T=m\}-\nu^0_{d',m}(t,X))}{\pi^0_{d',t}(X)}\right\}\cdot\frac{\pi_{d,1}(X)-\pi^0_{d,1}(X)}{P_{d,1}}\right]\notag\\
&-E\left[a^0_m(X)\cdot\left\{\frac{I\{D=d\}\cdot (1-T)\cdot \pi^0_{d,1}(X)\cdot (I\{M_T=m\}-\nu^0_{d,m}(0,X))}{\pi^0_{d,0}(X)\cdot P_{d,1}}\cdot\frac{\pi_{d,0}(X)-\pi^0_{d,0}(X)}{\pi^0_{d,0}(X)}\right.\right.\notag\\
&\qquad\left.\left.+\sum_{t\in\{0,1\}}(2t-1)\cdot\frac{I\{D=d'\}\cdot I\{T=t\}\cdot \pi^0_{d,1}(X)\cdot (I\{M_T=m\}-\nu^0_{d',m}(t,X))}{\pi^0_{d',t}(X)\cdot P_{d,1}}\right.\right.\notag\\
&\qquad\quad\left.\left.{}\cdot\frac{\pi_{d',t}(X)-\pi^0_{d',t}(X)}{\pi^0_{d',t}(X)}\vphantom{\frac{I\{D=d'\}}{\pi^0_{d',t}(X)}}\right\}\right]=0,
\end{align}
where $a_m(X)-a^0_m(X)=\sum_{t\in\{0,1\}}(2t-1)\cdot[\mu_{d',m}(t,X)-\mu^0_{d',m}(t,X)]$ and $b_m(X)-b^0_m(X)=[\nu_{d,m}(0,X)-\nu^0_{d,m}(0,X)]+\sum_{t\in\{0,1\}}(2t-1)\cdot[\nu_{d',m}(t,X)-\nu^0_{d',m}(t,X)]$. The first and second terms stem from the first term of the score function, the third to sixth terms from its second term, and the seventh to tenth terms from its third and fourth terms.

To see that the Gateaux derivative is zero, we first consider the first term and apply the law of iterated expectations, using $E[I\{D=d\}\cdot T|X]=\pi^0_{d,1}(X)$:
\begin{align}
&E\left[\frac{I\{D=d\}\cdot T\cdot [a_m(X)-a^0_m(X)]\cdot b^0_m(X)}{P_{d,1}}\right]\notag\\
&=E\left[\frac{E[I\{D=d\}\cdot T|X]\cdot [a_m(X)-a^0_m(X)]\cdot b^0_m(X)}{P_{d,1}}\right]\notag\\
&=\sum_{t\in\{0,1\}}(2t-1)\cdot E\left[\frac{\pi^0_{d,1}(X)\cdot b^0_m(X)\cdot [\mu_{d',m}(t,X)-\mu^0_{d',m}(t,X)]}{P_{d,1}}\right].\notag
\end{align}
Concerning the fourth term, we use $E[I\{D=d'\}\cdot I\{M=m\}\cdot I\{T=t\}|X]=\rho^0_{d',m,t}(X)$ to obtain
\begin{align}
&-E\left[\sum_{t\in\{0,1\}}(2t-1)\cdot\frac{b^0_m(X)\cdot\pi^0_{d,1}(X)\cdot [\mu_{d',m}(t,X)-\mu^0_{d',m}(t,X)]}{P_{d,1}}\right.\notag\\
&\qquad\quad\left.{}\cdot\frac{I\{D=d'\}\cdot I\{M=m\}\cdot I\{T=t\}}{\rho^0_{d',m,t}(X)}\right]\notag\\
&=-\sum_{t\in\{0,1\}}(2t-1)\cdot E\left[\frac{\pi^0_{d,1}(X)\cdot b^0_m(X)\cdot [\mu_{d',m}(t,X)-\mu^0_{d',m}(t,X)]}{P_{d,1}}\right].\notag
\end{align}
Since the first and fourth terms involve identical expressions but with opposite signs, they cancel out. In the same way, $E[I\{D=d\}\cdot T|X]=\pi^0_{d,1}(X)$ implies for the second term
\begin{align}
&E\left[\frac{I\{D=d\}\cdot T\cdot a^0_m(X)\cdot [b_m(X)-b^0_m(X)]}{P_{d,1}}\right]=E\left[\frac{\pi^0_{d,1}(X)\cdot a^0_m(X)\cdot [\nu_{d,m}(0,X)-\nu^0_{d,m}(0,X)]}{P_{d,1}}\right]\notag\\
&+\sum_{t\in\{0,1\}}(2t-1)\cdot E\left[\frac{\pi^0_{d,1}(X)\cdot a^0_m(X)\cdot [\nu_{d',m}(t,X)-\nu^0_{d',m}(t,X)]}{P_{d,1}}\right],\notag
\end{align}
and $E[I\{D=d\}\cdot (1-T)|X]=\pi^0_{d,0}(X)$ and $E[I\{D=d'\}\cdot I\{T=t\}|X]=\pi^0_{d',t}(X)$ imply for the eighth term
\begin{align}
&-E\left[a^0_m(X)\cdot\left\{\frac{I\{D=d\}\cdot (1-T)\cdot \pi^0_{d,1}(X)\cdot [\nu_{d,m}(0,X)-\nu^0_{d,m}(0,X)]}{\pi^0_{d,0}(X)\cdot P_{d,1}}\right.\right.\notag\\
&\qquad\left.\left.+\sum_{t\in\{0,1\}}(2t-1)\cdot\frac{I\{D=d'\}\cdot I\{T=t\}\cdot \pi^0_{d,1}(X)\cdot [\nu_{d',m}(t,X)-\nu^0_{d',m}(t,X)]}{\pi^0_{d',t}(X)\cdot P_{d,1}}\right\}\right]\notag\\
&=-E\left[\frac{\pi^0_{d,1}(X)\cdot a^0_m(X)\cdot [\nu_{d,m}(0,X)-\nu^0_{d,m}(0,X)]}{P_{d,1}}\right]\notag\\
&-\sum_{t\in\{0,1\}}(2t-1)\cdot E\left[\frac{\pi^0_{d,1}(X)\cdot a^0_m(X)\cdot [\nu_{d',m}(t,X)-\nu^0_{d',m}(t,X)]}{P_{d,1}}\right],\notag
\end{align}
such that the second and eighth terms cancel out, too.

The third, fifth, and sixth terms are sums over $t$ of expectations of a function of $X$ multiplied by $I\{D=d'\}\cdot I\{M=m\}\cdot I\{T=t\}\cdot (Y_T-\mu^0_{d',m}(t,X))$. Applying the law of iterated expectations to the third term and using \eqref{resid1_doubl}, we obtain
\begin{align}
&E\left[\sum_{t\in\{0,1\}}(2t-1)\cdot\frac{[b_m(X)-b^0_m(X)]\cdot\pi^0_{d,1}(X)}{P_{d,1}}\right.\notag\\
&\qquad\quad\left.{}\cdot\frac{I\{D=d'\}\cdot I\{M=m\}\cdot I\{T=t\}\cdot (Y_T-\mu^0_{d',m}(t,X))}{\rho^0_{d',m,t}(X)}\right]\notag\\
&=\sum_{t\in\{0,1\}}(2t-1)\cdot E\left[\frac{[b_m(X)-b^0_m(X)]\cdot\pi^0_{d,1}(X)}{\rho^0_{d',m,t}(X)\cdot P_{d,1}}\right.\notag\\
&\qquad\quad\left.{}\cdot E[I\{D=d'\}\cdot I\{M=m\}\cdot I\{T=t\}\cdot (Y_T-\mu^0_{d',m}(t,X))|X]\vphantom{\frac{b^0_m(X)\cdot\pi^0_{d,1}(X)}{\rho^0_{d',m,t}(X)\cdot P_{d,1}}}\right]=0.\notag
\end{align}
The same reasoning applies to the fifth and sixth terms, which differ from the third term only in the first fraction. In the fifth term, this fraction is $b^0_m(X)\cdot[\pi_{d,1}(X)-\pi^0_{d,1}(X)]/P_{d,1}$. In the sixth term, it is $b^0_m(X)\cdot\pi^0_{d,1}(X)\cdot[\rho_{d',m,t}(X)-\rho^0_{d',m,t}(X)]/[\rho^0_{d',m,t}(X)\cdot P_{d,1}]$. Both fractions are functions of $X$, such that both terms are zero by \eqref{resid1_doubl}. Analogously, the seventh, ninth, and tenth terms are sums of expectations of a function of $X$ multiplied by $I\{D=d^*\}\cdot I\{T=t\}\cdot (I\{M_T=m\}-\nu^0_{d^*,m}(t,X))$ for $(d^*,t)\in\{(d,0),(d',1),(d',0)\}$. For the seventh term, \eqref{resid2_doubl} implies
\begin{align}
&E\left[[a_m(X)-a^0_m(X)]\cdot\left\{\frac{I\{D=d\}\cdot (1-T)\cdot \pi^0_{d,1}(X)\cdot (I\{M_T=m\}-\nu^0_{d,m}(0,X))}{\pi^0_{d,0}(X)\cdot P_{d,1}}\right.\right.\notag\\
&\qquad\left.\left.+\sum_{t\in\{0,1\}}(2t-1)\cdot\frac{I\{D=d'\}\cdot I\{T=t\}\cdot \pi^0_{d,1}(X)\cdot (I\{M_T=m\}-\nu^0_{d',m}(t,X))}{\pi^0_{d',t}(X)\cdot P_{d,1}}\right\}\right]\notag\\
&=E\left[\frac{[a_m(X)-a^0_m(X)]\cdot\pi^0_{d,1}(X)}{P_{d,1}}\cdot\left\{\frac{E[I\{D=d\}\cdot (1-T)\cdot (I\{M_T=m\}-\nu^0_{d,m}(0,X))|X]}{\pi^0_{d,0}(X)}\right.\right.\notag\\
&\qquad\left.\left.+\sum_{t\in\{0,1\}}(2t-1)\cdot\frac{E[I\{D=d'\}\cdot I\{T=t\}\cdot (I\{M_T=m\}-\nu^0_{d',m}(t,X))|X]}{\pi^0_{d',t}(X)}\right\}\right]=0.\notag
\end{align}
In the ninth term, the conditional expectations in \eqref{resid2_doubl} are multiplied by $a^0_m(X)\cdot[\pi_{d,1}(X)-\pi^0_{d,1}(X)]/[\pi^0_{d,0}(X)\cdot P_{d,1}]$ and $(2t-1)\cdot a^0_m(X)\cdot[\pi_{d,1}(X)-\pi^0_{d,1}(X)]/[\pi^0_{d',t}(X)\cdot P_{d,1}]$. In the tenth term, they are multiplied by $-a^0_m(X)\cdot\pi^0_{d,1}(X)\cdot[\pi_{d,0}(X)-\pi^0_{d,0}(X)]/[\pi^0_{d,0}(X)^2\cdot P_{d,1}]$ and $-(2t-1)\cdot a^0_m(X)\cdot\pi^0_{d,1}(X)\cdot[\pi_{d',t}(X)-\pi^0_{d',t}(X)]/[\pi^0_{d',t}(X)^2\cdot P_{d,1}]$. Both terms are therefore zero, too.

For this reason, $\partial E[\phi(m)(W,\eta^0,\Psi(m))] [\eta-\eta^0]=0$, such that the Gateaux derivative of $E[\phi(m)(W,\eta^0,\Psi(m))]$ is equal to zero, which satisfies Assumption 3.1(d) of \cite{Chetal2018}. Together with the Neyman orthogonality of the score function for $\Psi_{d,0}$, it follows that expression \eqref{psi_simple_doubl}, which is a linear combination of these terms, satisfies Neyman orthogonality. Assumption 3.1(e) of \cite{Chetal2018} holds trivially because the factor multiplying $\Psi(m)$ in the score function $\phi(m)$ is a constant, specifically $-1$. 

Finally, we show that the score function underlying DR expression \eqref{DiDidentDRATE} for repeated cross sections satisfies Assumption 3.1(a)--(e) of \cite{Chetal2018}, implying the identification of the ATE $\Delta(d,m,d',m')=E[Y_1(d,m)-Y_1(d',m')\mid T=1]$ under Assumptions \ref{ass1} to \ref{ass4} and satisfies Neyman orthogonality.
To this end, we define the nuisance parameters 
$$\eta=(\mu_{d^*,m^*}(t^*,X),\rho_{d^*,m^*,t^*}(X), p_1(X))$$ 
and denote their true values by 
$$\eta^0=(\mu^0_{d^*,m^*}(t^*,X),\rho^0_{d^*,m^*,t^*}(X), p^0_1(X)),$$
for $d^* \in \{d,d'\}$, $m^* \in \{m,m'\}$, and $t^*\in\{1,0\}$.
Furthermore, we denote by $\Psi_{d,m,t}=E[\mu_{d,m}(t,X)|T=1]$ the average of the conditional mean $\mu_{d,m}(t,X)$ given $T=1$, in order to express the ATE provided in equation \eqref{effectATE} as a function of $\Psi_{d',m',t}$:
\begin{align}\label{nuisancepar_mod2}
\Delta(d,m,d',m') &= \Psi_{d,m,1}  - \Psi_{d,m,0} - \Psi_{d',m',1} + \Psi_{d',m',0},
\end{align}
Let us denote by $\phi_{d,m,t}$ the Neyman-orthogonal score function for $\Psi_{d,m,t}$, which corresponds to the following expression:
\begin{align}\label{neymanscore_mod2}
\phi_{d,m,t}(W,\eta,\Psi_{d,m,t}) &= \frac{ T \cdot \mu_{d,m}(t,X)}{\Pr(T=1)} \\
&+ \frac{I\{D=d\}\cdot I\{M=m\}\cdot I\{T=t\}\cdot p_1(X)\cdot (Y_T - \mu_{d,m}(t,X))}{\rho_{d,m,t}(X)\cdot \Pr(T=1)} - \Psi_{d,m,t}. \notag
\end{align}
We note that the DR expression \eqref{DiDidentDRATE} for ATE identification is obtained by identifying $\Psi_{d,m,t}$ as the solution to the moment condition $E[\phi_{d,m,t} (W,\eta^0,\Psi_{d,m,t}) ]=0$ and plugging it into equation \eqref{nuisancepar_mod2}. To show that $E[\phi_{d,m,t}(W,\eta^0,\Psi_{d,m,t})]=0$ holds, we consider the expectation of the second term in equation \eqref{neymanscore_mod2}, evaluated at $\eta^0$, and note that it is zero. This follows from applying the law of iterated expectations and basic probability theory: 
\begin{align}
&E\left[\frac{I\{D=d\}\cdot I\{M=m\}\cdot I\{T=t\}\cdot p^0_{1}(X)\cdot (Y_T - \mu^0_{d,m}(t,X))}{\rho^0_{d,m,t}(X)\cdot \Pr(T=1)}\right]\notag\\
&=E\left[ \frac{p^0_{1}(X)}{\Pr(T=1)} \cdot  \frac{E\left[ I\{D=d\}\cdot I\{M=m\}\cdot I\{T=t\}\cdot (Y_T-\mu^0_{D,M}(T,X))| X\right] }{\rho^0_{d,m,t}(X)}  \right].\notag
\end{align}
We have by basic probability theory and iterated expectations that
\begin{align}
\frac{E [I\{D=d\} \cdot I\{M=m\} \cdot I\{T=t\} \cdot Y_T|X]}{\rho^0_{d,m,t}(X)} &= E[Y_t|D=d, M=m, X] = \mu^0_{d,m}(t,X) \notag
\end{align}
and
\begin{align}
\frac{E [I\{D=d\} \cdot I\{M=m\} \cdot I\{T=t\} \cdot \mu^0_{d,m}(t,X)|X]}{\rho^0_{d,m,t}(X)} &= \frac{\rho^0_{d,m,t}(X) \cdot \mu^0_{d,m}(t,X)}{\rho^0_{d,m,t}(X)} = \mu^0_{d,m}(t,X), \notag
\end{align}
and thus, the expression corresponds to 
\begin{align}
E \left[\frac{p^0_{1}(X)}{\Pr(T=1)} \cdot[\mu^0_{d,m}(t,X)-\mu^0_{d,m}(t,X)]\right]=0.
\end{align}
For this reason, 
\begin{align}
E[\phi_{d,m,t}(W,\eta^0,\Psi_{d,m,t})] &= E\left[\frac{ T \cdot \mu^0_{d,m}(t,X)}{\Pr(T=1)}\right] - \Psi_{d,m,t}\notag\\
&= E\left[ \mu^0_{d,m}(t,X) | T=1\right] - \Psi_{d,m,t} = \Psi_{d,m,t}-\Psi_{d,m,t}=0.
\end{align}
The score function in equation \eqref{neymanscore_mod2} thus satisfies Assumption 3.1(a) of \cite{Chetal2018}. It also satisfies Assumptions 3.1(b) and (c) of \cite{Chetal2018} 
because the score function is linear in $\Psi_{d,m,t}$ and 
the second Gateaux derivative 
$\eta \mapsto E[\phi_{d,m,t}(W,\eta,\Psi_{d,m,t})]$ is continuous. 

The Gateaux derivative of $E[\phi_{d,m,t}(W,\eta^0,\Psi_{d,m,t})]$ in the direction of $[\eta - \eta^0]$ is given by:
\begin{align}\label{derivscore_mod2}
&\partial E[\phi_{d,m,t}(W,\eta^0,\Psi_{d,m,t})] [\eta-\eta^0]  \\
&=  E \left[\frac{ T \cdot [\mu_{d,m}(t,X) - \mu^0_{d,m}(t,X)]}{\Pr(T=1)}\right] \notag\\
&-  E \left[\frac{I\{D=d\}\cdot I\{M=m\}\cdot I\{T=t\}\cdot p^0_{1}(X) \cdot [\mu_{d,m}(t,X) - \mu^0_{d,m}(t,X)]}{\rho^0_{d,m,t}(X) \cdot \Pr(T=1)}\right] \notag\\
&+  E \left[\frac{I\{D=d\}\cdot I\{M=m\}\cdot I\{T=t\}\cdot [Y_T - \mu^0_{d,m}(t,X)]}{\rho^0_{d,m,t}(X)} \cdot \frac{[p_{1}(X) - p^0_{1}(X)]}{\Pr(T=1)}\right] \notag\\
&-  E \left[\frac{I\{D=d\}\cdot I\{M=m\}\cdot I\{T=t\} \cdot p^0_{1}(X)\cdot [Y_T - \mu^0_{d,m}(t,X)]}{\rho^0_{d,m,t}(X)\cdot \Pr(T=1)} \cdot \frac{\rho_{d,m,t}(X) - \rho^0_{d,m,t}(X)}{\rho^0_{d,m,t}(X)}\right] = 0.\notag
\end{align}
To see that the Gateaux derivative is zero, we first consider the first term and apply the law of iterated expectations:
\begin{align}
& E \left[\frac{T \cdot [\mu_{d,m}(t,X) - \mu^0_{d,m}(t,X)]}{\Pr(T=1)}\right] \\
& = E\left[ \frac{p^0_{ 1}(X) \cdot [\mu_{d,m}(t,X) - \mu^0_{d,m}(t,X)]}{\Pr(T=1)}\right].\notag
\end{align}
Concerning the second term, we have 
\begin{align}
 &- E \left[\frac{I\{D=d\}\cdot I\{M=m\}\cdot I\{T=t\} \cdot p^0_{1}(X) \cdot [\mu_{d,m}(t,X) - \mu^0_{d,m}(t,X)]}{\rho^0_{d,m,t}(X) \cdot \Pr(T=1)}\right]\notag\\
&= - E \left[\frac{p^0_{1}(X) \cdot [\mu_{d,m}(t,X) - \mu^0_{d,m}(t,X)]}{\Pr(T=1)}\right].
\end{align}
Since the first and second terms involve identical expressions but with opposite signs, they cancel out.
Applying the law of iterated expectations to the third term, we obtain
\begin{align}
& E \left[\frac{I\{D=d\}\cdot I\{M=m\}\cdot I\{T=t\}\cdot [Y_T - \mu^0_{d,m}(t,X)]}{\rho^0_{d,m,t}(X)} \cdot \frac{[p_{1}(X) - p^0_{1}(X)]}{\Pr(T=1)}\right] \\
&= E \left[\frac{E\left[ I\{D=d\}\cdot I\{M=m\}\cdot I\{T=t\}\cdot [Y_T - \mu^0_{d,m}(t,X)] \mid X \right]}{\rho^0_{d,m,t}(X)} \cdot \frac{[p_{1}(X) - p^0_{1}(X)]}{\Pr(T=1)}\right].\notag
\end{align}
Concerning the first factor, we have by basic probability theory and iterated expectations that
\begin{align}
\frac{E [I\{D=d\} \cdot I\{M=m\} \cdot I\{T=t\} \cdot Y_T|X]}{\rho^0_{d,m,t}(X)} &= E[Y_t|D=d, M=m, X] = \mu^0_{d,m}(t,X) \notag
\end{align}
and
\begin{align}
\frac{E [I\{D=d\} \cdot I\{M=m\} \cdot I\{T=t\} \cdot \mu^0_{d,m}(t,X)|X]}{\rho^0_{d,m,t}(X)} &= \frac{\rho^0_{d,m,t}(X) \cdot \mu^0_{d,m}(t,X)}{\rho^0_{d,m,t}(X)} = \mu^0_{d,m}(t,X). \notag
\end{align}
Therefore, the third term corresponds to 
\begin{align}
E \left[  [\mu^0_{d,m}(t,X) - \mu^0_{d,m}(t,X)]  \cdot \frac{[p_{1}(X) - p^0_{1}(X)]}{\Pr(T=1)}\right]=0.
\end{align}
Finally, we apply the same reasoning to the fourth term to obtain
\begin{align}
&- E \left[\frac{I\{D=d\}\cdot I\{M=m\}\cdot I\{T=t\} \cdot p^0_{1}(X)\cdot [Y_T - \mu^0_{d,m}(t,X)]}{\rho^0_{d,m,t}(X)\cdot \Pr(T=1)} \cdot \frac{\rho_{d,m,t}(X) - \rho^0_{d,m,t}(X)}{\rho^0_{d,m,t}(X)}\right]\notag\\
&= - E \left[\frac{E \left[I\{D=d\}\cdot I\{M=m\}\cdot I\{T=t\} \cdot p^0_{1}(X)\cdot [Y_T - \mu^0_{d,m}(t,X)]|X\right]}{\rho^0_{d,m,t}(X)\cdot \Pr(T=1)} \cdot \frac{\rho_{d,m,t}(X) - \rho^0_{d,m,t}(X)}{\rho^0_{d,m,t}(X)}\right]\notag\\
&= - E \left[\frac{ p^0_{1}(X)\cdot [\mu^0_{d,m}(t,X) - \mu^0_{d,m}(t,X)]}{ \Pr(T=1)} \cdot \frac{\rho_{d,m,t}(X) - \rho^0_{d,m,t}(X)}{\rho^0_{d,m,t}(X)}\right]=0.
\end{align}
For this reason, $\partial E\left[\phi_{d,m,t}\left(W,\eta^0,\Psi_{d,m,t}\right)\right] \left[\eta-\eta^0\right]=0$, such that the Gateaux derivative of $E\left[\phi_{d,m,t}\left(W,\eta^0,\Psi_{d,m,t}\right)\right]$ is equal to zero, which satisfies Assumption 3.1(d) of \cite{Chetal2018}. For this reason, expression \eqref{nuisancepar_mod2}, which is a linear combination of such terms, satisfies Neyman orthogonality, too. Assumption 3.1(e) of \cite{Chetal2018}, which states that the factor multiplying $\Psi_{d,m,t}$ in the definition of the score function $\phi_{d,m,t}$ must be bounded in expectation, holds trivially because the factor is a constant, specifically $-1$.

\subsection{Identification and Neyman orthogonality in panel data}\label{Neymanpanel}

We first show that the score function underlying DR expression \eqref{DiDidentDRpanel} for panel data satisfies Assumption 3.1(a)--(e) of \cite{Chetal2018}, implying that expression \eqref{DiDidentDRpanel}  identifies the ATET $\Delta_{d,m}(d,m,d',m')=E[Y_1(d,m)-Y_1(d',m')|D=d,M=m]$ under Assumptions \ref{ass1} to \ref{ass4} and satisfies Neyman orthogonality. In the subsequent discussion, all bounds hold uniformly over all probability laws $P \in \mathcal{P}$, where $\mathcal{P}$ is the set of all possible probability laws, and we omit $P$ for brevity. We define the nuisance parameters 
$$\eta=(\mu_{d',m'}(X),\rho_{d,m}(X), \rho_{d',m'}(X))$$ 
and denote their true values by 
$$\eta^0=(\mu^0_{d',m'}(X), \rho^0_{d,m}(X), \rho^0_{d',m'}(X)).$$ 
We also define $\Psi_{d',m'}=E[\mu_{d',m'}(X)|D=d,M=m]$ to be the average of the conditional mean $\mu_{d',m'}(X)$ given $D=d$, $M=m$, in order to express the ATET as a function of $\Psi_{d',m'}$:
\begin{align}\label{nuisancepar_modpanel}
\Delta_{d,m}(d,m,d',m') &= E[Y_1-Y_0|D=d,M=m] - \Psi_{d',m'}.
\end{align}

Let us denote by $\phi_{d',m'}$ the Neyman-orthogonal score function for $\Psi_{d',m'}$, which corresponds to the following expression for $W=(Y_1,Y_0,D,M,X)$:
\begin{align}\label{neymanscore_modpanel}
\phi_{d',m'}(W,\eta,\Psi_{d',m'}) &= \frac{I\{D=d\}\cdot I\{M=m\} \cdot \mu_{d',m'}(X)}{\Pi_{d,m}} \\
&+ \frac{I\{D=d'\}\cdot I\{M=m'\}\cdot \rho_{d,m}(X)\cdot (Y_1-Y_0 - \mu_{d',m'}(X))}{\rho_{d',m'}(X)\cdot \Pi_{d,m}} - \Psi_{d',m'}.\notag
\end{align}
DR expression \eqref{DiDidentDRpanel} for ATET identification is obtained by identifying $\Psi_{d',m'}$ as the solution to the moment condition $E[\phi_{d',m'}(W,\eta^0,\Psi_{d',m'}) ]=0$ and plugging it into equation \eqref{nuisancepar_modpanel}. To show that $E[\phi_{d',m'}(W,\eta^0,\Psi_{d',m'})]=0$ holds, we consider the expectation of the second term in equation \eqref{neymanscore_modpanel}, evaluated at $\eta^0$, and note that it is zero. This follows from applying the law of iterated expectations and basic probability theory: 
\begin{align}
&E\left[\frac{I\{D=d'\}\cdot I\{M=m'\}\cdot \rho^0_{d,m}(X)\cdot (Y_1 - Y_0 - \mu^0_{d',m'}(X))}{\rho^0_{d',m'}(X)\cdot \Pi_{d,m}}\right]\notag\\
&=E\left[ \frac{\rho^0_{d,m}(X)}{\Pi_{d,m}} \cdot \frac{E\left[ I\{D=d'\}\cdot I\{M=m'\}\cdot (Y_1-Y_0-\mu^0_{D,M}(X))| X\right] }{\rho^0_{d',m'}(X)}  \right].\notag
\end{align}
We have by basic probability theory and iterated expectations that
\begin{align}
\frac{E [I\{D=d'\} \cdot I\{M=m'\}  \cdot (Y_1-Y_0)|X]}{\rho^0_{d',m'}(X)} &= E[Y_1-Y_0|D=d', M=m', X] = \mu^0_{d',m'}(X) \notag
\end{align}
and
\begin{align}
\frac{E [I\{D=d'\} \cdot I\{M=m'\}  \cdot \mu^0_{d',m'}(X)|X]}{\rho^0_{d',m'}(X)} &= \frac{\rho^0_{d',m'}(X) \cdot \mu^0_{d',m'}(X)}{\rho^0_{d',m'}(X)} = \mu^0_{d',m'}(X), \notag
\end{align}
and thus, the expression corresponds to 
\begin{align}
E \left[\frac{\rho^0_{d,m}(X)}{\Pi_{d,m}} \cdot[\mu^0_{d',m'}(X)-\mu^0_{d',m'}(X)]\right]=0.
\end{align}
For this reason, 
\begin{align}
E[\phi_{d',m'}(W,\eta^0,\Psi_{d',m'})] &= E\left[\frac{I\{D=d\}\cdot I\{M=m\} \cdot \mu^0_{d',m'}(X)}{\Pi_{d,m}}\right] - \Psi_{d',m'}\notag\\
&= E\left[ \mu^0_{d',m'}(X) | D=d, M=m\right] - \Psi_{d',m'} = \Psi_{d',m'}-\Psi_{d',m'}=0.
\end{align}
The score function in equation \eqref{neymanscore_modpanel} thus satisfies Assumption 3.1(a) of \cite{Chetal2018}. It also satisfies Assumptions 3.1(b) and (c) of \cite{Chetal2018} because the score function is linear in $\Psi_{d',m'}$ and the second Gateaux derivative $\eta \mapsto E[\phi_{d',m'}(W,\eta,\Psi_{d',m'})]$ is continuous. 

The Gateaux derivative of $E[\phi_{d',m'}(W,\eta^0,\Psi_{d',m'})]$ in the direction of $[\eta - \eta^0]$ is given by:
\begin{align}\label{derivscore_modpanel}
&\partial E[\phi_{d',m'}(W,\eta^0,\Psi_{d',m'})] [\eta-\eta^0] \\
&=  E \left[\frac{I\{D=d\}\cdot I\{M=m\}\cdot [\mu_{d',m'}(X) - \mu^0_{d',m'}(X)]}{\Pi_{d,m}}\right]\notag\\
&-  E \left[\frac{I\{D=d'\}\cdot I\{M=m'\}\cdot \rho^0_{d,m}(X) \cdot [\mu_{d',m'}(X) - \mu^0_{d',m'}(X)]}{\rho^0_{d',m'}(X) \cdot \Pi_{d,m}}\right] \notag\\
&+  E \left[\frac{I\{D=d'\}\cdot I\{M=m'\}\cdot [Y_1-Y_0 - \mu^0_{d',m'}(X)]}{\rho^0_{d',m'}(X)} \cdot \frac{[\rho_{d,m}(X) - \rho^0_{d,m}(X)]}{\Pi_{d,m}}\right] \notag\\
&-  E \left[\frac{I\{D=d'\}\cdot I\{M=m'\} \cdot \rho^0_{d,m}(X)\cdot [Y_1-Y_0 - \mu^0_{d',m'}(X)]}{\rho^0_{d',m'}(X)\cdot \Pi_{d,m}} \cdot \frac{\rho_{d',m'}(X) - \rho^0_{d',m'}(X)}{\rho^0_{d',m'}(X)}\right] = 0.\notag
\end{align}
To see that the Gateaux derivative is zero, we first consider the first term and apply the law of iterated expectations:
\begin{align}
& E \left[\frac{I\{D=d\}\cdot I\{M=m\} \cdot [\mu_{d',m'}(X) - \mu^0_{d',m'}(X)]}{\Pi_{d,m}}\right] \\
& = E \left[\frac{E[I\{D=d\}\cdot I\{M=m\}|X] \cdot [\mu_{d',m'}(X) - \mu^0_{d',m'}(X)]}{\Pi_{d,m}}\right] \notag\\
& = E\left[ \frac{\rho^0_{d,m}(X) \cdot [\mu_{d',m'}(X) - \mu^0_{d',m'}(X)]}{\Pi_{d,m}}\right].\notag
\end{align}
Concerning the second term, we have 
\begin{align}
 &- E \left[\frac{I\{D=d'\}\cdot I\{M=m'\} \cdot \rho^0_{d,m}(X) \cdot [\mu_{d',m'}(X) - \mu^0_{d',m'}(X)]}{\rho^0_{d',m'}(X) \cdot \Pi_{d,m}}\right]\notag\\
&= - E \left[\frac{\rho^0_{d,m}(X) \cdot [\mu_{d',m'}(X) - \mu^0_{d',m'}(X)]}{\Pi_{d,m}}\right].
\end{align}
Since the first and second terms involve identical expressions but with opposite signs, they cancel out.
Applying the law of iterated expectations to the third term, we obtain
\begin{align}
& E \left[\frac{I\{D=d'\}\cdot I\{M=m'\}\cdot [Y_1-Y_0 - \mu^0_{d',m'}(X)]}{\rho^0_{d',m'}(X)} \cdot \frac{[\rho_{d,m}(X) - \rho^0_{d,m}(X)]}{\Pi_{d,m}}\right] \\
& E \left[E \left[\frac{I\{D=d'\}\cdot I\{M=m'\}\cdot [Y_1-Y_0 - \mu^0_{d',m'}(X)]}{\rho^0_{d',m'}(X)} \cdot \frac{[\rho_{d,m}(X) - \rho^0_{d,m}(X)]}{\Pi_{d,m}} \Bigg| X\right]\right] \notag\\
&= E \left[\frac{E\left[ I\{D=d'\}\cdot I\{M=m'\}\cdot [Y_1-Y_0 - \mu^0_{d',m'}(X)] \mid X \right]}{\rho^0_{d',m'}(X)} \cdot \frac{[\rho_{d,m}(X) - \rho^0_{d,m}(X)]}{\Pi_{d,m}}\right].\notag
\end{align}
Concerning the first factor, we have by basic probability theory and iterated expectations that
\begin{align}
\frac{E [I\{D=d'\} \cdot I\{M=m'\} \cdot (Y_1-Y_0)|X]}{\rho^0_{d',m'}(X)} &= E[Y_1-Y_0|D=d', M=m', X] = \mu^0_{d',m'}(X) \notag
\end{align}
and
\begin{align}
\frac{E [I\{D=d'\} \cdot I\{M=m'\} \cdot \mu^0_{d',m'}(X)|X]}{\rho^0_{d',m'}(X)} &= \frac{\rho^0_{d',m'}(X) \cdot \mu^0_{d',m'}(X)}{\rho^0_{d',m'}(X)} = \mu^0_{d',m'}(X). \notag
\end{align}
Therefore, the third term corresponds to 
\begin{align}
E \left[  [\mu^0_{d',m'}(X) - \mu^0_{d',m'}(X)]  \cdot \frac{[\rho_{d,m}(X) - \rho^0_{d,m}(X)]}{\Pi_{d,m}}\right]=0.
\end{align}
Finally, we apply the same reasoning to the fourth term to obtain
\begin{align}
&- E \left[\frac{I\{D=d'\}\cdot I\{M=m'\}\cdot \rho^0_{d,m}(X)\cdot [Y_1-Y_0 - \mu^0_{d',m'}(X)]}{\rho^0_{d',m'}(X)\cdot \Pi_{d,m}} \cdot \frac{\rho_{d',m'}(X) - \rho^0_{d',m'}(X)}{\rho^0_{d',m'}(X)}\right]\notag\\
&= - E \left[\frac{E \left[I\{D=d'\}\cdot I\{M=m'\} \cdot \rho^0_{d,m}(X)\cdot [Y_1-Y_0 - \mu^0_{d',m'}(X)]|X\right]}{\rho^0_{d',m'}(X)\cdot \Pi_{d,m}} \cdot \frac{\rho_{d',m'}(X) - \rho^0_{d',m'}(X)}{\rho^0_{d',m'}(X)}\right]\notag\\
&= - E \left[\frac{ \rho^0_{d,m}(X)\cdot [\mu^0_{d',m'}(X) - \mu^0_{d',m'}(X)]}{ \Pi_{d,m}} \cdot \frac{\rho_{d',m'}(X) - \rho^0_{d',m'}(X)}{\rho^0_{d',m'}(X)}\right]=0.
\end{align}
For this reason, $\partial E\left[\phi_{d',m'}\left(W,\eta^0,\Psi_{d',m'}\right)\right] \left[\eta-\eta^0\right]=0$, such that the Gateaux derivative of $E\left[\phi_{d',m'}\left(W,\eta^0,\Psi_{d',m'}\right)\right]$ is equal to zero, which satisfies Assumption 3.1(d) of \cite{Chetal2018}. The Gateaux derivative of $E[Y_1-Y_0|D=d,M=m]$ is trivially equal to zero, too. For this reason, expression \eqref{nuisancepar_modpanel}, which is a linear combination of these terms, satisfies Neyman orthogonality. Assumption 3.1(e) of \cite{Chetal2018}, which states that the factor multiplying $\Psi_{d',m'}$ in the definition of the score function $\phi_{d',m'}$ must be bounded in expectation, holds trivially because the factor is a constant, specifically $-1$.

Next, we consider the counterfactual $E[Y_1(d',M(d))|D=d]$, which is identified by the DR expression~\eqref{DiDidentDRcounterfacpanel2}. We define $\Psi_{d',M}=E[\mu_{d',M}(X)|D=d]$ to be the average of the conditional mean $\mu_{d',M}(X)$ given $D=d$, in order to express the counterfactual as a function of $\Psi_{d',M}$:
\begin{align}\label{decomp_noprobpanel}
    E[Y_1(d',M(d))|D=d]=E[Y_0|D=d]+\Psi_{d',M},
\end{align}
Let us denote the nuisance parameters by
$$\eta=(\mu_{d',M}(X),\pi_d(M,X),\pi_{d'}(M,X))$$
and by $\phi_{d',M}$ the Neyman-orthogonal score function of $\Psi_{d',M}$, which is defined as:
\begin{align}\label{neymanscore_noprobpanel}
\phi_{d',M}(W,\eta,\Psi_{d',M}) &= \frac{I\{D=d\}\cdot\mu_{d',M}(X)}{\Pr(D=d)}\notag\\
&+\frac{I\{D=d'\}\cdot\pi_d(M,X)\cdot(Y_1-Y_0-\mu_{d',M}(X))}{\Pr(D=d)\cdot\pi_{d'}(M,X)}-\Psi_{d',M}.
\end{align}
We note that the DR expression \eqref{DiDidentDRcounterfacpanel2} is obtained by identifying $\Psi_{d',M}$ as the solution to the moment condition $E[\phi_{d',M}(W,\eta^0,\Psi_{d',M})]=0$ and plugging it into equation \eqref{decomp_noprobpanel}. To show that $E[\phi_{d',M}(W,\eta^0,\Psi_{d',M})]=0$ holds, we consider the expectation of the second term in equation \eqref{neymanscore_noprobpanel}, evaluated at $\eta^0$, and note that it is zero. This follows from applying the law of iterated expectations and basic probability theory:
\begin{align}
&E\left[\frac{I\{D=d'\}\cdot\pi^0_d(M,X)\cdot(Y_1-Y_0-\mu^0_{d',M}(X))}{\Pr(D=d)\cdot\pi^0_{d'}(M,X)}\right]\notag\\
&=E\left[\frac{\pi^0_d(M,X)}{\Pr(D=d)}\cdot\frac{E\left[I\{D=d'\}\cdot(Y_1-Y_0-\mu^0_{d',M}(X))|M,X\right]}{\pi^0_{d'}(M,X)}\right].\notag
\end{align}
We have by basic probability theory and iterated expectations that
\begin{align}
\frac{E[I\{D=d'\}\cdot(Y_1-Y_0)|M,X]}{\pi^0_{d'}(M,X)} &= E[Y_1-Y_0|D=d',M,X]=\mu^0_{d',M}(X)\notag
\end{align}
and
\begin{align}
\frac{E[I\{D=d'\}\cdot\mu^0_{d',M}(X)|M,X]}{\pi^0_{d'}(M,X)} &= \frac{\pi^0_{d'}(M,X)\cdot\mu^0_{d',M}(X)}{\pi^0_{d'}(M,X)}=\mu^0_{d',M}(X),\notag
\end{align}
and thus, the expression corresponds to
\begin{align}
E\left[\frac{\pi^0_d(M,X)}{\Pr(D=d)}\cdot[\mu^0_{d',M}(X)-\mu^0_{d',M}(X)]\right]=0.
\end{align}
For this reason,
\begin{align}
E[\phi_{d',M}(W,\eta^0,\Psi_{d',M})] &= E\left[\frac{I\{D=d\}\cdot\mu^0_{d',M}(X)}{\Pr(D=d)}\right]-\Psi_{d',M}\notag\\
&=E[\mu^0_{d',M}(X)|D=d]-\Psi_{d',M}=\Psi_{d',M}-\Psi_{d',M}=0.
\end{align}
The score function in equation \eqref{neymanscore_noprobpanel} thus satisfies Assumption 3.1(a) of \cite{Chetal2018}. It also satisfies Assumptions 3.1(b) and (c) of \cite{Chetal2018} because the score function is linear in $\Psi_{d',M}$ and the second Gateaux derivative $\eta\mapsto E[\phi_{d',M}(W,\eta,\Psi_{d',M})]$ is continuous.

The Gateaux derivative of $E[\phi_{d',M}(W,\eta^0,\Psi_{d',M})]$ in the direction of $[\eta-\eta^0]$ is given by:
\begin{align}\label{derivscore_modpanel_m}
&\partial E[\phi_{d',M}(W,\eta^0,\Psi_{d',M})][\eta-\eta^0]\notag\\
&=E\left[\frac{I\{D=d\}\cdot[\mu_{d',M}(X)-\mu^0_{d',M}(X)]}{\Pr(D=d)}\right]\notag\\
&-E\left[\frac{I\{D=d'\}\cdot\pi^0_d(M,X)\cdot[\mu_{d',M}(X)-\mu^0_{d',M}(X)]}{\pi^0_{d'}(M,X)\cdot\Pr(D=d)}\right]\notag\\
&+E\left[\frac{I\{D=d'\}\cdot(Y_1-Y_0-\mu^0_{d',M}(X))}{\pi^0_{d'}(M,X)}\cdot\frac{\pi_d(M,X)-\pi^0_d(M,X)}{\Pr(D=d)}\right]\notag\\
&-E\left[\frac{I\{D=d'\}\cdot\pi^0_d(M,X)\cdot(Y_1-Y_0-\mu^0_{d',M}(X))}{\pi^0_{d'}(M,X)\cdot\Pr(D=d)}\cdot\frac{\pi_{d'}(M,X)-\pi^0_{d'}(M,X)}{\pi^0_{d'}(M,X)}\right]=0.
\end{align}
To see that the Gateaux derivative is zero, we first consider the first term and apply the law of iterated expectations:
\begin{align}
&E\left[\frac{I\{D=d\}\cdot[\mu_{d',M}(X)-\mu^0_{d',M}(X)]}{\Pr(D=d)}\right]\notag\\
&=E\left[E\left[\frac{I\{D=d\}\cdot[\mu_{d',M}(X)-\mu^0_{d',M}(X)]}{\Pr(D=d)}\Big|M,X\right]\right]\notag\\
&=E\left[\frac{E[I\{D=d\}|M,X]\cdot[\mu_{d',M}(X)-\mu^0_{d',M}(X)]}{\Pr(D=d)}\right]\notag\\
&=E\left[\frac{\pi_d^0(M,X)\cdot[\mu_{d',M}(X)-\mu^0_{d',M}(X)]}{\Pr(D=d)}\right].
\end{align}
Concerning the second term, we have
\begin{align}
&-E\left[\frac{I\{D=d'\}\cdot\pi^0_d(M,X)\cdot[\mu_{d',M}(X)-\mu^0_{d',M}(X)]}{\pi^0_{d'}(M,X)\cdot\Pr(D=d)}\right]\notag\\
&=-E\left[\frac{\pi^0_{d'}(M,X)\cdot\pi^0_d(M,X)\cdot[\mu_{d',M}(X)-\mu^0_{d',M}(X)]}{\pi^0_{d'}(M,X)\cdot\Pr(D=d)}\right]\notag\\
&=-E\left[\frac{\pi^0_d(M,X)\cdot[\mu_{d',M}(X)-\mu^0_{d',M}(X)]}{\Pr(D=d)}\right].
\end{align}
Since the first and second terms involve identical expressions but with opposite signs, they cancel out. Applying the law of iterated expectations to the third term, we obtain
\begin{align}
&E\left[\frac{I\{D=d'\}\cdot(Y_1-Y_0-\mu^0_{d',M}(X))}{\pi^0_{d'}(M,X)}\cdot\frac{\pi_d(M,X)-\pi^0_d(M,X)}{\Pr(D=d)}\right]\notag\\
&=E\left[E\left[\frac{I\{D=d'\}\cdot(Y_1-Y_0-\mu^0_{d',M}(X))}{\pi^0_{d'}(M,X)}\cdot\frac{\pi_d(M,X)-\pi^0_d(M,X)}{\Pr(D=d)}\Big|M,X\right]\right]\notag\\
&=E\left[\frac{E\left[I\{D=d'\}\cdot(Y_1-Y_0-\mu^0_{d',M}(X))|M,X\right]}{\pi^0_{d'}(M,X)}\cdot\frac{\pi_d(M,X)-\pi^0_d(M,X)}{\Pr(D=d)}\right].
\end{align}
Concerning the first factor, we have by basic probability theory and iterated expectations that
\begin{align}
\frac{E[I\{D=d'\}\cdot (Y_1-Y_0)|M,X]}{\pi^0_{d'}(M,X)} &= E[Y_1-Y_0|D=d',M,X]=\mu^0_{d',M}(X)\notag
\end{align}
and
\begin{align}
\frac{E[I\{D=d'\}\cdot\mu^0_{d',M}(X)|M,X]}{\pi^0_{d'}(M,X)} &= \frac{\pi^0_{d'}(M,X)\cdot\mu^0_{d',M}(X)}{\pi^0_{d'}(M,X)}=\mu^0_{d',M}(X).\notag
\end{align}
Therefore, the third term corresponds to
\begin{align}
E\left[[\mu^0_{d',M}(X)-\mu^0_{d',M}(X)]\cdot\frac{\pi_d(M,X)-\pi^0_d(M,X)}{\Pr(D=d)}\right]=0.
\end{align}
Finally, we apply the same reasoning to the fourth term to obtain
\begin{align}
&-E\left[\frac{I\{D=d'\}\cdot\pi^0_d(M,X)\cdot(Y_1-Y_0-\mu^0_{d',M}(X))}{\pi^0_{d'}(M,X)\cdot\Pr(D=d)}\cdot\frac{\pi_{d'}(M,X)-\pi^0_{d'}(M,X)}{\pi^0_{d'}(M,X)}\right]\notag\\
&=-E\left[\frac{E\left[I\{D=d'\}\cdot\pi^0_d(M,X)\cdot(Y_1-Y_0-\mu^0_{d',M}(X))|M,X\right]}{\pi^0_{d'}(M,X)\cdot\Pr(D=d)}\cdot\frac{\pi_{d'}(M,X)-\pi^0_{d'}(M,X)}{\pi^0_{d'}(M,X)}\right]\notag\\
&=-E\left[\frac{\pi^0_d(M,X)\cdot[\mu^0_{d',M}(X)-\mu^0_{d',M}(X)]}{\Pr(D=d)}\cdot\frac{\pi_{d'}(M,X)-\pi^0_{d'}(M,X)}{\pi^0_{d'}(M,X)}\right]=0.
\end{align}
For this reason, $\partial E\left[\phi_{d',M}\left(W,\eta^0,\Psi_{d',M}\right)\right]\left[\eta-\eta^0\right]=0$, such that the Gateaux derivative of \linebreak $E\left[\phi_{d',M}\left(W,\eta^0,\Psi_{d',M}\right)\right]$ is equal to zero, which satisfies Assumption 3.1(d) of \cite{Chetal2018}. The Gateaux derivative of $E[Y_0|D=d]$ is trivially equal to zero, too. For this reason, expression \eqref{decomp_noprobpanel}, which is a linear combination of these terms, satisfies Neyman orthogonality. Assumption 3.1(e) of \cite{Chetal2018}, which states that the factor multiplying $\Psi_{d',M}$ in the definition of the score function $\phi_{d',M}$ must be bounded in expectation, holds trivially because the factor is a constant, specifically $-1$.

In a next step, we consider the counterfactual $E[Y_1(d',M(d'))|D=d]$, which is identified by the DR expression~\eqref{DiDidentDRsimplepanel}. We define $\Psi_{d'}=E[\mu_{d'}(X)|D=d]$ to be the average of the conditional mean $\mu_{d'}(X)$ given $D=d$, in order to express the counterfactual as a function of $\Psi_{d'}$:
\begin{align}\label{psi_simplepanel}
E[Y_1(d',M(d'))|D=d]= E[ Y_0 | D=d ] + \Psi_{d'} 
\end{align}
Let us denote the nuisance parameters by
$$\eta=(\mu_{d'}(X),\pi_{d}(X),\pi_{d'}(X))$$
and by $\phi_{d'}$ the Neyman-orthogonal score function of $\Psi_{d'}$, which is defined as:
\begin{align}\label{neymanscore_modsimplepanel}
\phi_{d'}(W,\eta,\Psi_{d'}) &= \frac{I\{D=d\}\cdot \mu_{d'}(X)}{\Pr(D=d)} \\
&+ \frac{I\{D=d'\}\cdot \pi_{d}(X)\cdot (Y_1-Y_0- \mu_{d'}(X))}{\pi_{d'}(X)\cdot \Pr(D=d)} - \Psi_{d'}.\notag
\end{align}
We note that the DR expression \eqref{DiDidentDRsimplepanel} is obtained by identifying $\Psi_{d'}$ as the solution to the moment condition $E[\phi_{d'}(W,\eta^0,\Psi_{d'}) ]=0$ and plugging it into equation \eqref{psi_simplepanel}. Showing that Assumptions 3.1(a)--(e) of \cite{Chetal2018} hold for \(\phi_{d'}(W,\eta,\Psi_{d'})\) proceeds analogously to the case of \(\phi_{d',M}(W,\eta,\Psi_{d',M})\) and is therefore omitted.

Next, we consider the alternative identification of the counterfactual based on equation \eqref{doubleparalleltrendspanel}, which corresponds to
\begin{align}\label{psi_simple_doublpanel}
E[Y_1(d',M(d'))|D=d]=E[Y_0|D=d]+\sum_{m \in \mathcal{M}}\Psi(m)
\end{align}
when defining $\Psi(m)=E[\mu_{d',m}(X)\cdot b_m(X)|D=d]$ with $b_m(X)=\nu_{d,m}(0,X)+\nu_{d',m}(X)$. Let us denote the nuisance parameters by
$$\eta=(\mu_{d',m}(X),\nu_{d,m}(0,X),\nu_{d',m}(X),\rho_{d',m}(X),\pi_{d}(X),\pi_{d'}(X))$$
and their true values by
$$\eta^0=(\mu^0_{d',m}(X),\nu^0_{d,m}(0,X),\nu^0_{d',m}(X),\rho^0_{d',m}(X),\pi^0_{d}(X),\pi^0_{d'}(X)),$$
and let $b^0_m(X)=\nu^0_{d,m}(0,X)+\nu^0_{d',m}(X)$. We denote by $\phi(m)$ the Neyman-orthogonal score function of $\Psi(m)$, which is defined as:
\begin{align}\label{neymanscore_doublpanel}
\phi(m)(W,\eta,\Psi(m))&=\frac{I\{D=d\}\cdot \mu_{d',m}(X)\cdot \{I\{M_0=m\}+\nu_{d',m}(X)\}}{\Pr(D=d)}-\Psi(m)\\
&+b_m(X)\cdot\frac{I\{D=d'\}\cdot I\{M=m\}\cdot \pi_{d}(X)\cdot (Y_1-Y_0-\mu_{d',m}(X))}{\rho_{d',m}(X)\cdot \Pr(D=d)}\notag\\
&+\mu_{d',m}(X)\cdot\frac{I\{D=d'\}\cdot \pi_{d}(X)\cdot (I\{M=m\}-I\{M_0=m\}-\nu_{d',m}(X))}{\pi_{d'}(X)\cdot \Pr(D=d)}.\notag
\end{align}
DR expression \eqref{DiDidentDRsimple2doublpanel} is obtained by identifying $\Psi(m)$ as the solution to the moment condition $E[\phi(m)(W,\eta^0,\Psi(m))]=0$ for each $m$ and plugging it into equation \eqref{psi_simple_doublpanel}. To show that this moment condition holds, we first derive two conditional moment restrictions. We have by basic probability theory and iterated expectations that
\begin{align}
E[I\{D=d'\}\cdot I\{M=m\}\cdot (Y_1-Y_0)|X]&=\rho^0_{d',m}(X)\cdot E[Y_1-Y_0|D=d',M=m,X]\notag\\
&=\rho^0_{d',m}(X)\cdot\mu^0_{d',m}(X),\notag\\
E[I\{D=d'\}\cdot I\{M=m\}\cdot\mu^0_{d',m}(X)|X]&=\rho^0_{d',m}(X)\cdot\mu^0_{d',m}(X),\notag
\end{align}
such that
\begin{align}\label{resid1_doublpanel}
E[I\{D=d'\}\cdot I\{M=m\}\cdot (Y_1-Y_0-\mu^0_{d',m}(X))|X]=0.
\end{align}
In the same way, we have that
\begin{align}
E[I\{D=d'\}\cdot (I\{M=m\}-I\{M_0=m\})|X]&=\pi^0_{d'}(X)\cdot E[I\{M=m\}-I\{M_0=m\}|D=d',X]\notag\\
&=\pi^0_{d'}(X)\cdot\nu^0_{d',m}(X),\notag\\
E[I\{D=d'\}\cdot\nu^0_{d',m}(X)|X]&=\pi^0_{d'}(X)\cdot\nu^0_{d',m}(X),\notag
\end{align}
such that
\begin{align}\label{resid2_doublpanel}
E[I\{D=d'\}\cdot (I\{M=m\}-I\{M_0=m\}-\nu^0_{d',m}(X))|X]=0.
\end{align}
Applying the law of iterated expectations to the second term in equation \eqref{neymanscore_doublpanel}, evaluated at $\eta^0$, and using \eqref{resid1_doublpanel}, we obtain
\begin{align}
&E\left[b^0_m(X)\cdot\frac{I\{D=d'\}\cdot I\{M=m\}\cdot \pi^0_{d}(X)\cdot (Y_1-Y_0-\mu^0_{d',m}(X))}{\rho^0_{d',m}(X)\cdot \Pr(D=d)}\right]\notag\\
&=E\left[\frac{b^0_m(X)\cdot\pi^0_{d}(X)}{\rho^0_{d',m}(X)\cdot \Pr(D=d)}\cdot E[I\{D=d'\}\cdot I\{M=m\}\cdot (Y_1-Y_0-\mu^0_{d',m}(X))|X]\right]=0.\notag
\end{align}
Likewise, the expectation of the third term evaluated at $\eta^0$ is zero by \eqref{resid2_doublpanel}:
\begin{align}
&E\left[\mu^0_{d',m}(X)\cdot\frac{I\{D=d'\}\cdot \pi^0_{d}(X)\cdot (I\{M=m\}-I\{M_0=m\}-\nu^0_{d',m}(X))}{\pi^0_{d'}(X)\cdot \Pr(D=d)}\right]\notag\\
&=E\left[\frac{\mu^0_{d',m}(X)\cdot\pi^0_{d}(X)}{\pi^0_{d'}(X)\cdot \Pr(D=d)}\cdot E[I\{D=d'\}\cdot (I\{M=m\}-I\{M_0=m\}-\nu^0_{d',m}(X))|X]\right]=0.\notag
\end{align}
For the first term, we use $E[I\{M_0=m\}|D=d,X]=\nu^0_{d,m}(0,X)$ to obtain
\begin{align}
&E[\phi(m)(W,\eta^0,\Psi(m))]=E\left[\frac{I\{D=d\}\cdot \mu^0_{d',m}(X)\cdot \{I\{M_0=m\}+\nu^0_{d',m}(X)\}}{\Pr(D=d)}\right]-\Psi(m)\notag\\
&=E[\mu^0_{d',m}(X)\cdot \{I\{M_0=m\}+\nu^0_{d',m}(X)\}|D=d]-\Psi(m)\notag\\
&=E[\mu^0_{d',m}(X)\cdot \{\nu^0_{d,m}(0,X)+\nu^0_{d',m}(X)\}|D=d]-\Psi(m)=\Psi(m)-\Psi(m)=0.\notag
\end{align}
The score function in equation \eqref{neymanscore_doublpanel} thus satisfies Assumption 3.1(a) of \cite{Chetal2018}. It also satisfies Assumptions 3.1(b) and (c) of \cite{Chetal2018} because the score function is linear in $\Psi(m)$ and the second Gateaux derivative $\eta \mapsto E[\phi(m)(W,\eta,\Psi(m))]$ is continuous.

The Gateaux derivative of $E[\phi(m)(W,\eta^0,\Psi(m))]$ in the direction of $[\eta - \eta^0]$ is given by:
\begin{align}\label{derivscore_doublpanel}
&\partial E[\phi(m)(W,\eta^0,\Psi(m))] [\eta-\eta^0]\notag\\
&=E\left[\frac{I\{D=d\}\cdot [\mu_{d',m}(X)-\mu^0_{d',m}(X)]\cdot \{I\{M_0=m\}+\nu^0_{d',m}(X)\}}{\Pr(D=d)}\right]\notag\\
&+E\left[\frac{I\{D=d\}\cdot \mu^0_{d',m}(X)\cdot [\nu_{d',m}(X)-\nu^0_{d',m}(X)]}{\Pr(D=d)}\right]\notag\\
&+E\left[[b_m(X)-b^0_m(X)]\cdot\frac{I\{D=d'\}\cdot I\{M=m\}\cdot \pi^0_{d}(X)\cdot (Y_1-Y_0-\mu^0_{d',m}(X))}{\rho^0_{d',m}(X)\cdot \Pr(D=d)}\right]\notag\\
&-E\left[b^0_m(X)\cdot\frac{I\{D=d'\}\cdot I\{M=m\}\cdot \pi^0_{d}(X)\cdot [\mu_{d',m}(X)-\mu^0_{d',m}(X)]}{\rho^0_{d',m}(X)\cdot \Pr(D=d)}\right]\notag\\
&+E\left[b^0_m(X)\cdot\frac{I\{D=d'\}\cdot I\{M=m\}\cdot (Y_1-Y_0-\mu^0_{d',m}(X))}{\rho^0_{d',m}(X)}\cdot\frac{\pi_{d}(X)-\pi^0_{d}(X)}{\Pr(D=d)}\right]\notag\\
&-E\left[b^0_m(X)\cdot\frac{I\{D=d'\}\cdot I\{M=m\}\cdot \pi^0_{d}(X)\cdot (Y_1-Y_0-\mu^0_{d',m}(X))}{\rho^0_{d',m}(X)\cdot \Pr(D=d)}\cdot\frac{\rho_{d',m}(X)-\rho^0_{d',m}(X)}{\rho^0_{d',m}(X)}\right]\notag\\
&+E\left[[\mu_{d',m}(X)-\mu^0_{d',m}(X)]\cdot\frac{I\{D=d'\}\cdot \pi^0_{d}(X)\cdot (I\{M=m\}-I\{M_0=m\}-\nu^0_{d',m}(X))}{\pi^0_{d'}(X)\cdot \Pr(D=d)}\right]\notag\\
&-E\left[\mu^0_{d',m}(X)\cdot\frac{I\{D=d'\}\cdot \pi^0_{d}(X)\cdot [\nu_{d',m}(X)-\nu^0_{d',m}(X)]}{\pi^0_{d'}(X)\cdot \Pr(D=d)}\right]\notag\\
&+E\left[\mu^0_{d',m}(X)\cdot\frac{I\{D=d'\}\cdot (I\{M=m\}-I\{M_0=m\}-\nu^0_{d',m}(X))}{\pi^0_{d'}(X)}\cdot\frac{\pi_{d}(X)-\pi^0_{d}(X)}{\Pr(D=d)}\right]\notag\\
&-E\left[\mu^0_{d',m}(X)\cdot\frac{I\{D=d'\}\cdot \pi^0_{d}(X)\cdot (I\{M=m\}-I\{M_0=m\}-\nu^0_{d',m}(X))}{\pi^0_{d'}(X)\cdot \Pr(D=d)}\right.\notag\\
&\left.\cdot\frac{\pi_{d'}(X)-\pi^0_{d'}(X)}{\pi^0_{d'}(X)}\right]=0,
\end{align}
where $b_m(X)-b^0_m(X)=[\nu_{d,m}(0,X)-\nu^0_{d,m}(0,X)]+[\nu_{d',m}(X)-\nu^0_{d',m}(X)]$. The first and second terms stem from the first term of the score function, the third to sixth terms from its second term, and the seventh to tenth terms from its third term.

To see that the Gateaux derivative is zero, we first consider the first term and apply the law of iterated expectations, using $E[I\{D=d\}\cdot I\{M_0=m\}|X]=\pi^0_{d}(X)\cdot\nu^0_{d,m}(0,X)$ and $E[I\{D=d\}|X]=\pi^0_{d}(X)$:
\begin{align}
&E\left[\frac{I\{D=d\}\cdot [\mu_{d',m}(X)-\mu^0_{d',m}(X)]\cdot \{I\{M_0=m\}+\nu^0_{d',m}(X)\}}{\Pr(D=d)}\right]\notag\\
&=E\left[\frac{\{E[I\{D=d\}\cdot I\{M_0=m\}|X]+E[I\{D=d\}|X]\cdot\nu^0_{d',m}(X)\}\cdot [\mu_{d',m}(X)-\mu^0_{d',m}(X)]}{\Pr(D=d)}\right]\notag\\
&=E\left[\frac{\pi^0_{d}(X)\cdot b^0_m(X)\cdot [\mu_{d',m}(X)-\mu^0_{d',m}(X)]}{\Pr(D=d)}\right].\notag
\end{align}
Concerning the fourth term, we use $E[I\{D=d'\}\cdot I\{M=m\}|X]=\rho^0_{d',m}(X)$ to obtain
\begin{align}
&-E\left[b^0_m(X)\cdot\frac{I\{D=d'\}\cdot I\{M=m\}\cdot \pi^0_{d}(X)\cdot [\mu_{d',m}(X)-\mu^0_{d',m}(X)]}{\rho^0_{d',m}(X)\cdot \Pr(D=d)}\right]\notag\\
&=-E\left[\frac{\pi^0_{d}(X)\cdot b^0_m(X)\cdot [\mu_{d',m}(X)-\mu^0_{d',m}(X)]}{\Pr(D=d)}\right].\notag
\end{align}
Since the first and fourth terms involve identical expressions but with opposite signs, they cancel out. In the same way, $E[I\{D=d\}|X]=\pi^0_{d}(X)$ implies for the second term
\begin{align}
&E\left[\frac{I\{D=d\}\cdot \mu^0_{d',m}(X)\cdot [\nu_{d',m}(X)-\nu^0_{d',m}(X)]}{\Pr(D=d)}\right]\notag\\
&=E\left[\frac{\pi^0_{d}(X)\cdot \mu^0_{d',m}(X)\cdot [\nu_{d',m}(X)-\nu^0_{d',m}(X)]}{\Pr(D=d)}\right],\notag
\end{align}
and $E[I\{D=d'\}|X]=\pi^0_{d'}(X)$ implies for the eighth term
\begin{align}
&-E\left[\mu^0_{d',m}(X)\cdot\frac{I\{D=d'\}\cdot \pi^0_{d}(X)\cdot [\nu_{d',m}(X)-\nu^0_{d',m}(X)]}{\pi^0_{d'}(X)\cdot \Pr(D=d)}\right]\notag\\
&=-E\left[\frac{\pi^0_{d}(X)\cdot \mu^0_{d',m}(X)\cdot [\nu_{d',m}(X)-\nu^0_{d',m}(X)]}{\Pr(D=d)}\right],\notag
\end{align}
such that the second and eighth terms cancel out, too.

The third, fifth, and sixth terms are expectations of a function of $X$ multiplied by $I\{D=d'\}\cdot I\{M=m\}\cdot (Y_1-Y_0-\mu^0_{d',m}(X))$. Applying the law of iterated expectations to the third term and using \eqref{resid1_doublpanel}, we obtain
\begin{align}
&E\left[[b_m(X)-b^0_m(X)]\cdot\frac{I\{D=d'\}\cdot I\{M=m\}\cdot \pi^0_{d}(X)\cdot (Y_1-Y_0-\mu^0_{d',m}(X))}{\rho^0_{d',m}(X)\cdot \Pr(D=d)}\right]\notag\\
&=E\left[\frac{[b_m(X)-b^0_m(X)]\cdot\pi^0_{d}(X)}{\rho^0_{d',m}(X)\cdot \Pr(D=d)}\cdot E[I\{D=d'\}\cdot I\{M=m\}\cdot (Y_1-Y_0-\mu^0_{d',m}(X))|X]\right]=0.\notag
\end{align}
The same reasoning applies to the fifth and sixth terms, in which the conditional expectation in \eqref{resid1_doublpanel} is multiplied by $b^0_m(X)\cdot[\pi_{d}(X)-\pi^0_{d}(X)]/[\rho^0_{d',m}(X)\cdot\Pr(D=d)]$ and $-b^0_m(X)\cdot\pi^0_{d}(X)\cdot[\rho_{d',m}(X)-\rho^0_{d',m}(X)]/[\rho^0_{d',m}(X)^2\cdot\Pr(D=d)]$, respectively. Both terms are therefore zero. Analogously, the seventh, ninth, and tenth terms are expectations of a function of $X$ multiplied by $I\{D=d'\}\cdot (I\{M=m\}-I\{M_0=m\}-\nu^0_{d',m}(X))$. For the seventh term, \eqref{resid2_doublpanel} implies
\begin{align}
&E\left[[\mu_{d',m}(X)-\mu^0_{d',m}(X)]\cdot\frac{I\{D=d'\}\cdot \pi^0_{d}(X)\cdot (I\{M=m\}-I\{M_0=m\}-\nu^0_{d',m}(X))}{\pi^0_{d'}(X)\cdot \Pr(D=d)}\right]\notag\\
&=E\left[\frac{[\mu_{d',m}(X)-\mu^0_{d',m}(X)]\!\cdot\!\pi^0_{d}(X)}{\pi^0_{d'}(X)\!\cdot\! \Pr(D=d)}\!\cdot\! E[I\{D=d'\}\!\cdot\! (I\{M=m\}\!-\!I\{M_0=m\}\!-\!\nu^0_{d',m}(X))|X]\right]=0.\notag
\end{align}
In the ninth and tenth terms, the conditional expectation in \eqref{resid2_doublpanel} is multiplied by $\mu^0_{d',m}(X)\cdot[\pi_{d}(X)-\pi^0_{d}(X)]/[\pi^0_{d'}(X)\cdot\Pr(D=d)]$ and $-\mu^0_{d',m}(X)\cdot\pi^0_{d}(X)\cdot[\pi_{d'}(X)-\pi^0_{d'}(X)]/[\pi^0_{d'}(X)^2\cdot\Pr(D=d)]$, respectively, such that both terms are zero, too.

For this reason, $\partial E[\phi(m)(W,\eta^0,\Psi(m))] [\eta-\eta^0]=0$, such that the Gateaux derivative of $E[\phi(m)(W,\eta^0,\Psi(m))]$ is equal to zero, which satisfies Assumption 3.1(d) of \cite{Chetal2018}. The Gateaux derivative of $E[Y_0|D=d]$ is trivially equal to zero, too. For this reason, expression \eqref{psi_simple_doublpanel}, which is a linear combination of these terms, satisfies Neyman orthogonality. Assumption 3.1(e) of \cite{Chetal2018} holds trivially because the factor multiplying $\Psi(m)$ in the score function $\phi(m)$ is a constant, specifically $-1$. 

Finally, we show that the score function underlying DR expression \eqref{DiDidentDRATEpanel} for panel data satisfies Assumption 3.1(a)--(e) of \cite{Chetal2018}, implying the identification of the ATE $\Delta(d,m,d',m')=E[Y_1(d,m)-Y_1(d',m')]$ under Assumptions \ref{ass1} to \ref{ass4} and satisfies Neyman orthogonality.
To this end, we define the nuisance parameters 
$$\eta=(\mu_{d^*,m^*}(X),\rho_{d^*,m^*}(X))$$ 
and denote their true values by 
$$\eta^0=(\mu^0_{d^*,m^*}(X),\rho^0_{d^*,m^*}(X)),$$
for $d^* \in \{d,d'\}$ and $m^* \in \{m,m'\}$.
Furthermore, we denote by $\Psi_{d,m}=E[\mu_{d,m}(X)]$ the average of the conditional mean difference $\mu_{d,m}(X)$, in order to express the ATE as a function of $\Psi_{d',m'}$:
\begin{align}\label{nuisancepar_mod2panel}
\Delta(d,m,d',m') &= \Psi_{d,m}  - \Psi_{d',m'}.
\end{align}
Let us denote by $\phi_{d,m}$ the Neyman-orthogonal score function for $\Psi_{d,m}$, which corresponds to the following expression:
\begin{align}\label{neymanscore_mod2panel}
\phi_{d,m}(W,\eta,\Psi_{d,m}) =  \mu_{d,m}(X)+ \frac{I\{D=d\}\cdot I\{M=m\}\cdot (Y_1-Y_0 - \mu_{d,m}(X))}{\rho_{d,m}(X)} - \Psi_{d,m}.
\end{align}
We note that the DR expression \eqref{DiDidentDRATEpanel} for ATE identification is
obtained by identifying $\Psi_{d,m}$ as the solution to the moment condition $E[\phi_{d,m} (W,\eta^0,\Psi_{d,m}) ]=0$ and plugging it into equation \eqref{nuisancepar_mod2panel}. To show that $E[\phi_{d,m}(W,\eta^0,\Psi_{d,m})]=0$ holds, we consider the expectation of the second term in equation \eqref{neymanscore_mod2panel}, evaluated at $\eta^0$, and note that it is zero. This follows from applying the law of iterated expectations and basic probability theory: 
\begin{align}
&E\left[\frac{I\{D=d\}\cdot I\{M=m\}\cdot (Y_1-Y_0 - \mu^0_{d,m}(X))}{\rho^0_{d,m}(X)}\right]\notag\\
&=E\left[ \frac{E\left[ I\{D=d\}\cdot I\{M=m\}\cdot (Y_1-Y_0-\mu^0_{D,M}(X))| X\right] }{\rho^0_{d,m}(X)}  \right].\notag
\end{align}
We have by basic probability theory and iterated expectations that
\begin{align}
\frac{E [I\{D=d\} \cdot I\{M=m\} \cdot (Y_1-Y_0)|X]}{\rho^0_{d,m}(X)} &= E[Y_1-Y_0|D=d, M=m, X] = \mu^0_{d,m}(X) \notag
\end{align}
and
\begin{align}
\frac{E [I\{D=d\} \cdot I\{M=m\} \cdot \mu^0_{d,m}(X)|X]}{\rho^0_{d,m}(X)} &= \frac{\rho^0_{d,m}(X) \cdot \mu^0_{d,m}(X)}{\rho^0_{d,m}(X)} = \mu^0_{d,m}(X), \notag
\end{align}
and thus, the expression corresponds to 
\begin{align}
E [\mu^0_{d,m}(X)-\mu^0_{d,m}(X)]=0.
\end{align}
For this reason, 
\begin{align}
E[\phi_{d,m}(W,\eta^0,\Psi_{d,m})] = E\left[ \mu^0_{d,m}(X)\right] - \Psi_{d,m}= \Psi_{d,m}-\Psi_{d,m}=0.
\end{align}
The score function in equation \eqref{neymanscore_mod2panel} thus satisfies Assumption 3.1(a) of \cite{Chetal2018}. It also satisfies Assumptions 3.1(b) and (c) of \cite{Chetal2018} 
because the score function is linear in $\Psi_{d,m}$ and 
the second Gateaux derivative 
$\eta \mapsto E[\phi_{d,m}(W,\eta,\Psi_{d,m})]$ is continuous. 

The Gateaux derivative of $E[\phi_{d,m}(W,\eta^0,\Psi_{d,m})]$ in the direction of $[\eta - \eta^0]$ is given by:
\begin{align}\label{derivscore_modpanel2}
&\partial E[\phi_{d,m}(W,\eta^0,\Psi_{d,m})] [\eta-\eta^0]  \notag\\
&=  E \left[\mu_{d,m}(X) - \mu^0_{d,m}(X)\right] \notag\\
&-  E \left[\frac{I\{D=d\}\cdot I\{M=m\} \cdot [\mu_{d,m}(X) - \mu^0_{d,m}(X)]}{\rho^0_{d,m}(X)}\right] \notag\\
&-  E \left[\frac{I\{D=d\}\cdot I\{M=m\}\cdot [Y_1 - Y_0- \mu^0_{d,m}(X)]}{\rho^0_{d,m}(X)} \cdot \frac{\rho_{d,m}(X) - \rho^0_{d,m}(X)}{\rho^0_{d,m}(X)}\right] = 0.
\end{align}
To see that the Gateaux derivative is zero, we first consider the first term, $E \left[\mu_{d,m}(X) - \mu^0_{d,m}(X)\right]$, as well as the 
 second term, which is 
\begin{align}
 &- E \left[\frac{I\{D=d\}\cdot I\{M=m\}\cdot [\mu_{d,m}(X) - \mu^0_{d,m}(X)]}{\rho^0_{d,m}(X)}\right]\notag\\
& = - E \left[\frac{ \rho^0_{d,m}(X) \cdot [\mu_{d,m}(X) - \mu^0_{d,m}(X)]}{\rho^0_{d,m}(X)}\right]\notag\\
&= - E \left[\mu_{d,m}(X) - \mu^0_{d,m}(X)\right].
\end{align}
Therefore, the first and second terms cancel out.
Applying the law of iterated expectations to the third term, we have
\begin{align}
&- E \left[\frac{I\{D=d\}\cdot I\{M=m\}\cdot [Y_1-Y_0 - \mu^0_{d,m}(X)]}{\rho^0_{d,m}(X)} \cdot \frac{\rho_{d,m}(X) - \rho^0_{d,m}(X)}{\rho^0_{d,m}(X)}\right]\notag\\
&= - E \left[\frac{E \left[I\{D=d\}\cdot I\{M=m\}\cdot [Y_1-Y_0 - \mu^0_{d,m}(X)]|X\right]}{\rho^0_{d,m}(X)} \cdot \frac{\rho_{d,m}(X) - \rho^0_{d,m}(X)}{\rho^0_{d,m}(X)}\right]\notag\\
&= - E \left[ [\mu^0_{d,m}(X) - \mu^0_{d,m}(X)] \cdot \frac{\rho_{d,m}(X) - \rho^0_{d,m}(X)}{\rho^0_{d,m}(X)}\right]=0.
\end{align}
For this reason, $\partial E\left[\phi_{d,m}\left(W,\eta^0,\Psi_{d,m}\right)\right] \left[\eta-\eta^0\right]=0$, such that the Gateaux derivative of \linebreak $E\left[\phi_{d,m}\left(W,\eta^0,\Psi_{d,m}\right)\right]$ is equal to zero, which satisfies Assumption 3.1(d) of \cite{Chetal2018}. For this reason, expression \eqref{nuisancepar_mod2panel}, which is a linear combination of such terms, satisfies Neyman orthogonality, too. Assumption 3.1(e) of \cite{Chetal2018}, which states that the factor multiplying $\Psi_{d,m}$ in the definition of the score function $\phi_{d,m}$ must be bounded in expectation, holds trivially because the factor is a constant, specifically $-1$.

\subsection{Descriptive statistics for the empirical application}\label{app:desc}

Table \ref{tab:desc} reports descriptive statistics for selected variables by treatment and mediator status. Statistically significant differences between treated and untreated respondents arise only for a limited number of characteristics, such as gender, urban residence, and highest completed grade. When comparing respondents by mediator status, gender, highest completed grade, certain ethnicity categories (`Black' and `White or Other'), the number of household members under age 18, as well as height and weight differ significantly. The latter differences may partly reflect the higher share of women among respondents who underwent a routine checkup.
}

\begin{sidewaystable}
\centering
\caption{Descriptive statistics}
\label{tab:desc}
\renewcommand{\arraystretch}{1.05}
\begin{tabular*}{\linewidth}{@{\extracolsep{\fill}}lccccccccc}
  \hline
 & overall & $D=1$ & $D=0$ & diff & $p$-val & $M=1$ & $M=0$ & diff & $p$-val \\ 
 $n$ & 1020 & 245 & 775 & & & 347 & 673 & &\\
  \hline
Female & 0.36 & 0.49 & 0.32 & 0.16 & 0 & 0.53 & 0.28 & 0.25 & 0 \\ 
  Age & 22.67 & 22.67 & 22.66 & 0.01 & 0.92 & 22.73 & 22.63 & 0.1 & 0.27 \\ 
  Ethnicity \\
  \quad \textit{Black} & 0.3 & 0.29 & 0.3 & -0.01 & 0.83 & 0.38 & 0.25 & 0.13 & 0 \\ 
  \quad \textit{Hispanic} & 0.28 & 0.26 & 0.28 & -0.03 & 0.43 & 0.31 & 0.26 & 0.04 & 0.14 \\ 
  \quad \textit{Mixed} & 0.01 & 0.01 & 0.01 & 0 & 0.79 & 0.01 & 0.01 & 0 & 0.75 \\ 
  \quad \textit{White or Other} & 0.42 & 0.44 & 0.41 & 0.03 & 0.4 & 0.31 & 0.48 & -0.18 & 0 \\ 
  Relationship/marriage \\
  \quad \textit{Not Cohabiting} & 0.62 & 0.61 & 0.62 & -0.02 & 0.65 & 0.62 & 0.62 & 0 & 0.96 \\ 
  \quad \textit{Cohabiting} & 0.2 & 0.19 & 0.2 & -0.01 & 0.64 & 0.19 & 0.2 & -0.02 & 0.53 \\ 
  \quad \textit{Married} & 0.15 & 0.17 & 0.14 & 0.03 & 0.22 & 0.15 & 0.14 & 0.01 & 0.67 \\ 
  \quad \textit{Separated/Divorced/Widowed} & 0.03 & 0.03 & 0.03 & 0 & 0.85 & 0.03 & 0.03 & 0.01 & 0.59 \\ 
  \quad \textit{Missing} & 0 & 0 & 0.01 & 0 & 0.82 & 0.01 & 0 & 0 & 0.79 \\ 
  Urban & 0.79 & 0.84 & 0.77 & 0.07 & 0.02 & 0.79 & 0.78 & 0.01 & 0.85 \\ 
  \quad \textit{Missing} & 0.02 & 0.01 & 0.02 & -0.01 & 0.34 & 0.02 & 0.02 & 0.01 & 0.48 \\ 
  HH Income & 29,830 & 30,672 & 29,564 & 1107 & 0.71 & 28,872 & 30,325 & -1453 & 0.58 \\ 
  \quad \textit{Missing} & 0.2 & 0.19 & 0.21 & -0.02 & 0.52 & 0.21 & 0.2 & 0.01 & 0.63 \\ 
  HH Size & 3.44 & 3.4 & 3.45 & -0.05 & 0.73 & 3.56 & 3.37 & 0.19 & 0.13 \\ 
  \quad \textit{Missing} & 0 & 0 & 0 & 0 & 0.32 & 0 & 0 & 0 & 0.32 \\ 
  HH Members under 18 & 0.81 & 0.76 & 0.82 & -0.06 & 0.49 & 0.95 & 0.73 & 0.21 & 0.01 \\ 
  \quad \textit{Missing} & 0 & 0 & 0 & 0 & 0.74 & 0 & 0 & 0 & 0.98 \\ 
  Biological Children & 0.57 & 0.51 & 0.59 & -0.08 & 0.2 & 0.6 & 0.56 & 0.04 & 0.56 \\  
   \hline
\end{tabular*}
\end{sidewaystable}

\begin{sidewaystable}\ContinuedFloat
\centering
\caption{Continued}
\renewcommand{\arraystretch}{1.05}
\begin{tabular*}{\linewidth}{@{\extracolsep{\fill}}lccccccccc}
  \hline
 & overall & $D=1$ & $D=0$ & diff & $p$-val & $M=1$ & $M=0$ & diff & $p$-val \\ 
 $n$ & 1020 & 245 & 775 & & & 347 & 673 & &\\
  \hline
  Highest Completed Grade & 11.44 & 12.05 & 11.25 & 0.8 & 0 & 11.71 & 11.31 & 0.4 & 0.01 \\ 
  \quad \textit{Missing} & 0.01 & 0 & 0.01 & -0.01 & 0 & 0 & 0.01 & -0.01 & 0.12 \\
  Employment \\
  \quad \textit{Employed} & 0.66 & 0.71 & 0.64 & 0.06 & 0.06 & 0.63 & 0.67 & -0.04 & 0.25 \\ 
  \quad \textit{Unemployed} & 0.09 & 0.07 & 0.09 & -0.02 & 0.2 & 0.1 & 0.08 & 0.01 & 0.44 \\ 
  \quad \textit{Out of Labor Force} & 0.25 & 0.21 & 0.26 & -0.05 & 0.11 & 0.25 & 0.24 & 0.01 & 0.77 \\ 
  \quad \textit{Military }& 0 & 0.01 & 0 & 0.01 & 0.16 & 0 & 0 & 0 & 0.67 \\ 
  \quad \textit{Missing} & 0.01 & 0.01 & 0.01 & 0 & 0.79 & 0.01 & 0 & 0.01 & 0.09 \\ 
  Working Hours (per Week) & 24.89 & 25.84 & 24.59 & 1.25 & 0.41 & 23.25 & 25.73 & -2.48 & 0.08 \\ 
  Weight (pounds) & 168 & 166 & 168 & -2 & 0.61 & 163 & 170 & -7 & 0.05 \\ 
  \quad \textit{Missing} & 0.04 & 0.06 & 0.03 & 0.03 & 0.07 & 0.04 & 0.03 & 0.01 & 0.54 \\ 
  Height (feet) & 5.41 & 5.39 & 5.41 & -0.03 & 0.75 & 5.27 & 5.48 & -0.21 & 0.01 \\ 
  \quad \textit{Missing} & 0.04 & 0.04 & 0.05 & 0 & 0.77 & 0.05 & 0.04 & 0.01 & 0.41 \\ 
  Days with 5+ drinks (per months) & 1.91 & 1.67 & 1.99 & -0.32 & 0.29 & 1.61 & 2.07 & -0.46 & 0.13 \\ 
  \quad \textit{Missing} & 0.04 & 0.05 & 0.04 & 0.01 & 0.46 & 0.04 & 0.05 & -0.01 & 0.67 \\ 
  Days of Exercise (per week) & 2.23 & 1.99 & 2.3 & -0.32 & 0.07 & 2.12 & 2.29 & -0.17 & 0.31 \\ 
  \quad \textit{Missing} & 0.09 & 0.09 & 0.09 & 0 & 0.92 & 0.07 & 0.1 & -0.02 & 0.2 \\ 
  Depressed/down \\
  \quad \textit{Never} & 0.31 & 0.28 & 0.32 & -0.03 & 0.3 & 0.29 & 0.32 & -0.02 & 0.49 \\ 
  \quad \textit{Sometimes} & 0.46 & 0.47 & 0.46 & 0 & 0.9 & 0.46 & 0.46 & -0.01 & 0.87 \\ 
  \quad \textit{Mostly} & 0.1 & 0.14 & 0.09 & 0.05 & 0.03 & 0.11 & 0.09 & 0.01 & 0.52 \\ 
  \quad \textit{Always} & 0.03 & 0.03 & 0.03 & 0 & 0.99 & 0.03 & 0.03 & 0 & 0.73 \\ 
  \quad \textit{Missing} & 0.1 & 0.09 & 0.11 & -0.02 & 0.26 & 0.12 & 0.1 & 0.02 & 0.41 \\ 
   \hline
\end{tabular*}
\par
\vspace{0.75em}
{\raggedright\scriptsize Notes: columns `overall', `$D=1$',`$D=0$',`$M=1$', and `$M=0$' report the mean of the respective variable for the full sample, the treated group, the untreated group, the mediated group, and the non-mediated group, respectively. `diff' and `$p$-val' report the mean difference (across treatment or mediator states) and the $p$-value from a two-sample t-test, respectively.\par}
\end{sidewaystable}
\pagebreak

\printbibliography

@article{sun2021estimating,
  title={Estimating dynamic treatment effects in event studies with heterogeneous treatment effects},
  author={Sun, Liyang and Abraham, Sarah},
  journal={Journal of Econometrics},
  volume={225},
  pages={175-199},
  year={2021}
}

@article{CallawayLi,
author = {Callaway, Brantly and Li, Tong},
title = {Quantile treatment effects in difference in differences models with panel data},
journal = {Quantitative Economics},
volume = {10},
pages = {1579-1618},
year = {2019}
}

@inproceedings{LewisSyrgkanis2020,
 author = {Lewis, Greg and Syrgkanis, Vasilis},
 booktitle = {Advances in Neural Information Processing Systems},
 %editor = {M. Ranzato and A. Beygelzimer and Y. Dauphin and P.S. Liang and J. Wortman Vaughan},
 pages = {22695--22707},
 publisher = {Curran Associates, Inc.},
 title = {Double/Debiased Machine Learning for Dynamic Treatment Effects},
 volume = {34},
 year = {2021}
}

@article{renson2023identifying,
  title={Identifying and estimating effects of sustained interventions under parallel trends assumptions},
  author={Renson, Audrey and Hudgens, Michael G and Keil, Alexander P and Zivich, Paul N and Aiello, Allison E},
  journal={Biometrics},
  volume={79},
  number={4},
  pages={2998-3009},
  year={2023}
}

@article{shahn2022structural,
  title={Structural nested mean models under parallel trends assumptions},
  author={Shahn, Zach and Dukes, Oliver and Shamsunder, Meghana and Richardson, David and Tchetgen Tchetgen, Eric and Robins, James M.},
  journal={arXiv preprint 2204.10291},
  year={2022}
}

@techreport{Schenk2024,
  title={Mediation analysis in difference-in-differences designs},
  author={Schenk, Timo Daniel},
  year={2024},
  type= {Working Paper},
  institution= {Aarhus University}
}

@techreport{callaway2024difference,
%  title={Difference-in-differences with a continuous treatment},
%  author={Callaway, Brantly and Goodman-Bacon, Andrew and Sant'Anna, Pedro H C},
%  year={2024},
%  institution={National Bureau of Economic Research}
%}

@article{CALLAWAY2018395,
author={Brantly Callaway and Tong Li and Tatsushi Oka},
title = {Quantile treatment effects in difference in differences models under dependence restrictions and with only two time periods},
journal = {Journal of Econometrics},
volume = {206},
pages = {395-413},
year = {2018}
}

@ARTICLE{zhang2025,
      title={Continuous difference-in-differences with double/debiased machine learning}, 
      author={Zhang, Lucas Z.},
      year={2025},
      journal = {arXiv preprint 2408.10509v2}
}

@article{Chang2020,
    author = {Chang, Neng-Chieh},
    title = "{Double/debiased machine learning for difference-in-differences models}",
    journal = {The Econometrics Journal},
    volume = {23},
    pages = {177-191},
    year = {2020},
 }

@ARTICLE{Deuchertetal2019,
  author =       {Deuchert, E and Huber, M and Schelker, M},
  title =        {Direct and indirect effects based on difference-in-differences with an application to political preferences following the {V}ietnam draft lottery},
  journal =      {Journal of Business \& Economic Statistics},
  year =         {2019},
   volume =       {37},
  pages =        {710-720}
}

@article{farbmacher2022causal,
  title={Causal mediation analysis with double machine learning},
  author={Farbmacher, Helmut and Huber, Martin and Laff{\'e}rs, Luk{\'a}{\v{s}} and Langen, Henrika and Spindler, Martin},
  journal={The Econometrics Journal},
  volume={25},
  pages={277-300},
  year={2022},
  publisher={Oxford University Press}
}

@INBOOK{Neyman1959,
  author = {J Neyman},
  title = {Optimal asymptotic tests of composite statistical hypotheses},
  booktitle={Probability and Statistics},
  year = {1959},
  pages = {416-444},
  publisher = {Wiley},
  }

@inbook{HiranoImbens2005,
    author = {Hirano, Keisuke and Imbens, Guido W.},
    year = {2004},
    publisher = {Wiley},
    title = {The Propensity Score with Continuous Treatments},
    booktitle = {Applied {B}ayesian Modeling and Causal Inference from Incomplete‐Data Perspectives},
    chapter = {7},
    pages = {73-84},
    editor = {Gelman, Andrew and Meng, Xiao-Li}
}

@ARTICLE{Lechner2010,
  author =       {M. Lechner},
  title =        {The Estimation of Causal Effects by Difference-in-Difference Methods},
  journal =      {Foundations and Trends in Econometrics},
  year =         {2011},
  volume =       {4},
  pages =        {165-224},
}

@ARTICLE{Im04,
  AUTHOR =       {G. W. Imbens},
  TITLE =        {Nonparametric estimation of average treatment effects under exogeneity: a review},
  JOURNAL =      {The Review of Economics and Statistics},
  YEAR =         {2004},
  volume =       {86},
  pages =        {4-29},
}

@article{ImaivanDyk2004,
author = {Imai, Kosuke  and van Dyk, David A },
title = {Causal Inference With General Treatment Regimes},
journal = {Journal of the American Statistical Association},
volume = {99},
pages = {854-866},
year  = {2004}
}

@article{RoHeBr00,
	author = { J M Robins and M A Hernán and B Brumback},
	title = {Marginal Structural Models and Causal Inference in Epidemiology},
	year  ={2000},
	journal = {Epidemiology},
	volume={11},
	pages = {550-560}
}

@article{Chetal2018,
author = {Chernozhukov, Victor and Chetverikov, Denis and Demirer, Mert and Duflo, Esther and Hansen, Christian and Newey, Whitney and Robins, James M.},
title = {Double/debiased machine learning for treatment and structural parameters},
journal = {The Econometrics Journal},
volume = {21},
pages = {C1-C68},
year = {2018}
}

@ARTICLE{Rubin74,
	author    = "D. B. Rubin",
	title	  = "Estimating Causal Effects of Treatments in Randomized and Nonrandomized Studies",
	year      = 1974,
	journal   = "Journal of Educational Psychology",
	volume    = 66,
	pages     = "688-701"}

@ARTICLE{Angrist+96,
	author    = "J. Angrist and G. W. Imbens and D. B. Rubin",
	title	  = "Identification of Causal Effects using Instrumental Variables",
	year      = 1996,
	journal   = "Journal of the American Statistical Association",
	volume    = 91,
	pages     = "444-472 (with discussion)"
}

@article{Ashenfelter78,
 author = {Orley Ashenfelter},
 title = {Estimating the Effect of Training Programs on Earnings},
 year = {1978},
 journal = {The Review of Economics and Statistics},
 volume = {60},
 number = {1},
 pages = {47--57}
}

@ARTICLE{Zimmert2018,
  author =       {Michael Zimmert},
  title =        {Efficient Difference-in-Differences Estimation with High-Dimensional Common Trend Confounding},
  journal =      {arXiv preprint 1809.01643},
  year =         {2020},
}

@article{cha19,
    Author = {de Chaisemartin, C. and D'Haultf{\oe}uille, X. },
    Title = {Two-way fixed effects estimators with heterogeneous treatment effects},
    Journal = {American Economic Review},
        Volume = {110},
    pages = {2964-2996},
    Year = 2020,
}

@techreport{BodoryHuber2018,
  author =       {Hugo Bodory and Martin Huber},
  title =        {The causalweight package for causal inference in {R}},
  institution =  {University of Fribourg},
  type =         {SES Working Paper},
  number =       {493},
  year =         {2018},
}

@article { Tranetal2019,
	author = "Linh Tran and Constantin Yiannoutsos and Kara Wools-Kaloustian and Abraham Siika and Mark van der Laan and Maya Petersen",
	title = "Double Robust Efficient Estimators of Longitudinal Treatment Effects: Comparative Performance in Simulations and a Case Study",
	journal = "The International Journal of Biostatistics",
	year = "2019",
	publisher = "De Gruyter",
	address = "Berlin, Boston",
	volume = "15",
	number = "2",
	pages = "1-27",
}

@Article{caetano2022difference,
      title={Difference in Differences with Time-Varying Covariates},
      author={Carolina Caetano and Brantly Callaway and Stroud Payne and Hugo Sant'Anna Rodrigues},
      year={2022},
      journal={arXiv preprint 2202.02903}
}

@Manual{Rcore2025,
    title = {R: A Language and Environment for Statistical Computing},
  author = {{R Core Team}},
 organization = {R Foundation for Statistical Computing},
 %address = {Vienna, Austria},
 year = {2025},
 url = {https://www.R-project.org/}
   }

@ARTICLE{SantAnnaZhao2018,
  author = {Pedro H. C. Sant'Anna and Jun B. Zhao},
  title = {Doubly Robust Difference-in-Differences Estimators},
  journal = {Journal of Econometrics},
  volume={219},
  pages={101-122},
  year={2020}
}

@article{Hong2013,
	author = {Hong, Seung-Hyun},
	title = {Measuring the effect of {N}apster on recorded music sales: difference-in-differences estimates under compositional changes},
	journal = {Journal of Applied Econometrics},
	volume = {28},
	pages = {297-324},
	year = {2013}
}

@ARTICLE{GoodmanBacon2018,
  author =       {Goodman-Bacon, A},
  title =        {Difference-in-differences with variation in treatment timing},
journal = {Journal of Econometrics},
volume = {225},
pages = {254-277},
year = {2021}
}

@book{huber2023causal,
  title={Causal analysis: Impact evaluation and Causal Machine Learning with applications in {R}},
  author={Huber, Martin},
  year={2023},
  publisher={MIT Press}
}

@ARTICLE{borusyak2024revisiting,
    author = {Borusyak, Kirill and Jaravel, Xavier and Spiess, Jann},
    title = {Revisiting Event-Study Designs: Robust and Efficient Estimation},
    journal = {The Review of Economic Studies},
    volume = {91},
    number = {6},
    pages = {3253-3285},
    year = {2024},
}

@article{CallawaySantAnna2018,
title = {Difference-in-Differences with multiple time periods},
journal = {Journal of Econometrics},
volume = {225},
pages = {200-230},
year = {2021},
author = {Brantly Callaway and Pedro H. C. Sant'Anna},
}

@ARTICLE{abadie2005,
  author =       {Alberto Abadie},
  title =        {Semiparametric Difference-in-Differences Estimators},
  journal =      {The Review of Economic Studies},
  year =         {2005},
  volume =       {72},
  pages =        {1-19},
}

@INBOOK{Huber2019,
	author =       {M Huber},
	title =        {Mediation Analysis},
	booktitle =      {Handbook of Labor, Human Resources and Population Economics},
	year =         {2021},
              editor = {K F Zimmermann},
             publisher = {Springer Nature Switzerland AG}
}

@ARTICLE{Neyman23,
	author    = "J. Neyman",
	title	  = "On the Application of Probability Theory to Agricultural Experiments. {E}ssay on Principles.",
	year      = 1923,
	journal   = "Statistical Science",
	volume    = "Reprint, 5",
	pages     = "463-480"}

@ARTICLE{Rubin80,
	author    = "D. B. Rubin",
	title	  = "Comment on ``{Randomization} analysis of experimental data: {The} {Fisher} randomization test'' by {D. Basu}",
	year      = 1980,
	journal   = "Journal of the American Statistical Association",
	volume    = 75,
	pages     = "591-593"}

@book{Cox58,
	author 	  = "D. Cox",
	publisher = "Wiley",
	address   = "New York",
	title     = "Planning of Experiments",
	year      = "1958"}

@book{Pearl00,
	author 	  = "J. Pearl",
	publisher = "Cambridge University Press",
	address   = "Cambridge",
	title     = "Causality: Models, Reasoning, and Inference",
	year      = "2000"}

@ARTICLE{RoRo95,
  AUTHOR =       {J M Robins and Andrea Rotnitzky},
  TITLE =        {Semiparametric Efficiency in Multivariate Regression Models with Missing Data},
  JOURNAL =      {Journal of the American Statistical Association},
  YEAR =         {1995},
  volume =       {90},
  pages =        {122-129},
}

@article{Robins+94,
    author = {James M. Robins and Andrea Rotnitzky and Lue Ping Zhao},
    title = {Estimation of Regression Coefficients When Some Regressors are not Always Observed},
    journal = {Journal of the American Statistical Association},
    year = {1994},
    volume = {89},
    pages = {846--866}
}

@article{bradic2024high,
  title={High-dimensional inference for dynamic treatment effects},
  author={Bradic, Jelena and Ji, Weijie and Zhang, Yuqian},
  journal={The Annals of Statistics},
  volume={52},
  pages={415-440},
  year={2024},
  publisher={Institute of Mathematical Statistics}
}

@article{bodory2022evaluating,
  title={Evaluating (weighted) dynamic treatment effects by double machine learning},
  author={Bodory, Hugo and Huber, Martin and Laff{\'e}rs, Luk{\'a}{\v{s}}},
  journal={The Econometrics Journal},
  volume={25},
  pages={628-648},
  year={2022},
  publisher={Oxford University Press}
}

@ARTICLE{VanderLaanRubin2006,
  author =       {van der Laan, M. and Rubin, D.},
  title =        {Targeted Maximum Likelihood Learning},
  journal =      {The International Journal of Biostatistics},
  year =         {2006},
  volume =       {2},
  pages =        {1-38},
}

@ARTICLE{Ro86,
  AUTHOR =       {J M Robins},
  TITLE =        {A new approach to causal inference in mortality studies with sustained exposure periods -- application to control of the healthy worker survivor effect},
  JOURNAL =      {Mathematical Modelling},
  YEAR =         {1986},
  volume =       {7},
  pages =        {1393-1512},
}

@article{frangakis2002principal,
  title={Principal stratification in causal inference},
  author={Frangakis, Constantine E and Rubin, Donald B},
  journal={Biometrics},
  volume={58},
  number={1},
  pages={21--29},
  year={2002},
  publisher={Oxford University Press}
}

@INPROCEEDINGS{Pearl01,
  AUTHOR =       {J Pearl},
  TITLE =        {Direct and indirect effects},
  BOOKTITLE =    {Proceedings of the Seventeenth Conference on Uncertainty in Artificial Intelligence},
  YEAR =         {2001},
  pages =        {411-420},
  address =      {San Francisco},
  publisher =    {Morgan Kaufmann},
}

@ARTICLE{RoGr92,
  AUTHOR =       {J M Robins and Sander Greenland},
  TITLE =        {Identifiability and Exchangeability for Direct and Indirect Effects},
  JOURNAL =      {Epidemiology},
  YEAR =         {1992},
  volume =       {3},
  pages =        {143-155},
}

@ARTICLE{Bellonietal2014,
  author =       {Alexandre Belloni and Victor Chernozhukov and Christian Hansen},
  title =        {Inference on Treatment Effects after Selection among High-Dimensional Controls},
  journal =      {The Review of Economic Studies},
  year =         {2014},
  volume =       {81},
  pages =        {608-650},
}

@Book{Snow1855,
  Title = {On the Mode of Communication of Cholera},
  Author = {Snow, John},
  Year = {1855},
  Publisher = {John Churchill}
}

@report{blackwell2022difference,
  title={Difference-in-differences Designs for Controlled Direct Effects},
  author={Blackwell, Matthew and Glynn, Adam and Hilbig, Hanno and Phillips, Connor Halloran},
  type = {Working paper},
  institution = {Harvard University},
  year={2022}
}

@article{haddad2024,
      title={Difference-in-Differences with Time-varying Continuous Treatments using Double/Debiased Machine Learning}, 
      author={Michel F. C. Haddad and Martin Huber and Lucas Z. Zhang},
      year={2024},
      journal={arXiv preprint 2410.21105}
}

@article{dechaisemartin2023differenceindifferences,
  title={Difference-in-Differences for Continuous Treatments and Instruments with Stayers},
  author={de Chaisemartin, Cl{\'e}ment and D'Haultf{\oe}uille, Xavier and Pasquier, F{\'e}lix and  Sow, Doulo and Vazquez-Bare, Gonzalo},
  journal={arXiv preprint 2201.06898},
  year={2024}
}

@article{fan1996estimation,
 author = {Jianqing Fan and Qiwei Yao and Howell Tong},
 journal = {Biometrika},
 number = {1},
 pages = {189--206},
 publisher = {[Oxford University Press, Biometrika Trust]},
 title = {Estimation of Conditional Densities and Sensitivity Measures in Nonlinear Dynamical Systems},
 volume = {83},
 year = {1996}
}

@dataset{nlsy97,
  author    = {{Bureau of Labor Statistics, U.S. Department of Labor}},
  title     = {{National Longitudinal Survey of Youth} 1997 cohort, 1997--2017 (rounds 1--18)},
  publisher = {Center for Human Resource Research (CHRR), The Ohio State University},
  year      = {2019},
  url       = {https://www.nlsinfo.org/content/cohorts/nlsy97}
}

@article{deng2026did,
    author = {Deng, Yuhao and Wei, Haoyu and Ouyang, Zhongzhe},
    title = {Difference-in-differences with a mediator},
    journal = {arXiv preprint 2604.24049}, 
    year = {2026},
}

@article{ban2026did,
    author = {Ban, Kyunghoon and Chen, Zhengrun and K\'{e}dagni, D\'{e}sir\'{e}},
    title = {Difference-in-Differences Models in the Presence of Time-Varying Mediators},
    journal = {arXiv preprint 2608.18375},
    year = {2026},
}

@article{hsia2025mediation,
    author = {Hsia, Pei Hsuan and Tai, An Shun and Kao, Chu Lan Michael and Lin, Yu Hsuan and Lin, Sheng Hsuan},
    title = {Causal Mediation Analysis for Difference-in-Difference Design and Panel Data},
    journal = {Epidemiologic Methods},
    volume = {14},
    number = {1},
    pages = {20240025},
    year = {2025}
}

@article{andersen2026,
author = {Andersen, Robert and Holm, Anders},
title = {Uncovering the Mechanisms of Policy Interventions: A Difference-In-Differences Framework for Mediation Analysis},
journal = {Human Resource Management Journal},
year = {2026},
}
\end{document}

A DR expression for the counterfactual, which under the identifying assumptions is numerically equivalent to the fourth line of equation \eqref{mediationcounterfactual} and satisfies Neyman orthogonality (as formally shown in Appendix \ref{Neymanrepeated}), is given in Proposition \ref{prop:DRidentificationcount}:
\begin{proposition}\label{prop:DRidentificationcount}
Under Assumptions \ref{ass1}--\ref{ass4}, holding for any $m$ in the support of $M(d)$ conditional on $D=d$ and $T=1$, the counterfactual $E[Y_1(d',M(d))|D=d, T=1]$ is identified by the following expression, which satisfies Neyman orthogonality:
\begin{align}\label{DiDidentDRcounterfac}
&E[Y_1(d',M(d))|D=d, T=1]=\sum_{m \in \mathcal{M}} \Pr(M=m|D=d,T=1)\notag\\
&\times E\left[  \frac{I\{D=d\}\cdot I\{M=m\}\cdot T \cdot [\mu_{d,m}(0,X)+\mu_{d',m}(1,X)-\mu_{d',m}(0,X)]}{\Pi_{d,m,1}}\right.\notag\\
&+ \frac{(Y_T-\mu_{D,M}(T,X))\cdot \rho_{d,m,1}(X)}{\Pi_{d,m,1}} \cdot \left\{\frac{I\{D=d\}\cdot I\{M=m\}\cdot (1-T)}{ \rho_{d,m,0}(X)}\right.\\
&+ \left.\left.\frac{I\{D=d'\}\cdot I\{M=m\}\cdot  T}{ \rho_{d',m,1}(X)} - \frac{I\{D=d'\}\cdot I\{M=m\}\cdot (1-T)}{ \rho_{d',m,0}(X)} \right\}   \right].\notag
\end{align}
\end{proposition}

Rather than conditioning on each specific mediator value $m$ and averaging over mediator values among the treated in the post-treatment period using the weights $\Pr(M=m \mid D=d, T=1)$, we may instead average directly over the distribution of the realized mediator $M$ among the treated in the post-treatment period. That is, instead of computing for any conditional mean outcome
\begin{align*}
&E\left[
  \frac{I\{D=d\} \cdot I\{M=m\} \cdot T}{\Pi_{d,m,1}} \cdot
  \mu_{d',m}(t,X)  \right] \cdot
\Pr(M=m \mid D=d, T=1)\\
=& E\left[
  \mu_{d',m}(t,X) | D=d, M=m, T=1 \right] \cdot
\Pr(M=m \mid D=d, T=1),
\end{align*}
we may consider the equivalent expression
\begin{align*}
&E\left[
 \mu_{d',M}(t,X) \big| D=d, T=1
\right]=E\left[
  \frac{I\{D=d\}  \cdot T}{P_{d,1}}\cdot \mu_{d',M}(t,X) \right],
\end{align*}
which integrates over the observed distribution of $M$ given $D=d$ and $T=1$. Here, $P_{d,t}=\Pr(D=d,T=t)$ denotes the joint probability of treatment and period. Analogously, for any debiasing term based on inverse weighting by the propensity score, instead of
\[
E\!\left[
  \frac{(Y_T - \mu_{D,M}(T,X)) \cdot \rho_{d,m,1}(X)}{\Pi_{d,m,1}}
  \cdot
  \frac{I\{D=d'\} \cdot I\{M=m\} \cdot I\{T=t\}}{\rho_{d',m,t}(X)}
\right]\cdot
\Pr(M=m \mid D=d, T=1),
\]
we may consider
\[
E\!\left[
  \frac{(Y_T - \mu_{D,M}(T,X)) \cdot \pi_{d,1}(M,X)}{P_{d,1}}
  \cdot
  \frac{I\{D=d'\} \cdot I\{T=t\}}{\pi_{d',t}(M,X)}
\right],
\] 
where $\pi_{d,t}(M,X) = \Pr(D=d, T=t \mid M, X)$ is the joint propensity score of treatment and period given $M$ and $X$. Therefore, the identification result in equation \eqref{DiDidentDRcounterfac} is equivalent to the following expression, which avoids conditional mediator probabilities:
\begin{align}
E[Y_1(d',M(d))|D=d]&=E\left[ \frac{I\{D=d\}\cdot T \cdot [\mu_{d,M}(0,X)+\mu_{d',M}(1,X)-\mu_{d',M}(0,X)]}{P_{d,1}}\right.\notag\\
&+ \frac{(Y_T-\mu_{D,M}(T,X))\cdot \pi_{d,1}(M,X)}{P_{d,1}} \cdot \left\{\frac{I\{D=d\} \cdot (1-T)}{ \pi_{d,0}(M,X)}\right.\label{noprobidentification2}\\
&+ \left.\left.\frac{I\{D=d'\} \cdot  T}{ \pi_{d',1}(M,X)} - \frac{I\{D=d'\} \cdot (1-T)}{ \pi_{d',0}(M,X)} \right\} \right].\notag
\end{align}

\begin{align}\label{DiDidentDRcounterfacpanel}
&E[Y_1(d',M(d))|D=d]=\sum_{m \in \mathcal{M}} \Pr(M=m|D=d)\times E\left[  \frac{I\{D=d\}\cdot I\{M=m\} \cdot [Y_0+\mu_{d',m}(X)]}{\Pi_{d,m}}\right.\notag\\
&+ \left. \frac{I\{D=d'\}\cdot I\{M=m\}\cdot[Y_1-Y_0-\mu_{D,M}(X)]\cdot \rho_{d,m}(X)}{\Pi_{d,m}\cdot \rho_{d',m}(X)}    \right].
\end{align}
\end{proposition}
In analogy to \eqref{noprobidentification2} for repeated cross sections, an alternative (but numerically equivalent) DR expression to that in Proposition~\ref{prop:DRidentificationcountpanel}, which, however, avoids conditional mediator probabilities, is given by:
\begin{align}\label{DiDidentDRcounterfacpanel2}
&E[Y_1(d',M(d))|D=d]\notag\\
&= E\left[ \frac{I\{D=d\}\cdot [Y_0+\mu_{d',M}(X)]}{\Pr(D=d)}+ \frac{I\{D=d'\}\cdot[Y_1-Y_0-\mu_{D,M}(X)]\cdot \pi_{d}(M,X)}{\Pr(D=d)\cdot \pi_{d'}(M,X)}  \right],
\end{align}
where $\pi_{d}(M,X)=\Pr(D=d|M,X)$.

Next, we consider the identification of counterfactual $E[Y_1(d',M(d))|D=d]$ based on the DR expression \eqref{DiDidentDRcounterfac} and note that the latter corresponds to 
\begin{align}\label{psi_counter}
E[Y_1(d',M(d))|D=d] =  \sum_{m \in \mathcal{M}} \Pr(M=m|D=d,T=1)  \cdot (\Psi_{d,m,0} + \Psi_{d',m,1} - \Psi_{d',m,0}).
\end{align}
Therefore, given the previous results concerning identification and Neyman orthogonality of $\Psi_{d',m',t}$ and the fact that the Gateaux derivative of $\Pr(M=m|D=d,T=1)$ in the direction of $[\eta - \eta^0]$ is trivially zero, too, we have that Assumptions 3.1(a)--(e) of \cite{Chetal2018} hold with respect to expression  \eqref{psi_counter}. 

Next, we consider the identification of counterfactual $E[Y_1(d',M(d))|D=d]$ based on the DR expression \eqref{DiDidentDRcounterfacpanel} and note that the latter corresponds to 
\begin{align}\label{psi_counterpanel}
E[Y_1(d',M(d))|D=d] =  \sum_{m \in \mathcal{M}} \Pr(M=m|D=d)  \cdot (E[Y_0|D=d,M=m] + \Psi_{d',m'}).
\end{align}
Therefore, given the previous results concerning identification and Neyman orthogonality of $\Psi_{d',m'}$  and the fact that the Gateaux derivatives of $E[Y_0|D=d,M=m]$ and $\Pr(M=m|D=d)$ in the direction of $[\eta - \eta^0]$ are trivially zero, too, we have that Assumptions 3.1(a)--(e) of \cite{Chetal2018} hold with respect to expression  \eqref{psi_counterpanel}.